%% file: main.tex
\documentclass[aps,pra,nobalancelastpage,superscriptaddress,reprint, twocolumns]{revtex4-2}

\usepackage{standalone, tensor, amsmath,amsfonts,amssymb,amsthm,bbm,bm, braket}
\usepackage[T1]{fontenc}
\usepackage[inline]{enumitem}
\usepackage{scalerel, physics}
\usepackage{wasysym}
\usepackage{comment}
\usepackage{enumerate}
\usepackage{graphicx}
\usepackage{mathtools}
\usepackage{soul}
\usepackage{xparse}
\usepackage{dsfont}
\usepackage{ifthen}
\usepackage{tensor}
\usepackage{tikz}

\graphicspath{{./}{./images/}}

\usepackage{tikz-network}
\usepackage{simplewick} 
\usetikzlibrary{patterns,decorations.pathreplacing}
\makeatletter
\long\def\@makecaption#1#2{
  \par
  \vskip\abovecaptionskip
  \begingroup
    \small\rmfamily
    \samepage
    \flushing
    \let\footnote\@footnotemark@gobble
    \noindent\@make@capt@title{#1}{#2}\par
  \endgroup
  \vskip\belowcaptionskip
}
\makeatother
\theoremstyle{plain}        

\theoremstyle{plain}

\usepackage{color}
\definecolor{OliveGreen}{RGB}{85,107,47}
\definecolor{NavyBlue}{RGB}{0,0,128}
\definecolor{ReminderOrange}{RGB}{220,120,20}

\setstcolor{ReminderOrange}

\usepackage{tensor}

\theoremstyle{plain}

\theoremstyle{plain}

\newtheorem{conjecture}{Conjecture}

\usepackage[colorlinks,bookmarks=false,citecolor=NavyBlue,linkcolor=OliveGreen,urlcolor=blue]{hyperref}

\newcommand{\mcirc}{\mathbin{\scalerel*{\fullmoon}{G}}}
\newcommand{\msquare}{\mathord{\scalerel*{\square}{G}}}
\newcommand{\mcircf}{\mathbin{\scalerel*{\newmoon}{G}}}

\newcommand{\I}{\mathds{1}}

\newcommand{\prlsection}[1]{{\em {#1}.---}}
\newcommand{\be}{\begin{equation}}
\newcommand{\ee}{\end{equation}}
\newcommand{\ba}{\begin{aligned}}
\newcommand{\ea}{\end{aligned}}
\newcommand{\bw}{\begin{widetext}}
\newcommand{\ew}{\end{widetext}}

\input{tikz_lib}

\newcommand{\rtmequationscale}{0.9}

\begin{document}
	\title{Low Rank Structure of the Reduced Transition Matrix}
    \author{Cathy Li}
     \affiliation{Department of Physics, Virginia Tech, Blacksburg, Virginia 24061, USA}
	\author{Bruno Bertini}
	\author{Katja Klobas}
	\affiliation{School of Physics and Astronomy, University of Birmingham, Birmingham B15 2TT, United Kingdom}
    \author{Tianci Zhou}
	\affiliation{Department of Physics, Virginia Tech, Blacksburg, Virginia 24061, USA}

\begin{abstract}
The influence-matrix formalism provides an alternative route to the classical simulation of quantum dynamics. Because influence matrices retain information only about the effective bath seen by local observables, they are expected to be easier to simulate than the full wavefunction. Recent work, however, has shown that they carry strong temporal correlations even in maximally chaotic systems, making them difficult to represent efficiently. Here we show that the \emph{reduced transition matrix}, a suitable combination of influence matrices that directly determines local expectation values, can nevertheless be efficiently approximated. We first show that the truncation error is controlled by its singular-value spectrum, which naturally motivates a low-rank approximation. We then prove that, for chaotic dual-unitary circuits, the associated entropy grows at most logarithmically in time. Our conclusions follow from exact results for random dual-unitary circuits and are further supported by numerical results for fixed instances of both dual-unitary and random circuits.
\end{abstract}	

\maketitle

\prlsection{Introduction} The fundamental challenge of non-equilibrium quantum many-body dynamics is that the full time evolution of a quantum state is generically hard to simulate classically~\cite{vidal2003efficient,schollwock2011density}. This remains true even for `simple', low-entangled initial states that admit an efficient matrix product state (MPS) representation~\cite{cirac2021matrix}. At short times, when the entanglement is still low, one can continue to approximate the time-evolved state efficiently by an MPS~\cite{vidal2004efficient,daley2004time,white2004real,schollwock2011density}. As time increases, however, the growth of entanglement causes both memory and runtime costs to increase exponentially preventing any efficient simulation~\cite{potter2022entanglement,fisher2022random, bertini2025exactly}. 

\begin{figure}[t]
\centering
\includegraphics[width=\columnwidth]{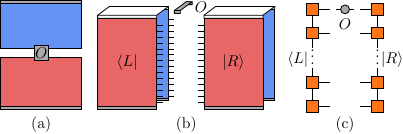}
  \caption{The influence matrices. (a) Schematic representation of a one-point function, where the forward and backward time sheets are shown in red and blue, respectively. (b) Folded representation of the one-point function, the left $\langle L |$ and right $|R \rangle$ influence matrices. (c) Tensor-network representation of the influence matrices. Partial contraction of the lower legs defines the reduced transition matrix. \label{fig:infl_mat}}
\end{figure}

The `folding algorithm' was proposed as a practical workaround in cases where one is interested only in expectation values of local observables~\cite{banuls2009matrix,muellerhermes2012tensor,hastings2015connecting,lerose2021influence,lerose2023overcoming}. Instead of simulating the full wavefunction, this algorithm targets the `influence matrices', denoted here by $|L\rangle$ and $|R\rangle$, see Fig.~\ref{fig:infl_mat}~(a--b). These objects encode the action exerted on the observable by all degrees of freedom away from its support, in a way that is conceptually similar to the Feynman-Vernon influence functional theory~\cite{feynman1963the,leggett1987dynamics}.

The influence matrices are expected to be more economical to store because it retains less information than the full many-body wave function by encoding only local properties. A central question is whether they actually have low entanglement~\cite{hastings2015connecting,lerose2021influence,lerose2023overcoming,foligno2023temporal}. The answer is provably affirmative in several important settings, including free fermions~\cite{lerose2021scaling}, certain interacting integrable systems~\cite{klobas2021exact, klobas2021exactrelaxation, giudice2022temporal}, dual-unitary circuits with suitably chosen initial states~\cite{bertini2019entanglement, piroli2020exact, foligno2025nonequilibrium}, and several quantum impurity problems~\cite{leggett1987dynamics,segal2010numerically,segal2011nonequilibrium,magazzu2022feynman,thoenniss2023nonequilibrium,thoenniss2023efficient,ng2023realtime,chen2024grassmann,chen2024realtime,park2024continuous,nayak2025steadystate,sonner2025semigroup}. However, previous work by Foligno {\it et al}~\cite{foligno2023temporal} showed that the von Neumann entropy $S_1$ of the influence matrices follows a volume law in chaotic quantum circuits. This finding is surprising because chaotic systems are expected to generate an almost memoryless Markovian environment, implying that influence matrices should be simple to simulate. Yet, the volume law entanglement suggests that a direct MPS encoding of the influence matrices is impossible for large enough times. While Ref.~\cite{vilkoviskiy2025temporal} (see also Ref.~\cite{odonovan2026diagnosing}) recently argued that such volume law entanglement can be removed by performing an appropriate coarse graining procedure, the link between coarse grained influence matrices and the computation of (bulk) observables is still not clear. 

We address this tension by showing that the entanglement of the normalized influence matrix is \emph{not} the relevant complexity measure for the folding algorithm. The usual rule of thumb that $\exp(S(A))$ sets the bond dimension across a bipartition $A|\bar A$ applies to $L^2$ normalized wavefunctions with $\|\psi(t) \rangle \|= 1$~\cite{vidal2003efficient,schuch2008entropy,schollwock2011density}. However, for influence matrices, the natural normalization is ${\langle L | R \rangle=1}$ and one-point functions are written as $\langle L | O | R \rangle$, where $O$ is a local operator. Since the normalized vectors ${|L\rangle}/\|\smash{|L\rangle}\|$ and ${|R\rangle}/\|\smash{|R\rangle}\|$ are never used, their entanglement should not play a role in the approximation cost. Instead, the appropriate compression strategy must be directly tied to the bilinear structure of ${\langle L | R \rangle=1}$ and $\langle L | O | R \rangle$. This argument is in line with many practical implementations of the folding algorithm~\cite{hastings2015connecting,carignano2023temporal,carignano2026itransverse,carignano2025overcoming}, which compress $|L\rangle$ and $|R\rangle$ jointly, although separate compression schemes have also been adopted~\cite{lerose2021scaling, yadalam2026process}.

To make the argument precise, we first show that additive errors in observables, such as $\langle L | O | R \rangle$, are controlled by the singular-value decay of an appropriate `reduced transition matrix' (RTM)~\cite{hastings2015connecting,carignano2023temporal}. We then use the von Neumann entropy of the normalized singular values to quantify how rapidly this spectrum decays. We find that, in random dual-unitary circuits with generic dimerized initial states, this entropy grows only logarithmically, $S_1 \lesssim \ln t$, while the temporal entanglement is linear. We numerically verify that this behavior persists in disorder-free dual-unitary circuits, and we observe similar scaling in random unitary circuits. Therefore, despite the volume law entanglement of the influence matrices, one can still hope to compute local correlation functions of slowly decaying operators. 

\begin{figure}[t]
\centering
\includegraphics[width=\columnwidth]{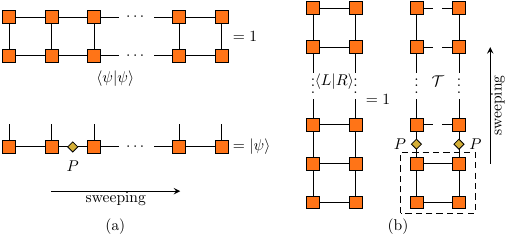}
\caption{The standard MPS compression and joint compression based on the reduced transition matrix. The diamond $P$ represents the projection required to truncate the singular value spectrum. (a) The standard truncation procedure applied to each link of an $L^2$ normalized MPS. (b) The equivalent computation through the influence matrices and the joint truncation procedure of the reduced transition matrix. \label{fig:compression}}
\end{figure}

\prlsection{Low-rank approximation of the RTM} We consider a quantum system prepared in a low entangled initial state $|\Psi_0\rangle$ and evolved in time by a unitary operator $\mathbb U(t)$, so that $|\Psi_t\rangle=\mathbb U(t)|\Psi_0\rangle$ is the state at time $t$. For the one point function of a local operator $O$, we can write
\begin{equation}
\langle  O({t}) \rangle =\langle \Psi_t | O | \Psi_t \rangle = \langle L | O | R \rangle,
\label{eq:directtime}
\end{equation}
in terms of the influence matrices $\langle L |$ and $|R\rangle$ (Fig.~\ref{fig:infl_mat}~(b)). Suppose the influence matrices are represented as MPS, see Fig.~\ref{fig:infl_mat}~(c). A standard MPS compression would sweep independently through each bond of the $L^2$ normalized wavefunction, perform a singular value decomposition (SVD), and discard the small singular values or equivalently insert low-rank projectors (Fig.~\ref{fig:compression}~(a)~\cite{schollwock2011density}).

As discussed above, this independent sweep is not aligned with the error control we need for approximating observables such as $\langle L | O | R \rangle$. The RTM, $\mathcal{T}_{t_0}$, is introduced to solve this problem by implementing a joint compression. It is obtained by contracting the first $t_0$ legs (from the bottom) of the two influence matrices
\begin{equation}
    \mathcal{T}_{t_0}  = \operatorname{tr}_{t_0} ( | R \rangle \langle L |) = \rtmtensorjcompnodiamondconn[\rtmequationscale]{3}{2}.
\end{equation}
Let $i$ and $j$ be the indices of free uncontracted legs, the one-point function can be written as $\sum_{i,j} (\mathcal{T}_{t_0})_{ij} O_{ji} = \operatorname{tr}( \mathcal{T}_{t_0} O)$.

A low-rank approximation of $\mathcal{T}_{t_0}$ achieves two goals at once: (i) it bounds the additive error of the observable, and (ii) it provides the projectors needed to reduce the bond dimensions of both $|L\rangle$ and $|R\rangle$. Concretely, we seek a rank $\chi$ matrix $\mathcal{T}_{t_0, \chi}$ such that the approximation error is bounded as
\begin{equation}
\label{eq:T-T_low_rank}
|\operatorname{tr}( (\mathcal{T}_{t_0} - \mathcal{T}_{{t_0},\chi} )O ) | \le \left\lVert O \right\rVert_{\infty} \left\lVert \mathcal{T}_{t_0} - \mathcal{T}_{{t_0},\chi} \right\rVert_{1} .
\end{equation}
Here $\left\lVert \, \right\rVert_p$ denotes the Schatten $p$ norm $\left\lVert M \right\rVert_{p} = \operatorname{tr}( (M M^\dagger)^{p/2})^{1/p}$. Relevant examples are $p = 1$, which gives the trace norm, and $p = \infty$, which gives the largest singular value. For local observables, such as Pauli operators, $\left\lVert O \right\rVert_{\infty}$ is of order $1$. 

\begin{figure*}[t]
  \includegraphics[width=\textwidth]{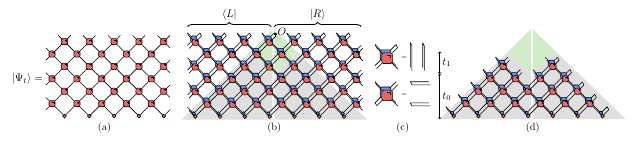}
\caption{Diagrammatic representation of the circuit geometry and the reduced transition matrix construction. We consider an initial state of the form $|\Psi_0\rangle=\prod_{i=1}^L |\psi_0\rangle$, where $|\psi_0\rangle$ is represented by a red circle with legs pointing up and, analogously, $\langle \psi_0|$ by a blue circle with legs pointing down. Red squares denote local gates and blue squares their Hermitian conjugates. (a) State $|\Psi_t\rangle$ after $t$ layers of a brickwork circuit. (b) Folded circuit representation of a one-point function in terms of the influence matrices $\langle L|$ and $|R\rangle$. Gates outside the light cones can be removed by unitarity; gates in the green region can be removed by dual unitarity without affecting the singular spectrum. (c) Local cancellation identities implied by unitarity and dual unitarity. (d) Reduced transition matrix after the simplifications implied by dual unitarity ($t_1=(t-t_0)/2$). \label{fig:Cdiag}}
\end{figure*}

By the Schmidt-Mirsky theorem~\cite{mirsky1960symmetric}, truncating the singular-value spectrum yields the optimal low-rank approximation in any unitarily invariant norm, including Schatten $p$ norms. Namely, we have 
\begin{equation}
\left\lVert \mathcal{T}_{t_0} - \mathcal{T}_{{t_0},\chi}  \right\rVert_{1} \geq \sum_{n>\chi} \tau_n,
\end{equation}
where $\{\tau_n\}$ are the singular values of $\mathcal T_{t_0}$, and the optimal approximation (the equality) is obtained by keeping the leading $\chi$ singular values.  

In practice, we iteratively contract the lower legs of the influence matrices from the bottom. Viewing the remaining $t-t_0$ left and right legs as rows and columns, we perform an SVD and retain only the largest $\chi$ singular values, which is equivalent to inserting rank-$\chi$ projectors on the left and right legs. We then place those projectors on the corresponding bonds of both $|L\rangle$ and $|R\rangle$ to compress the two influence matrices simultaneously. Repeating this procedure bond by bond from bottom to top defines a joint sweep (Fig.~\ref{fig:compression}~(b)) that reduces the bond dimensions of both $|L\rangle$ and $|R\rangle$. In this process, we introduce an error $\epsilon_i$ at the $i$th step
\begin{equation}
\label{eq:each_step_error}
\begin{aligned}[t]
\left\lVert
  \rtmtensorjcompconn[1]{3}{2}
  -
  \rtmtensorj[1]{3}
\right\rVert_{1}
\le
\epsilon_{j}
\!\implies\!
\left\lVert
  \rtmtensorjcompconn[1]{3}{5}
  -
  \rtmtensorjconn[1]{3}{5}
\right\rVert_{1}
\le
\epsilon_{j}.
\end{aligned}
\end{equation}
In the second of Eq.~\eqref{eq:each_step_error} we use the fact that the error cannot increase after taking a partial trace, so the bound also holds when all legs are connected but one. Summing and telescoping the resulting inequalities, we find that the accumulated error is at most 
\begin{equation}
\left\lVert
  \rtmtensorconn[\rtmequationscale]{0}{5}
  -
  \rtmtensorconn[\rtmequationscale]{5}{5}
\right\rVert_{1}
\le
\sum_{i=1}^{t-1} \epsilon_{i},
\end{equation}
and this is precisely the error of the low-rank approximation in Eq.~\eqref{eq:T-T_low_rank}. Choosing $\sum_i \epsilon_i < \epsilon$ keeps the target observable error under control.

The truncation error is controlled by the discarded tail of the singular spectra of the RTMs encountered during the sweep. A rigorous bound would require two ingredients that we do not delve into here. First, one should analyze the singular value decay of each intermediate RTM appearing in the first inequality of Eq.~\eqref{eq:each_step_error} after accounting for the projectors inserted at previous steps. Second, as in standard MPS analyses, one would ideally use entanglement measures that are more sensitive to the tail of the spectrum, such as R\'enyi entropies with index $n<1$. Both tasks are difficult in practice. We therefore follow the heuristic diagnostic commonly used to assess numerical compression schemes~\cite{hastings2015connecting,carignano2023temporal} and characterize the singular value distribution of $\mathcal{T}_{t_0}$ through a von Neumann entropy. We introduce the density operator 
\begin{equation}
\rho_{t_0} = \frac{\mathcal{T}^\dag_{t_0}\mathcal{T}_{t_0}}{\operatorname{tr}\smash{[\mathcal{T}^\dag_{t_0}\mathcal{T}_{t_0}]}}, 
\end{equation}
whose eigenvalues $\{\lambda_n\}$ are related to the singular values of $\mathcal T_{t_0}$ by $\lambda_n = {\tau^2_n}/{\sum_m \tau^2_m}$. A scaling slower than a volume law in $t$ of 
\begin{equation}
\label{eq:vonNeumann}
S(\rho_{t_0}) = -\sum_{n} \lambda_n \ln \lambda_n, 
\end{equation} 
implies a rapid decay of $\{\tau_n\}$.

\prlsection{Bound on $S(\rho_{t_0})$} To study the singular spectrum of $\mathcal T_{t_0}$, we now focus on the concrete setting of brickwork quantum circuits. We consider $2L$ qudits with $d$ internal states, initially prepared in a low-entangled state $|\Psi_0\rangle$ and evolved in time by discrete applications of unitary operators that couple only nearest neighbours, see Fig.~\ref{fig:Cdiag}~(a). 

Starting from the diagram for the one-point function, see Fig.~\ref{fig:Cdiag}~(b), we cut the network in the middle and obtain the left and right influence matrices. For a generic local unitary evolution, only the gates inside the past light cone of the operator can affect the one-point function. In a brickwork circuit, this light cone is exact, since the unitary gates outside cancel when multiplied by their Hermitian conjugates. The influence matrices therefore reduce to the triangular tensor networks shown in the shaded regions of Fig.~\ref{fig:Cdiag}~(b). Connecting the lower $t_0$ legs of these two tensors in each of the two layers defines the reduced transition matrix $\mathcal T_{t_0}$ for the dual unitary circuit (Fig.~\ref{fig:Cdiag}~(d)).

To bound the entropy of $\rho_{t_0}$, we use a general `decomposition trick' introduced in Ref.~\cite{foligno2023temporal}. Let $|\psi\rangle$ be a pure state on $A\cup \bar A$, and let $\{P_k\}$ be orthogonal projectors on $\bar A$ with $\sum_k P_k=\I_{\bar A}$. Then we can decompose the state by these projectors $|\psi\rangle = \sum_k P_k |\psi\rangle$, and due to their orthogonality in $\bar{A}$, the reduced density matrix on $A$ can be written as
\begin{equation}
\sigma_{A} = \operatorname{tr}_{\bar A} (|\psi\rangle\langle\psi|) = \sum_k p_k \sigma_k ,
\end{equation}
where
\begin{equation}
\label{eq:projeq}
p_k = \operatorname{tr} (P_k |\psi\rangle\langle\psi|), \qquad
\sigma_k = \frac{\operatorname{tr}_{\bar{A}} (P_k |\psi\rangle\langle\psi| P_k)}{p_k}.
\end{equation}
Using the standard entropy bound for mixing states (cf.\ e.g.\ Ref.~\cite{nielsen2010quantum}), this gives
\begin{equation}
\label{eq:Sbound}
\sum_k p_k S(\sigma_k) \le S(\sigma_{A}) \le \sum_k p_k S(\sigma_k) + H(\{p_k\}),
\end{equation}
where $H(\{p_k\})$ is the Shannon entropy of the probability distribution $\{p_k\}$.

\begin{figure*}[t]
    \centering
    \includegraphics[width=0.327\linewidth]{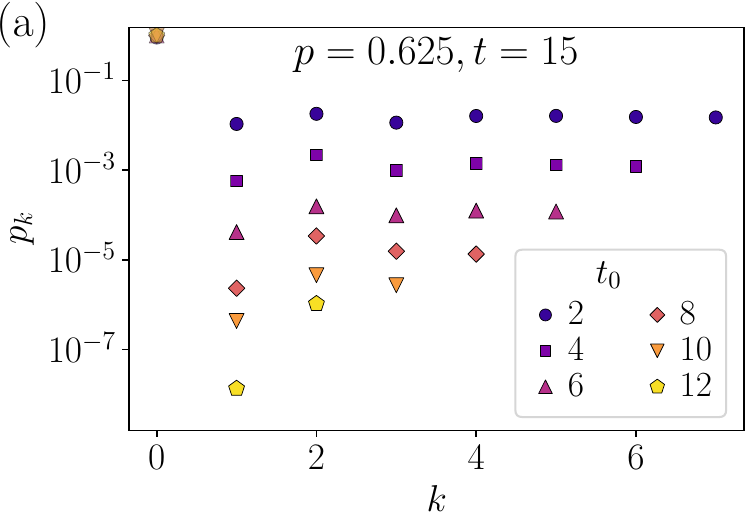}   
    \includegraphics[width=0.325\linewidth]{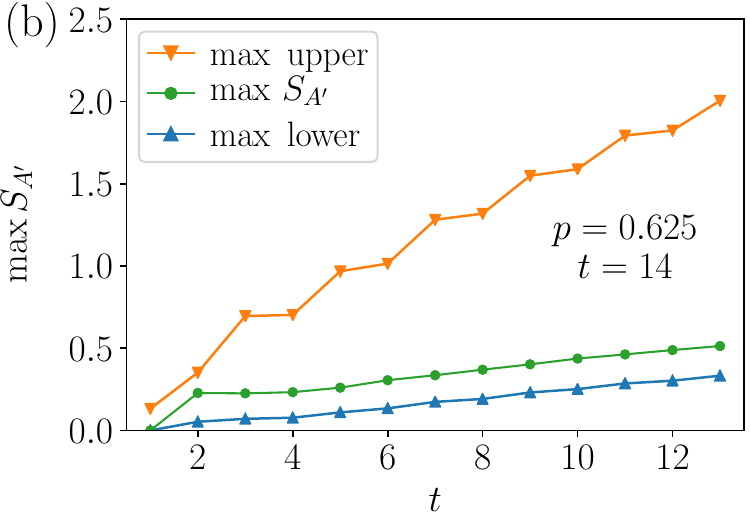}
    \includegraphics[width=0.325\linewidth]{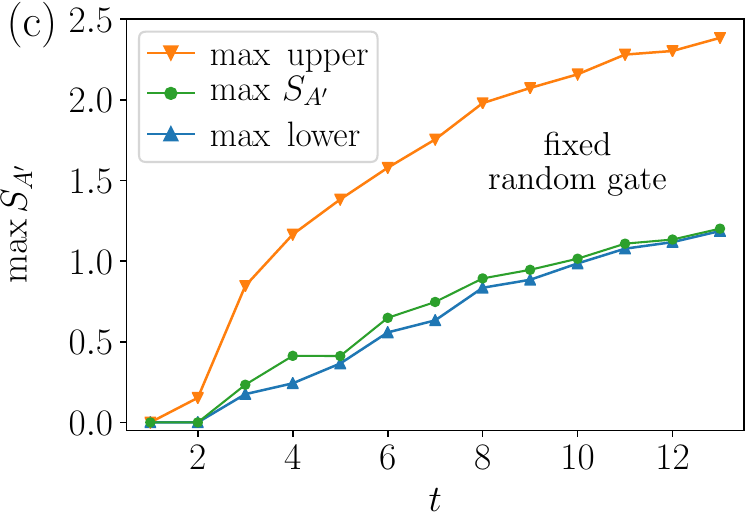}
\caption{\label{fig:pkfixedrealisation} (a) Probability distribution $\{p_k\}$ versus $k$ for a fixed dual-unitary circuit with $p=0.625$ and $t=15$. The distribution is almost flat for $k \ge 1$, especially with larger $t_0$. The decay at each fixed $k$ is well described by an exponential in $t_0$. (b) Exact numerical evaluation of $\max_{t_0} S(\rho_{t_0})$ together with the upper and lower bounds from Eq.~\eqref{eq:Sbound} for a non-random dual-unitary circuit with $p=0.625$. (c) Same comparison for a generic unitary circuit obtained from a Haar-random gate fixed in space and time.}
\end{figure*}

Specializing the above reasoning to the state  
\begin{equation}
|\psi\rangle = \sum_{i,j} (\mathcal T_{t_0})_{ij}|i\rangle_A\otimes|j\rangle_{\bar A},
\end{equation}
with $A$ and $\bar{A}$ denoting the systems of $t-t_0$ top legs on left and right respectively,
we find a bound on the desired von-Neumann entropy. The effectiveness of this bound relies on choosing projectors adapted to the specific physical setting. Here we consider dual-unitary circuits~\cite{bertini2019exact} evolving from a generic dimerized initial state, which is known to produce influence matrices with volume law entanglement~\cite{foligno2023temporal}. 

Dual-unitary circuits are defined by the property that they remain unitary when the roles of space and time are exchanged, see Fig.~\ref{fig:Cdiag}~(c). Using this property, the green region in Fig.~\ref{fig:Cdiag}~(b) is also unitary when viewed horizontally, which does not change the singular spectrum of the RTM. We thus remove these unitaries, see Fig.~\ref{fig:Cdiag}~(d) and the Supplemental Material (SM) for details. This way, $A$ and $\bar A$ are reduced to their halves $A'$ and $\bar A'$. We choose the projectors $\{P_k\}_{k=0}^\ell$ in the bound~\eqref{eq:Sbound} as
\begin{equation}
\begin{aligned}
\!\!\!\! P_0 &\!=\! \left[\frac{1}{d} |\mcirc\rangle\langle \mcirc|\right]^{\otimes \ell},\\
\!\!\!\! P_{k>0} &\!=\!  \left[\frac{1}{d} |\mcirc\rangle\langle \mcirc|\right]^{\!\!\otimes  \ell -k} \!\!\!\!\!\!\!\otimes \left[ \I -  \frac{1}{d} |\mcirc\rangle\langle \mcirc|\right] \otimes \I^{\otimes(k-1)}. 
\end{aligned}
\label{eq:duproj}
\end{equation}
Here, $\ell$ is the size of $A'$, and $|\mcirc\rangle= \sum_i |i,i\rangle$. These projectors sum to the identity on $\bar A'$; they pick out the sector where the infinite-temperature portion of the state increases in size $k$, mimicking the structure observed for solvable states~\cite{piroli2020exact}. Using Eq.~\eqref{eq:duproj} in Eq.~\eqref{eq:projeq}, we find~\cite{Note50} 
\begin{equation}
p_k \simeq  
\left\lbrace\begin{aligned}
  &  \text{poly}(t,t_0) e^{ - \alpha t_0 }  & \quad k > 0 \\
  &  1 - \text{poly}(t,t_0) e^{- \alpha t_0}   & \quad k = 0 \\
\end{aligned} \right. \,. 
\label{eq:pkform}
\end{equation}
Here, $\text{poly}(t,t_0)$ are polynomials in $t$ and $t_0$. The rate $\alpha$ is determined by the competition between emergent modes, such as domain wall and magnon in random circuits~\cite{jonay2024physical,jonay2025twostage}. In the SM, we show that this form holds exactly if $p_k$ is averaged over random dual-unitary circuits~\cite{bertini2020scrambling, foligno2022growth} with `critical' entangling power $p = p_c \equiv 1-{1}/{d^2}$~\footnote{The entangling power $p$ measures the ability of the local gate to entangle two qubits in a random product state~\cite{zanardi2000entangling}.}. In the general case, its validity can be argued from the entanglement membrane approach~\cite{jonay2018coarsegrained, zhou2020entanglement}. See the representative example in Fig.~\ref{fig:pkfixedrealisation}~(a) and the SM~\cite{Note50} for further numerical evidence. 

The upper bound in Eq.~\eqref{eq:Sbound} has two contributions, the classical entropy $H(\{p_k\})$ and the convex sum of entanglement $\sum_k p_k S(\sigma_k)$. The probability distribution $\{p_k\}$ has at most $t$ terms, so the classical entropy is upper bounded by $\ln t$. We then focus on the convex sum of entanglement. We note that $S(\sigma_k) \le 2k \ln d$ by the dimension of $k$-qudit state $\sigma_k$. From Eq.~\eqref{eq:pkform}, the sum $\sum_k p_k S(\sigma_k) = \text{poly}(t,t_0) e^{- \alpha t_0}$. We find that for $t_0 \geq \mathcal{O}(\ln t)$, the exponential decay in $t_0$ from Eq.~\eqref{eq:pkform} suppresses any polynomial in $t$, so the convex sum of entanglement is $\mathcal{O}(1)$ in this regime. On the other hand, since the number of nonzero singular values of $\mathcal T_{t_0}$ cannot exceed $d^{2 t_0}$, the convex sum of entanglement is upper bounded by $S(\sigma_A)$, which in this case is upper bounded by $2 t_0 \ln d$. Therefore, for $t_0 \lesssim \mathcal{O}(\ln t)$, the entropy is bounded by $\mathcal{O}(\ln t)$. Putting everything together, we obtain
\begin{equation}
\label{eq:mainresult}
S(\rho_{t_0})  \lesssim \mathcal{O}( \ln t ) ,\qquad \forall t_0. 
\end{equation}
This suggests that, in dual-unitary circuits evolving from generic states, the RTM can be efficiently approximated despite the large temporal entanglement of the influence matrices. 

To test the validity of this result beyond the exactly solvable cases we resort to numerical experiments. First, we consider the behavior of $\max _{t_0}(S(\rho_{t_0}))$ versus $t$ for non-random dual-unitary circuits. A representative sample of our findings is reported in Fig.~\ref{fig:pkfixedrealisation}~(b), where we compare the exact numerical evaluation with the bounds in Eq.~\eqref{eq:Sbound}. We find that the upper bound is satisfied but not tight, while the lower bound appears to become tight for sufficiently large $t$. All three curves, however, are compatible with logarithmic growth in time, albeit with different prefactors, in agreement with Eq.~\eqref{eq:mainresult}. Second, we consider the same quantities in generic unitary circuits, obtained by drawing a gate from the Haar distribution and keeping it fixed in the space-time evolution. To produce the bounds, we again use the projectors in Eq.~\eqref{eq:duproj}, but the temporal lattice cannot be reduced due to the lack of dual-unitarity. In this case, the probability distribution $p_k$ does not reproduce the form in Eq.~\eqref{eq:pkform}, and we are unable to evaluate the bounds analytically. Nevertheless, the representative examples in Fig.~\ref{fig:pkfixedrealisation}~(c) suggest that the lower bound again becomes tight at large times. Most importantly, the upper bound again shows logarithmic growth in time, suggesting that the RTM can be efficiently approximated in generic (chaotic) quantum circuits.

\prlsection{Conclusion and discussion} In this Letter, we studied the complexity of using influence matrices to evaluate local correlation functions in generic quantum systems. Combining analytical and numerical arguments, we showed that although the influence matrices are not efficiently representable because of their large temporal entanglement, a suitable combination of them, the `reduced transition matrix', is sufficient to evaluate local expectation values. In a nutshell, the entropy associated with RTM grows only logarithmically in time, which leads to favorable control of the absolute error in the calculation of expectation values. We provided a rigorous proof of the logarithmic bound for random dual-unitary circuits and numerical evidence for both non-random dual-unitary circuits and generic unitary circuits.

Our logarithmic bound for $S_1$ does not contradict the volume-law $S_0$ bound in Ref.~\cite{carignano2023temporal} as $S_0$ measures the truncation-free bond dimension (see~\cite{Note50} for the singular value distribution we obtain in the cases we studied). In the boundary correlator setup, one of the influence matrices has a constant bond dimension, and our RTM sweep reduces the other to have the same constant bond dimension independently of the dynamics. We also note a comparable $\ln t$ scaling has been found by Refs.~\cite{carignano2025overcoming,carignano2025loschmidt} for the generalized entropy of the eigenvalues of a (forward-copy only) reduced transition matrix arising from the Loschmidt echo.

These results are largely a proof of principle, and many questions remain open. What we have shown is given an exact influence matrices, we can produce an economic approximation via a joint sweeping in Fig.~\ref{fig:compression}~(b) with controlled error. But how to obtain the influence matrix in the first place, or how to evolve this approximation in time without generating the exact influence matrices every time step, is unclear.  A natural approach is to evolve in time the low-rank approximations of the influence matrices used to construct the reduced transition matrix.  This strategy has been implemented numerically in several recent works~\cite{hastings2015connecting,carignano2023temporal,carignano2026itransverse,carignano2025overcoming}, but a quantitative error analysis is still missing.

Another immediate question concerns the `signal to noise ratio' of this approach. The entanglement scaling we observed suggests that the absolute error in expectation values is small, but in the chaotic systems we considered the expectation values themselves are also exponentially small. A compelling question is then whether a similar result can hold in systems with conservation laws in either continuous or discrete time. In such settings, one expects larger signals, since correlation functions can decay as power laws, so the bounds on the absolute error should become more informative. We leave these questions for future work.

\footnotetext[50]{See Supplemental Material at [URL will be inserted by publisher]
for the derivation of the entropy bounds for dual-unitary circuits, the proof of the conjectured form of $p_k$ for random dual-unitary circuits, the membrane calculation away from the critical point, additional numerical data for the decay of $p_k$ and the singular values of the reduced transition matrix, and the explicit gate matrices used in the numerics.}

\begin{acknowledgments}
We thank J\'er{\^o}me Dubail and Alessandro Foligno for collaborating at an early stage of this project and for many valuable discussions. We also thank Toma{\v z} Prosen, Pavel Kos, Dmitri Abanin, Wen Wei Ho and Jacopo De Nardis, Alessio Lerose and Mark Mitchson for insightful discussions and/or comments on the manuscript. We acknowledge the hospitality and support of the Simons Center for Geometry and Physics, Stony Brook University, where the work was initiated during the program `Fluctuations, Entanglements, and Chaos: Exact Results'. We also acknowledge the hospitality of the Kavli Institute for Theoretical Physics program `Learning the Fine Structure of Quantum Dynamics in Programmable Quantum Matter', where stimulating discussions with colleagues benefited this work. This research was supported in part by grant NSF PHY-2309135 to the Kavli Institute for Theoretical Physics (KITP). This work was supported by the Royal Society, through the University Research Fellowship No. 201101 (B.\ B.). We acknowledge Advanced Research Computing at Virginia Tech for providing computational resources and technical support that have contributed to the results reported within this paper. 
\end{acknowledgments}

\bibliography{bibliography}

\pagebreak

\onecolumngrid

\newcounter{equationSM}
\newcounter{figureSM}
\newcounter{tableSM}
\stepcounter{equationSM}
\setcounter{equation}{0}
\setcounter{figure}{0}
\setcounter{table}{0}
\setcounter{section}{0}
\makeatletter
\renewcommand{\theequation}{\textsc{sm}-\arabic{equation}}
\renewcommand{\thefigure}{\textsc{sm}-\arabic{figure}}
\renewcommand{\thetable}{\textsc{sm}-\arabic{table}}

{\bf \center\large  Supplemental Material: Low Rank Structure of the Reduced Transition Matrix}\\

Here we report some useful information complementing the main text. In particular
\begin{itemize}
  \item[-] In Sec.~\ref{app:SinDU} we explicitly compute the bounds in Eq.~\eqref{eq:Sbound} for dual unitary circuits with the set of projectors given in Eq.~\eqref{eq:duproj}. 
  \item[-] In Sec.~\ref{app:proof} we explicitly derive the form of $p_k$ in random dual-unitary circuits. 
  \item[-] In Sec.~\ref{app:membrane} we present we a membrane calculation to argue for the validity of Eq.~\eqref{eq:pkform} away from the critical point.
  \item[-] In Sec.~\ref{app:pk} we provide additional numerical data on the distribution $p_k$ in non-random circuits. 
  \end{itemize}

\section{Entanglement entropy of $\rho_A$ for dual-unitary circuits}
\label{app:SinDU}

In this appendix we prove the bounds in Eq.~\eqref{eq:Sbound}. We begin by considering the diagram in Fig.~\ref{fig:Cdiag} for the reduced transition matrix, from which we can immediately write 
\begin{equation}
\langle i | \mathcal{T}^{\phantom{\dag}}_{t_0} \mathcal{T}_{t_0}^\dagger | j \rangle  = 
\fineq[-0.8ex][.65]{
\eigenVR[1.25][0][r][7][tr][init][bertiniorange][2]
\draw[thick] (-.75-.5,7.5+.5) -- (-.35-.5,7.15+.5);
\draw[fill=white] (-1.25,7.5+0.5) circle (0.15cm);
\eigenVL[-1.25][0][l][6][tr][init][bertiniorange][2]
\node[scale=2]  at (-.125,7.575) {\scalebox{0.7}{$i_1j_1$}};
\node[scale=2]  at (-.125,5.575) {\scalebox{0.7}{$\vdots$}};
\node[scale=2]  at (-.15,3.25) {\scalebox{0.7}{$i_{|\!A\!|}j_{|\!A\!|}$}};
\foreach \i in {0,...,2}{
\draw[thick, black] (-.75,\i+.5) -- (.75,\i+.5);}
\foreach \i in {0,...,4}{
 \sqrstate[.75][\i+3.5]}
 \foreach \i in {-2,...,2}{
\draw[very thick, dashed, white] (-2.75-.625-1+\i+0.4,2.75-.625+1+\i-0.4) -- (-1.5-.625-1+\i-0.4,1.5-.625+1+\i+0.4);
}
 \foreach \i in {0,...,3}{
\draw[very thick, dashed,white] (4.6-\i,.35+\i) -- (5-\i,.75+\i);
}
}.
\label{eq:rhoA}
\end{equation}
Here $i,j\in\{1,2,\ldots, d^{2|A|+2}-1\}$ and $i_k$ denotes $i$'s $k$-th digit in base $d^2$. Moreover we introduced a diagrammatic representation for the multi-folded gates 
\begin{equation}
\label{eq:foldedgaten}
(U \otimes U^*)^{\otimes n} = \fineq[-0.8ex][0.75][1]{
    \roundgate[0][0][1][topright][orange][n]
},
\end{equation}
and the states 
\begin{equation}  
\begin{aligned}
&|\mcirc\rangle=\hspace{-.5cm}\sum_{i_1,\ldots, i_n=1}^d |i_1\rangle\otimes|i_1\rangle\otimes \cdots \otimes |i_n\rangle\otimes|i_n\rangle = \fineq[-0.8ex][0.6][1]{
    \draw(0, 0)--++(0, .5);
    \cstate[0][0]
    }\,, \\
&|\msquare\rangle=\hspace{-.5cm}\sum_{i_1,\ldots, i_n=1}^d |i_n\rangle\otimes|i_1\rangle\otimes \cdots \otimes |i_{n-1}\rangle\otimes|i_n\rangle = \fineq[-0.8ex][0.6][1]{
    \draw(0, 0)--++(0, .5);
    \sqrstate[0][0]}\,,
\end{aligned}
\end{equation}
representing connections of different replicas. In Eq.~\eqref{eq:rhoA} we dropped their $n$ dependence to lighten the notation and, similarly, we have also set
\begin{equation}
(|\psi_0\rangle\otimes|\psi_0\rangle^*)^{\otimes n} = \fineq[-0.8ex][0.6][1]{\thetastatevar[0][0][1.5][orange];},
\end{equation}
without including a replica index. Finally, we used the notation 
\begin{equation}
 \fineq[-0.8ex][0.6][1]{
 \draw[thick] (-.5,0) -- (.5,0);
 \node[scale=2]  at (1,0) {\scalebox{0.7}{$ij$}};} = 
  \fineq[-0.8ex][0.6][1]{
  \draw[thick] (-.5+.25,0+.25) -- (.5+.25,0+.25);
 \node[scale=2]  at (1,0+.25) {\scalebox{0.7}{$j$}};
  \draw[thick] (-.5,0) -- (.5,0);
 \node[scale=2]  at (1-0.25,0) {\scalebox{0.7}{$i$}};
 }.
\end{equation} 

Specialising the treatment to dual unitary circuits, where the square tensors act as unitary matrices also from left to right, we can remove the two isosceles triangles of gates with basis (of length $|A|$) along the time axis. Indeed, they implement a unitary transformation within the subsystem $A$ which does not affect the entanglement. Namely we have 
\begin{equation}
S(\rho_A) =S(\sigma_{A'}),
\end{equation}
where we introduced
\begin{equation}
\sigma_{A'}=  \frac{\widetilde{\mathcal{T}}^{\phantom{\dag}}_{t_0} \widetilde{\mathcal{T}}_{t_0}^\dagger }{\operatorname{tr}[\widetilde{\mathcal{T}}^{\phantom{\dag}}_{t_0} \widetilde{\mathcal{T}}_{t_0}^\dagger ]}, 
\end{equation}
with  
\begin{equation}
\langle i | \widetilde{\mathcal{T}}^{\phantom{\dag}}_{t_0} \widetilde{\mathcal{T}}_{t_0}^\dagger | j \rangle = \hspace{-1cm}
\fineq[-0.8ex][.65]{
 \roundgate[1.25+2][5][1][topright][orange][2]
 \roundgate[-1.25-2][4][1][topright][orange][2]
 \roundgate[1.25+1][4][1][topright][orange][2]
 \roundgate[1.25+3][4][1][topright][orange][2]
 \roundgate[-1.25-1][3][1][topright][orange][2]
 \roundgate[-1.25-3][3][1][topright][orange][2]
 \roundgate[1.25][3][1][topright][orange][2]
 \roundgate[1.25+2][3][1][topright][orange][2]
 \roundgate[1.25+4][3][1][topright][orange][2]
 \roundgate[-1.25][2][1][topright][orange][2]
 \roundgate[-1.25-2][2][1][topright][orange][2]
 \roundgate[-1.25-4][2][1][topright][orange][2]
  \roundgate[1.25+1][2][1][topright][orange][2]
 \roundgate[1.25+3][2][1][topright][orange][2]
 \roundgate[1.25+5][2][1][topright][orange][2]
 \roundgate[-1.25-1][1][1][topright][orange][2]
 \roundgate[-1.25-3][1][1][topright][orange][2]
 \roundgate[-1.25-5][1][1][topright][orange][2]
 \roundgate[1.25][1][1][topright][orange][2]
 \roundgate[1.25+2][1][1][topright][orange][2]
 \roundgate[1.25+4][1][1][topright][orange][2]
\roundgate[1.25+6][1][1][topright][orange][2]
\foreach \i in {0, ..., 3}{
\thetastatevar[-7.25+2*\i][0][1.5][orange]
}
\foreach \i in {0, ..., 3}{
\thetastatevar[2*\i+2.25][0][1.5][orange]
}
\foreach \i in {0,...,2}{
\draw[thick, black] (-.75,\i+.5) -- (.75,\i+.5);}
\foreach \i in {0,...,2}{
\sqrstate[\i+.75][\i+3.5]
}
\foreach \i in {-1,...,4}{
\cstate[7.75-\i][\i+1.5]
}
\foreach \i in {-1,...,3}{
\cstate[\i-6.75][\i+1.5]
}
\node[scale=2]  at (-2.25,5) {\scalebox{0.7}{$(ij)_1$}};
\node[scale=2]  at (-1.85,4.6) {\scalebox{0.7}{$\ddots$}};
\node[scale=2]  at (-1.15,3.9) {\scalebox{0.7}{$(ij)_{|A'|}$}};
\foreach \i in {-2,...,1}{
\draw[very thick, dashed, white] (-2.75-.625-1+\i+0.4,2.75-.625+1+\i-0.4) -- (-1.5-.625-1+\i-0.4,1.5-.625+1+\i+0.4);
}
 \foreach \i in {0,...,3}{
\draw[very thick, dashed,white] (4.6-\i,.35+\i) -- (5-\i,.75+\i);
}
},
\end{equation}
and $|A'| = \lfloor|A|/2\rfloor$. In fact one can conveniently set 
\begin{equation}
\widetilde{\mathcal{T}}^{\phantom{\dag}}_{t_0} \widetilde{\mathcal{T}}_{t_0}^\dagger = \operatorname{tr}_{\bar A'}[|\Psi_{{A'}{\bar A'}}\rangle\langle \Psi_{{A'}{\bar A'}}|], 
\end{equation}
where we defined 
\begin{equation}
\hspace{-2cm} |\Psi_{{A'}{\bar A'}}\rangle \!\!= \hspace{-1cm}
\fineq[-0.8ex][.65]{
 \roundgate[1.25+2][5][1][topright][orange][1]
 \roundgate[-1.25-2][4][1][topright][orange][1]
 \roundgate[1.25+1][4][1][topright][orange][1]
 \roundgate[1.25+3][4][1][topright][orange][1]
 \roundgate[-1.25-1][3][1][topright][orange][1]
 \roundgate[-1.25-3][3][1][topright][orange][1]
 \roundgate[1.25][3][1][topright][orange][1]
 \roundgate[1.25+2][3][1][topright][orange][1]
 \roundgate[1.25+4][3][1][topright][orange][1]
 \roundgate[-1.25][2][1][topright][orange][1]
 \roundgate[-1.25-2][2][1][topright][orange][1]
 \roundgate[-1.25-4][2][1][topright][orange][1]
  \roundgate[1.25+1][2][1][topright][orange][1]
 \roundgate[1.25+3][2][1][topright][orange][1]
 \roundgate[1.25+5][2][1][topright][orange][1]
 \roundgate[-1.25-1][1][1][topright][orange][1]
 \roundgate[-1.25-3][1][1][topright][orange][1]
 \roundgate[-1.25-5][1][1][topright][orange][1]
 \roundgate[1.25][1][1][topright][orange][1]
 \roundgate[1.25+2][1][1][topright][orange][1]
 \roundgate[1.25+4][1][1][topright][orange][1]
\roundgate[1.25+6][1][1][topright][orange][1]
\foreach \i in {0, ..., 3}{
\thetastatevar[-7.25+2*\i][0][1.5][orange]
}
\foreach \i in {0, ..., 3}{
\thetastatevar[2*\i+2.25][0][1.5][orange]
}
\foreach \i in {0,...,2}{
\draw[thick, black] (-.75,\i+.5) -- (.75,\i+.5);}
\foreach \i in {-1,...,4}{
\cstate[7.75-\i][\i+1.5]
}
\foreach \i in {-1,...,3}{
\cstate[\i-6.75][\i+1.5]
}
\draw [decorate, decoration = {brace}, thick,shift={(0,2.)}]  (-2.75,2.75) -- (-1.5,1.5);
\draw [decorate, decoration = {brace}, thick,shift={(0,2.)}]  (.5,1.5) -- (2.75,3.75);
\node[scale=2]  at (-1.75,4.5) {\scalebox{0.7}{$A'$}};
\node[scale=2]  at (1.25,5) {\scalebox{0.7}{$\bar A'$}};
\foreach \i in {-2,...,1}{
\draw[very thick, dashed, white] (-2.75-.625-1+\i+0.4,2.75-.625+1+\i-0.4) -- (-1.5-.625-1+\i-0.4,1.5-.625+1+\i+0.4);
}
 \foreach \i in {0,...,3}{
\draw[very thick, dashed,white] (4.6-\i,.35+\i) -- (5-\i,.75+\i);
}
},
\end{equation}
and $|\bar A'|=\lceil|A|/2\rceil \equiv t_1$. Having defined the relevant quantities we are now in a position to prove the bounds on $S(\rho_A)$ by following the decomposition method in the main text and Ref.~\cite{foligno2023temporal}. We begin by defining the following set of projectors acting on $\bar A'$
\begin{equation}
\begin{aligned}
P_0 &= \left(\frac{1}{d} |\mcirc\rangle\langle \mcirc|\right)^{\otimes  t_1}\\
P_k &=  \left(\frac{1}{d} |\mcirc\rangle\langle \mcirc|\right)^{\otimes  t_1 -k} \otimes \left( \I -  \frac{1}{d} |\mcirc\rangle\langle \mcirc|\right) \otimes \I^{\otimes(k-1)}, & & k = 1,\ldots,  t_1.  
\end{aligned}
\end{equation}
These projectors are orthogonal and complete, i.e., 
\begin{equation} 
P_i P_j  = \delta_{ij} P_i, \qquad \sum_{k=0}^{t_1} P_k =  \I^{\otimes  t_1}\,.
\end{equation}
We can then write 
\begin{equation}
\begin{aligned}
  \sigma_{A'}&= \sum_{k=0}^{ t_1}   \frac{\operatorname{tr}_{\bar A'}( P_k |\Psi_{A'\bar A'}\rangle\langle \Psi_{A'\bar A'}|)}{\langle \Psi_{A'\bar A'} | \Psi_{A'\bar A'} \rangle}=\sum_{k=0}^{t_1} \frac{\operatorname{tr}_{\bar A'}( P^2_k |\Psi_{A'\bar A'}\rangle\langle \Psi_{A'\bar A'}|)}{\langle \Psi_{A'\bar A'} | \Psi_{A'\bar A'} \rangle} =  \sum_{k=0}^{ t_1}\frac{\operatorname{tr}_{\bar A'}( P_k |\Psi_{A'\bar A'}\rangle\langle \Psi_{A'\bar A'}| P_k)}{\langle \Psi_{A'\bar A'} | \Psi_{A'\bar A'} \rangle}. 
\end{aligned}
\end{equation}
The above expression is directly rewritten as 
\begin{equation}
\sigma_{A'} = \sum_{k=0}^{t_1} p_k \sigma_k,
\label{eq:decomposition}
\end{equation}
where we set 
\begin{equation}
p_k = \frac{\operatorname{tr}( P_k |\Psi_{A'\bar A'}\rangle\langle \Psi_{A'\bar A'}| P_k)}{\langle \Psi_{A'\bar A'} | \Psi_{A'\bar A'} \rangle}\,,
\qquad
\sigma_k = \frac{\operatorname{tr}_{\bar A'}( P_k |\Psi_{A'\bar A'}\rangle\langle \Psi_{A'\bar A'}| P_k)}{\operatorname{tr}( P_k |\Psi_{A'\bar A'}\rangle\langle \Psi_{A'\bar A'}| P_k)}\,.
\end{equation}
Therefore, using concavity lower bound and mixing upper bound (cf., e.g., Ref.~\cite{nielsen2010quantum}) for the decomposition in Eq.~\eqref{eq:decomposition} we find 
\begin{equation}
\sum_{k=0}^{t_1} p_k S( \sigma_k ) \le S( \rho_{A})   \le \sum_{k=0}^{t_1} p_k S( \sigma_k ) + H( \{ p_k \}), 
\end{equation}
where $H( \{ p_k \})$ is the Shannon entropy of the distribution $\{ p_k \}$. 

To estimate upper and lower bound we begin considering the entanglement of $S(\sigma_k )$. Since 
\begin{equation}
P_k |\Psi_{A'\bar A'}\rangle, 
\end{equation}
is a pure state we have that 
\begin{equation}
S(\sigma_k)=S(\rho_k), 
\end{equation}
where we introduced 
\begin{equation}
\rho_k = \frac{\operatorname{tr}_{A'}( P_k |\Psi_{A'\bar A'}\rangle\langle \Psi_{A'\bar A'}| P_k)}{\operatorname{tr}( P_k |\Psi_{A'\bar A'}\rangle\langle \Psi_{A'\bar A'}| P_k)}.
\end{equation}
We now compute $\rho_k$ by noting that, due to the unitarity of the gates, we have 
\begin{equation}
\fineq[-0.8ex][.65]{
 \roundgate[0][0][1][topright][orange][2]
 \cstate[.5][.5]
 \cstate[-.5][.5]
} = \fineq[-0.8ex][.65]{
 \draw [thick] (.5,.5) -- (.5,-0.5); 
 \draw [thick] (-.5,.5) -- (-.5,-0.5); 
 \cstate[.5][.5]
 \cstate[-.5][.5]
},\qquad \fineq[-0.8ex][.65]{
 \roundgate[0][0][1][topright][orange][2]
 \cstate[.5][-.5]
 \cstate[-.5][-.5]
} = \fineq[-0.8ex][.65]{
 \draw [thick] (.5,.5) -- (.5,-0.5); 
 \draw [thick] (-.5,.5) -- (-.5,-0.5); 
 \cstate[.5][-.5]
 \cstate[-.5][-.5]
},
\end{equation}
and, therefore, 
\begin{equation}
\fineq[-0.8ex][.65]{
 \roundgate[-1.25-3][5][1][topright][orange][2]
  \roundgate[-1.25-2][4][1][topright][orange][2]
  \roundgate[-1.25-1][3][1][topright][orange][2]
  \roundgate[-1.25][2][1][topright][orange][2]
 \roundgate[-1.25+1][1][1][topright][orange][2]
\thetastatevar[2.75-2][0][1.5][orange]
\foreach \i in {-1,...,4}{
\cstate[.25-\i][\i+1.5]
}
\cstate[.25-5][5.5]
}=\fineq[-0.8ex][.65]{
\foreach \i in {-1,...,3}{
\draw [thick] (.25-\i,\i+1.5) -- (.25-\i-0.5,\i+1.5-0.5); 
\cstate[.25-\i][\i+1.5]
}
}. 
\end{equation}
This means that we can write
\begin{equation}
\label{eq:rhorhotilde}
\rho_k= \left(\frac{1}{d^2} |\mcirc\rangle\langle \mcirc|\right)^{\otimes t_1-k} \otimes \tilde \rho_k,
\end{equation}
where we defined
\begin{equation}
\tilde \rho_k = \frac{
\fineq[-0.8ex][.65]{
 \roundgate[1.25+2][5][1][topright][orange][2]
 \roundgate[-1.25-3][5][1][topright][orange][2]
 \roundgate[-1.25-2][4][1][topright][orange][2]
  \roundgate[-1.25-4][4][1][topright][orange][2]
 \roundgate[1.25+1][4][1][topright][orange][2]
 \roundgate[1.25+3][4][1][topright][orange][2]
 \roundgate[-1.25-1][3][1][topright][orange][2]
 \roundgate[-1.25-3][3][1][topright][orange][2]
  \roundgate[-1.25-5][3][1][topright][orange][2]
 \roundgate[1.25][3][1][topright][orange][2]
 \roundgate[1.25+2][3][1][topright][orange][2]
 \roundgate[1.25+4][3][1][topright][orange][2]
 \roundgate[-1.25][2][1][topright][orange][2]
 \roundgate[-1.25-2][2][1][topright][orange][2]
 \roundgate[-1.25-4][2][1][topright][orange][2]
  \roundgate[-1.25-6][2][1][topright][orange][2]
  \roundgate[1.25+1][2][1][topright][orange][2]
 \roundgate[1.25+3][2][1][topright][orange][2]
 \roundgate[1.25+5][2][1][topright][orange][2]
 \roundgate[-1.25-1][1][1][topright][orange][2]
 \roundgate[-1.25-3][1][1][topright][orange][2]
 \roundgate[-1.25-5][1][1][topright][orange][2]
 \roundgate[-1.25-7][1][1][topright][orange][2]
 \roundgate[1.25][1][1][topright][orange][2]
 \roundgate[1.25+2][1][1][topright][orange][2]
 \roundgate[1.25+4][1][1][topright][orange][2]
\roundgate[1.25+6][1][1][topright][orange][2]
\foreach \i in {-1, ..., 3}{
\thetastatevar[-7.25+2*\i][0][1.5][orange]
}
\foreach \i in {0, ..., 3}{
\thetastatevar[2*\i+2.25][0][1.5][orange]
}
\foreach \i in {0,...,2}{
\draw[thick, black] (-.75,\i+.5) -- (.75,\i+.5);}
\foreach \i in {0,...,2}{
\sqrstate[-1.75-\i][\i+3.5]
}
\foreach \i in {-1,...,4}{
\cstate[7.75-\i][\i+1.5]
}
\foreach \i in {-1,...,4}{
\cstate[\i-8.75][\i+1.5]
}
\cstate[3.75-1][4.5+1][][black]
\draw[thick, black] (3.75-1,5.5) -- (3.75-1-0.25,5.5+0.25);
\draw [decorate, decoration = {brace}, thick,shift={(0,2.)}] (0.5,1.5) -- (1.75,2.75);
\node[scale=2]  at (.35,4.5) {\scalebox{0.7}{$k-1$}};
\foreach \i in {-2,...,1}{
\draw[very thick, dashed, white] (-2.75-.625-1+\i+0.4,2.75-.625+1+\i-0.4) -- (-1.5-.625-1+\i-0.4,1.5-.625+1+\i+0.4);
}
 \foreach \i in {0,...,3}{
\draw[very thick, dashed,white] (4.6-\i,.35+\i) -- (5-\i,.75+\i);
}
}}{\fineq[-0.8ex][.65]{
 \roundgate[1.25+2][5][1][topright][orange][2]
 \roundgate[-1.25-3][5][1][topright][orange][2]
 \roundgate[-1.25-2][4][1][topright][orange][2]
  \roundgate[-1.25-4][4][1][topright][orange][2]
 \roundgate[1.25+1][4][1][topright][orange][2]
 \roundgate[1.25+3][4][1][topright][orange][2]
 \roundgate[-1.25-1][3][1][topright][orange][2]
 \roundgate[-1.25-3][3][1][topright][orange][2]
  \roundgate[-1.25-5][3][1][topright][orange][2]
 \roundgate[1.25][3][1][topright][orange][2]
 \roundgate[1.25+2][3][1][topright][orange][2]
 \roundgate[1.25+4][3][1][topright][orange][2]
 \roundgate[-1.25][2][1][topright][orange][2]
 \roundgate[-1.25-2][2][1][topright][orange][2]
 \roundgate[-1.25-4][2][1][topright][orange][2]
  \roundgate[-1.25-6][2][1][topright][orange][2]
  \roundgate[1.25+1][2][1][topright][orange][2]
 \roundgate[1.25+3][2][1][topright][orange][2]
 \roundgate[1.25+5][2][1][topright][orange][2]
 \roundgate[-1.25-1][1][1][topright][orange][2]
 \roundgate[-1.25-3][1][1][topright][orange][2]
 \roundgate[-1.25-5][1][1][topright][orange][2]
 \roundgate[-1.25-7][1][1][topright][orange][2]
 \roundgate[1.25][1][1][topright][orange][2]
 \roundgate[1.25+2][1][1][topright][orange][2]
 \roundgate[1.25+4][1][1][topright][orange][2]
\roundgate[1.25+6][1][1][topright][orange][2]
\foreach \i in {-1, ..., 3}{
\thetastatevar[-7.25+2*\i][0][1.5][orange]
}
\foreach \i in {0, ..., 3}{
\thetastatevar[2*\i+2.25][0][1.5][orange]
}
\foreach \i in {0,...,2}{
\draw[thick, black] (-.75,\i+.5) -- (.75,\i+.5);}
\foreach \i in {0,...,1}{
\sqrstate[\i+.75][\i+3.5]
}
\foreach \i in {-1,...,4}{
\cstate[7.75-\i][\i+1.5]
}
\foreach \i in {-1,...,4}{
\cstate[\i-8.75][\i+1.5]
}
\sqrstate[-1.75][3.5]
\sqrstate[-1.75-2][3.5+2]
\sqrstate[-1.75-1][3.5+1]
\cstate[3.75-1][4.5+1][][black]
\draw[thick, black] (3.75-1,5.5) -- (3.75-1-0.25,5.5+0.25);
\draw [decorate, decoration = {brace}, thick,shift={(0,2.)}] (0.5,1.5) -- (1.75,2.75);
\node[scale=2]  at (.35,4.5) {\scalebox{0.7}{$k-1$}};
\sqrstate[2.25+0.25][5.75]
\foreach \i in {-2,...,1}{
\draw[very thick, dashed, white] (-2.75-.625-1+\i+0.4,2.75-.625+1+\i-0.4) -- (-1.5-.625-1+\i-0.4,1.5-.625+1+\i+0.4);
}
 \foreach \i in {0,...,3}{
\draw[very thick, dashed,white] (4.6-\i,.35+\i) -- (5-\i,.75+\i);
}
}},
\end{equation}
and 
\begin{equation}
 \left( \I -  \frac{1}{d} |\mcirc\rangle\langle \mcirc|\right) \otimes_r \left( \I -  \frac{1}{d} |\mcirc\rangle\langle \mcirc|\right) = \fineq[-0.8ex][0.6][1]{
 \draw[thick] (-.5,0) -- (.5,0);
 \cstate[0][0][][black]
 }.
\end{equation}
Eq.~\eqref{eq:rhorhotilde} implies 
\begin{equation}
S(\sigma_k)=S(\rho_k ) = S(\tilde \rho_k ) \le 2k \ln d, 
\end{equation}
where in the last step we bounded the entropy by using the dimension of the space where $\tilde \rho_k$ acts. 

Next we look at the probability distribution $p_k$. To this end, we note 
\begin{equation}
\fineq[-0.8ex][0.6][1]{
 \draw[thick] (-.5,0) -- (.5,0);
 \cstate[0][0][][black]
\sqrstate[.5][0]
 } = \fineq[-0.8ex][0.6][1]{
 \draw[thick] (0,0) -- (.5,0);
\sqrstate[.5][0]
 } - \frac{1}{d} \fineq[-0.8ex][0.6][1]{
 \draw[thick] (-.5,0) -- (0,0);
 \cstate[0][0]
 },
\end{equation}
which can be seen representing the l.h.s.\ in unfolded notation
\begin{equation}
 \fineq[-0.8ex][0.6][1]{
 \draw[thick] (-.5,0) -- (.5,0);
 \cstate[0][0][][black]
\sqrstate[.5][0]
 } = \fineq[-0.8ex][0.8][1]{
     \draw (-.5,0.4) -- (0,0.4);
    \draw (-.5,0.6) -- (0,0.6);
    \draw (-.5,0) -- (0,0);
    \draw (-.5,-0.2) -- (0,-0.2);
    \draw (0,0) arc (-90:90:0.2);
    \draw (0,-0.2) arc (-90:90:0.4);
  }-\frac{1}{d}\fineq[-0.8ex][0.8][1]{
     \draw (0,0.6) arc (90:270:0.1);
     \draw (-0.3,0.4) arc (-90:90:0.1);
      \draw (-.5,0.4) -- (-0.3,0.4);
    \draw (-.5,0.6) -- (-0.3,0.6);
    \draw (-.5,0) -- (0,0);
    \draw (-.5,-0.2) -- (0,-0.2);
    \draw (0,0) arc (-90:90:0.2);
    \draw (0,-0.2) arc (-90:90:0.4);
  }-\frac{1}{d}\fineq[-0.8ex][0.8][1]{
     \draw (0,0) arc (90:270:0.1);
     \draw (-0.3,-0.2) arc (-90:90:0.1);
      \draw (-.5,0) -- (-0.3,0);
    \draw (-.5,-0.2) -- (-0.3,-0.2);
     \draw (-.5,0.4) -- (0,0.4);
    \draw (-.5,0.6) -- (0,0.6);
    \draw (0,0) arc (-90:90:0.2);
    \draw (0,-0.2) arc (-90:90:0.4);
  }+\frac{1}{d^2}\fineq[-0.8ex][0.8][1]{
  \draw (0,0) arc (90:270:0.1);
     \draw (-0.3,-0.2) arc (-90:90:0.1);
      \draw (0,0.6) arc (90:270:0.1);
     \draw (-0.3,0.4) arc (-90:90:0.1);
    \draw (0,0) arc (-90:90:0.2);
    \draw (0,-0.2) arc (-90:90:0.4);
    \draw (-.5,0.4) -- (-0.3,0.4);
    \draw (-.5,0.6) -- (-0.3,0.6);
     \draw (-.5,0) -- (-0.3,0);
    \draw (-.5,-0.2) -- (-0.3,-0.2);
  } = \fineq[-0.8ex][0.8][1]{
     \draw (-.5,0.4) -- (0,0.4);
    \draw (-.5,0.6) -- (0,0.6);
    \draw (-.5,0) -- (0,0);
    \draw (-.5,-0.2) -- (0,-0.2);
    \draw (0,0) arc (-90:90:0.2);
    \draw (0,-0.2) arc (-90:90:0.4);
  }-\frac{1}{d}\fineq[-0.8ex][0.8][1]{
     \draw (-0.3,-0.2) arc (-90:90:0.1);
     \draw (-0.3,0.4) arc (-90:90:0.1);
    \draw (-.5,0.4) -- (-0.3,0.4);
    \draw (-.5,0.6) -- (-0.3,0.6);
     \draw (-.5,0) -- (-0.3,0);
    \draw (-.5,-0.2) -- (-0.3,-0.2);
  }.
\end{equation}
Hence $p_k$ is the difference of two amplitudes
\begin{equation}
\label{eq:pk}
\langle \Psi_{A'\bar A'} | \Psi_{A'\bar A'} \rangle p_k = \mathcal A_k - \mathcal A_{k-1},
\end{equation}
where we introduced
\begin{equation}
\label{eq:amplitudeAk}
\mathcal A_k = \frac{1}{d^{|A'|-k}}
\fineq[-0.8ex][.65]{
 \roundgate[1.25+2][5][1][topright][orange][2]
 \roundgate[-1.25-3][5][1][topright][orange][2]
 \roundgate[-1.25-2][4][1][topright][orange][2]
  \roundgate[-1.25-4][4][1][topright][orange][2]
 \roundgate[1.25+1][4][1][topright][orange][2]
 \roundgate[1.25+3][4][1][topright][orange][2]
 \roundgate[-1.25-1][3][1][topright][orange][2]
 \roundgate[-1.25-3][3][1][topright][orange][2]
  \roundgate[-1.25-5][3][1][topright][orange][2]
 \roundgate[1.25][3][1][topright][orange][2]
 \roundgate[1.25+2][3][1][topright][orange][2]
 \roundgate[1.25+4][3][1][topright][orange][2]
 \roundgate[-1.25][2][1][topright][orange][2]
 \roundgate[-1.25-2][2][1][topright][orange][2]
 \roundgate[-1.25-4][2][1][topright][orange][2]
  \roundgate[-1.25-6][2][1][topright][orange][2]
  \roundgate[1.25+1][2][1][topright][orange][2]
 \roundgate[1.25+3][2][1][topright][orange][2]
 \roundgate[1.25+5][2][1][topright][orange][2]
 \roundgate[-1.25-1][1][1][topright][orange][2]
 \roundgate[-1.25-3][1][1][topright][orange][2]
 \roundgate[-1.25-5][1][1][topright][orange][2]
 \roundgate[-1.25-7][1][1][topright][orange][2]
 \roundgate[1.25][1][1][topright][orange][2]
 \roundgate[1.25+2][1][1][topright][orange][2]
 \roundgate[1.25+4][1][1][topright][orange][2]
\roundgate[1.25+6][1][1][topright][orange][2]
\foreach \i in {-1, ..., 3}{
\thetastatevar[-7.25+2*\i][0][1.5][orange]
}
\foreach \i in {0, ..., 3}{
\thetastatevar[2*\i+2.25][0][1.5][orange]
}
\foreach \i in {0,...,2}{
\draw[thick, black] (-.75,\i+.5) -- (.75,\i+.5);}
\foreach \i in {0,...,2}{
\sqrstate[\i+.75][\i+3.5]
}
\foreach \i in {-1,...,4}{
\cstate[7.75-\i][\i+1.5]
}
\foreach \i in {-1,...,4}{
\cstate[\i-8.75][\i+1.5]
}
\sqrstate[-1.75-2][5.5]
\sqrstate[-1.75][3.5]
\sqrstate[-1.75-1][4.5]
\draw [decorate, decoration = {brace}, thick,shift={(0,2.)}]  (-3.75+.05,3.75+.05) -- (-1.5+.05,1.5+.05);
\node[scale=2]  at (-2.3,5) {\scalebox{0.7}{$k$}};
\foreach \i in {-2,...,1}{
\draw[very thick, dashed, white] (-2.75-.625-1+\i+0.4,2.75-.625+1+\i-0.4) -- (-1.5-.625-1+\i-0.4,1.5-.625+1+\i+0.4);
}
 \foreach \i in {0,...,3}{
\draw[very thick, dashed,white] (4.6-\i,.35+\i) -- (5-\i,.75+\i);
}
}. 
\end{equation}
and set $\mathcal A_{-1}=0$. We now make the following conjecture on the behavior of form of the amplitude $\mathcal A_k$
\begin{conjecture}
\label{conj:doubletriangle}
For large $|A|$ and $|\bar A|$ the leading order form of $\mathcal A_k$ reads as  
\begin{equation}
\label{eq:Akconj}
\mathcal A_k \simeq  d^{-|A|}  ( 1 + (k q({|A|,t_0}) + h({| A|,t_0})) e^{-\alpha t_0 } ). 
\end{equation}
where $\alpha>0$ and $q({x,y})$ and $h({x,y})$ are polynomials in $x$ and $y$. 
\end{conjecture}
This conjecture is proven in Appendix~\ref{app:proof} in the case when $\mathcal A_k$ is averaged over random dual-unitary gates (cf.~\cite{bertini2020scrambling, foligno2022growth}) with `critical' entangling power 
\begin{equation}
\label{eq:criticalep}
p = p_c \equiv 1-\frac{1}{d^2}, 
\end{equation}
while its validity in the general case is argued through the entanglement membrane approach~\cite{jonay2018coarsegrained, zhou2020entanglement}. Using Conjecture~\ref{conj:doubletriangle} in Eq.~\eqref{eq:pk} we find
\begin{equation}
p_k \simeq  
\left\lbrace\begin{aligned}
  &  q({|A|,t_0}) e^{ -\alpha t_0 }  & \quad k > 0 \\
  &  1 -  q({|A|,t_0}) (|D|-1) e^{ -\alpha t_0 }    & \quad k = 0 \\
\end{aligned} \right. \,. 
\end{equation}
Putting all together we find that the entanglement is upper bounded by
\begin{equation}
\begin{aligned}
S(\rho_A)  &\le  \sum_{k=1}^{|\bar A '|}  2k p_k \ln d +H(\{p_k\})  \\
&=  p_1 {|\bar A '|} ( {|\bar A '|}+ 1) \ln d - ({|\bar A '|}-1) p_1\ln p_1 - p_0 \ln p_0\,.
\end{aligned}
\end{equation}
The above bound implies that there exists a $K>0$ such that for 
\begin{equation}
t_0 \geq \frac{K}{2\ln d} \ln |A|, 
\end{equation}
the entanglement is $\mathcal{O}(1)$. Instead, for 
\begin{equation}
t_0 \leq \frac{K}{2\ln d} \ln |A|
\end{equation}
we can use a simple dimensional bound: By observing that the number of non-zero singular values of $\mathcal T_{t_0}$ (and therefore of non-zero eigenvalues of $\rho_A$) cannot exceed $d^{2|\bar A|}$ we have 
\begin{equation}
S(\rho_A) \leq 2 |\bar A| \ln d. 
\end{equation}
This immediately implies that $S(\rho_A)$ is upper bounded by $K \ln |A|$. Putting all together we can therefore conclude  
\begin{equation}
S(\rho_A)  \le K \ln |A|, \qquad \forall |A|. 
\end{equation}

\section{Proof of Conjecture~\ref{conj:doubletriangle}}
\label{app:proof}

In this appendix we compute exactly the average of the amplitude $\mathcal A_{k}$ in Eq.~\eqref{eq:amplitudeAk} over random dual unitary gates with `critical entangling power' in Eq.~\eqref{eq:criticalep}. The special nature of the $p=p_c$ point was recently unveiled in Ref.~\cite{suzuki2025global}, where it was shown that at that particular point the evaluation of certain two-replica quantities can be mapped to a free-fermions calculation. Instead, here we show that by making this special choice one attains drastic simplifications directly in the diagrammatic calculations leading to the full contraction of the diagram in Eq.~\eqref{eq:amplitudeAk}. 

Following Refs.~\cite{bertini2020scrambling, foligno2022growth} we consider dual unitary gates of the form 
\begin{equation}
U = (u_1\otimes u_2)  W (u_3\otimes u_4), 
\end{equation}
where $W$ is a fixed dual-unitary gate and $\{u_j\}$ are drawn randomly and independently from the Haar measure at each space-time point. Denoting the average over $\{u_j\}$ as $\mathbb E[\cdot]$ we have 
\begin{equation}
\mathbb E\left[(U \otimes U^*)^{\otimes n}\right] = \fineq[-0.8ex][0.75][1]{
    \roundgate[0][0][1][topright][ngray][n]
}, 
\end{equation}
becomes a $4\times4$ matrix that is expressed as 
\begin{equation}
\fineq[-0.8ex][0.75][1]{
    \roundgate[0][0][1][topright][ngray][n]
} = \begin{pmatrix}
1 & 0 &0 &0 \\
0 & 0 & 1-p & \frac{p}{\sqrt{d^2-1}} \\
0 & 1-p &0 & \frac{p}{\sqrt{d^2-1}}\\
0 & \frac{p}{\sqrt{d^2-1}} & \frac{p}{\sqrt{d^2-1}} & 1- \frac{2p}{{d^2-1}} \\
\end{pmatrix},
\end{equation}
where $p\in[0,1]$ is the `entangling power'~\cite{zanardi2000entangling} of the gate $W$, in the basis formed by $|\mcirc\rangle$ and its orthogonal state 
\begin{equation}
|\mcircf\rangle= \frac{d |\msquare\rangle-|\mcirc\rangle}{\sqrt{d^2-1}}. 
\end{equation}
Note that we choose to normalise these states to $d^2$. 

Let us now proceed to compute 
\begin{equation}
\label{eq:amplitudeAkave}
\mathbb E\left[\mathcal A_k\right] = \frac{1}{d^{|A'|-k}}
\fineq[-0.8ex][.65]{
 \roundgate[1.25+2][5][1][topright][ngray][2]
 \roundgate[-1.25-3][5][1][topright][ngray][2]
 \roundgate[-1.25-2][4][1][topright][ngray][2]
  \roundgate[-1.25-4][4][1][topright][ngray][2]
 \roundgate[1.25+1][4][1][topright][ngray][2]
 \roundgate[1.25+3][4][1][topright][ngray][2]
 \roundgate[-1.25-1][3][1][topright][ngray][2]
 \roundgate[-1.25-3][3][1][topright][ngray][2]
  \roundgate[-1.25-5][3][1][topright][ngray][2]
 \roundgate[1.25][3][1][topright][ngray][2]
 \roundgate[1.25+2][3][1][topright][ngray][2]
 \roundgate[1.25+4][3][1][topright][ngray][2]
 \roundgate[-1.25][2][1][topright][ngray][2]
 \roundgate[-1.25-2][2][1][topright][ngray][2]
 \roundgate[-1.25-4][2][1][topright][ngray][2]
  \roundgate[-1.25-6][2][1][topright][ngray][2]
  \roundgate[1.25+1][2][1][topright][ngray][2]
 \roundgate[1.25+3][2][1][topright][ngray][2]
 \roundgate[1.25+5][2][1][topright][ngray][2]
 \roundgate[-1.25-1][1][1][topright][ngray][2]
 \roundgate[-1.25-3][1][1][topright][ngray][2]
 \roundgate[-1.25-5][1][1][topright][ngray][2]
 \roundgate[-1.25-7][1][1][topright][ngray][2]
 \roundgate[1.25][1][1][topright][ngray][2]
 \roundgate[1.25+2][1][1][topright][ngray][2]
 \roundgate[1.25+4][1][1][topright][ngray][2]
\roundgate[1.25+6][1][1][topright][ngray][2]
\foreach \i in {-1, ..., 3}{
\thetastatevar[-7.25+2*\i][0][1.5][ngray]
}
\foreach \i in {0, ..., 3}{
\thetastatevar[2*\i+2.25][0][1.5][ngray]
}
\foreach \i in {0,...,2}{
\draw[thick, black] (-.75,\i+.5) -- (.75,\i+.5);}
\foreach \i in {0,...,2}{
\sqrstate[\i+.75][\i+3.5]
}
\foreach \i in {-1,...,4}{
\cstate[7.75-\i][\i+1.5]
}
\foreach \i in {-1,...,4}{
\cstate[\i-8.75][\i+1.5]
}
\sqrstate[-1.75-2][5.5]
\sqrstate[-1.75][3.5]
\sqrstate[-1.75-1][4.5]
\draw [decorate, decoration = {brace}, thick,shift={(0,2.)}]  (-3.75+.05,3.75+.05) -- (-1.5+.05,1.5+.05);
\node[scale=2]  at (-2.3,5) {\scalebox{0.7}{$k$}};
\foreach \i in {-2,...,1}{
\draw[very thick, dashed, white] (-2.75-.625-1+\i+0.4,2.75-.625+1+\i-0.4) -- (-1.5-.625-1+\i-0.4,1.5-.625+1+\i+0.4);
}
 \foreach \i in {0,...,3}{
\draw[very thick, dashed,white] (4.6-\i,.35+\i) -- (5-\i,.75+\i);
}
},
\end{equation}
where we painted also the initial state in grey to stress that it is restricted to the subspace spanned by $\{|\mcirc\rangle, |\mcircf\rangle\}$. We begin considering the tensor
\begin{equation}
\fineq[-0.8ex][.65]{
 \roundgate[-1.25-3][5][1][topright][ngray][2]
  \roundgate[-1.25-4][4][1][topright][ngray][2]
  \roundgate[-1.25-5][3][1][topright][ngray][2]
  \roundgate[-1.25-6][2][1][topright][ngray][2]
 \roundgate[-1.25-7][1][1][topright][ngray][2]
\thetastatevar[-7.25-2][0][1.5][ngray]
\foreach \i in {-1,...,4}{
\cstate[\i-8.75][\i+1.5]
}
\sqrstate[-1.75-2][5.5]
\draw [decorate, decoration = {brace}, thick,shift={(0,2.)}]  (-4.6+1,3.65-1) -- (-4.6-4.25+1,3.65-4.25-1);
\node[scale=2]  at (-4.25,2.25) {\scalebox{0.7}{$|\bar A|+k-1$}};
}, 
\end{equation}
appearing on the left edge of Eq.~\eqref{eq:amplitudeAkave}. Evaluating explicitly the action of the initial state on the empty bullet state we find 
\begin{equation}
\fineq[-0.8ex][.65]{
\thetastatevar[-7.25-2][0][1.5][ngray]
\cstate[-9.75][.5]
} =\frac{1}{d^2} \fineq[-0.8ex][.65]{
\draw [thick] (-1,0) -- (0,0); 
\cstate[-1][0]
} +\frac{c-1}{d^2\sqrt{d^2-1}} \fineq[-0.8ex][.65]{
\draw [thick] (-1,0) -- (0,0); 
\cstate[-1][0][][black]
}, \qquad \qquad c\equiv d \fineq[-0.8ex][.65]{
\thetastatevar[-7.25-2][0][1.5][ngray]
\cstate[-9.75][.5]
\sqrstate[-8.75][.5]
}.  
\end{equation}
Substituting this decomposition we find 
\begin{equation}
\label{eq:statesimplification}
\begin{aligned}
\fineq[-0.8ex][.65]{
 \roundgate[-1.25-3][5][1][topright][ngray][2]
  \roundgate[-1.25-4][4][1][topright][ngray][2]
  \roundgate[-1.25-5][3][1][topright][ngray][2]
  \roundgate[-1.25-6][2][1][topright][ngray][2]
 \roundgate[-1.25-7][1][1][topright][ngray][2]
\thetastatevar[-7.25-2][0][1.5][ngray]
\foreach \i in {-1,...,4}{
\cstate[\i-8.75][\i+1.5]
}
\sqrstate[-1.75-2][5.5]
\draw [decorate, decoration = {brace}, thick,shift={(0,2.)}]  (-4.6+1,3.65-1) -- (-4.6-4.25+1,3.65-4.25-1);
\node[scale=2]  at (-4.75,2.25) {\scalebox{0.7}{$x$}};
} & = \frac{1}{d^2} \fineq[-0.8ex][.65]{
 \roundgate[-1.25-3][5][1][topright][ngray][2]
  \roundgate[-1.25-4][4][1][topright][ngray][2]
  \roundgate[-1.25-5][3][1][topright][ngray][2]
  \roundgate[-1.25-6][2][1][topright][ngray][2]
 \roundgate[-1.25-7][1][1][topright][ngray][2]
\foreach \i in {0,...,4}{
\cstate[\i-8.75][\i+1.5]
\cstate[-1.75-2-5][5.5-5]
}
\sqrstate[-1.75-2][5.5]
} + \frac{c-1}{d^2\sqrt{d^2-1}} \fineq[-0.8ex][.65]{
 \roundgate[-1.25-3][5][1][topright][ngray][2]
  \roundgate[-1.25-4][4][1][topright][ngray][2]
  \roundgate[-1.25-5][3][1][topright][ngray][2]
  \roundgate[-1.25-6][2][1][topright][ngray][2]
 \roundgate[-1.25-7][1][1][topright][ngray][2]
\foreach \i in {0,...,4}{
\cstate[\i-8.75][\i+1.5]
}
\cstate[-1.75-2-5][5.5-5][][black]
\sqrstate[-1.75-2][5.5]
} \\
& = \frac{1}{d} \fineq[-0.8ex][.65]{
\foreach \i in {0,...,4}{
\draw [thick] (\i-8.75,\i+1.5) -- (\i-8.75+0.5,\i+1.5-0.5); 
\cstate[\i-8.75][\i+1.5]
}
} + \frac{c-1}{d^{1+2x}} \fineq[-0.8ex][.65]{
\foreach \i in {0,...,4}{
 \roundgate[1.25*\i-8.25][1.25*\i+1][1][topright][ngray][2]
\cstate[1.25*\i-8.75][1.25*\i+1.5]
\cstate[1.25*\i-8.75][1.25*\i+.5][][black]
\cstate[1.25*\i-8.75+1][1.25*\i+.5+1][][black]
}
}
\end{aligned}
\end{equation}
where in the second step we used the dual-unitary relations
\begin{equation}
\fineq[-0.8ex][.65]{
 \roundgate[0][0][1][topright][ngray][2]
 \cstate[-.5][.5]
 \cstate[-.5][-.5]
} = \fineq[-0.8ex][.65]{
 \draw [thick] (.5,.5) -- (-.5,0.5); 
 \draw [thick] (.5,-.5) -- (-.5,-0.5); 
 \cstate[-.5][.5]
 \cstate[-.5][-.5]
},\qquad \fineq[-0.8ex][.65]{
 \roundgate[0][0][1][topright][ngray][2]
 \cstate[.5][.5]
 \cstate[.5][-.5]
} = \fineq[-0.8ex][.65]{
 \draw [thick] (.5,.5) -- (-.5,0.5); 
 \draw [thick] (.5,-.5) -- (-.5,-0.5); 
 \cstate[.5][.5]
 \cstate[.5][-.5]
}.
\end{equation}
to simplify the first term. Instead in the second term we made repeated use of the identity 
\begin{equation}
 \fineq[-0.8ex][.65]{
 \draw [thick] (.5,0) -- (-.5,0);
}= \frac{1}{d^2}\fineq[-0.8ex][.65]{
 \draw [thick] (.75,0) -- (0.25,0); 
 \draw [thick] (-0.25,0) -- (-.75,0); 
 \cstate[.25][0]
 \cstate[-.25][0]
}+ \frac{1}{d^2}\fineq[-0.8ex][.65]{
 \draw [thick] (.75,0) -- (0.25,0); 
 \draw [thick] (-0.25,0) -- (-.75,0); 
 \cstate[.25][0][][black]
 \cstate[-.25][0][][black]
}
\end{equation}
and dual-unitarity conditions to simplify the white bullets against their orthogonal state. 

We now observe 
\begin{equation}
\frac{1}{d^2} \fineq[-0.8ex][.65]{
\begin{scope}[rotate around={45:(0,0)}]
\roundgate[0][0][1][topright][ngray][2]
 \cstate[.5][0.5][][black]
 \cstate[-.5][-0.5][][black]
  \cstate[-.5][0.5]
\end{scope}
} = \frac{p}{\sqrt{d^2-1}}\fineq[-0.8ex][.65]{
 \draw [thick] (.75,0) -- (0.25,0); 
 \cstate[.25][0][][black]
}+(1-p)\fineq[-0.8ex][.65]{
 \draw [thick] (.75,0) -- (0.25,0); 
 \cstate[.25][0]
}=\frac{1}{d}\fineq[-0.8ex][.65]{
 \draw [thick] (.75,0) -- (0.25,0); 
 \sqrstate[.25][0] 
}+\frac{d}{\sqrt{d^2-1}}\left(1-\frac{1}{d^2}-p\right)\fineq[-0.8ex][.65]{
 \draw [thick] (.75,0) -- (0.25,0); 
 \sqrstate[.25][0][][black] 
},
\end{equation}
which means that at the critical entangling power we have 
\begin{equation}
\frac{1}{d^2} \fineq[-0.8ex][.65]{
\begin{scope}[rotate around={45:(0,0)}]
\roundgate[0][0][1][topright][ngray][2]
 \cstate[.5][0.5][][black]
 \cstate[-.5][-0.5][][black]
  \cstate[-.5][0.5]
\end{scope}
} =\frac{1}{d}\fineq[-0.8ex][.65]{
 \draw [thick] (.75,0) -- (0.25,0); 
 \sqrstate[.25][0] 
},
\end{equation}
and Eq.~\eqref{eq:statesimplification} becomes
\begin{equation}
\label{eq:statesimplificationpc}
\begin{aligned}
\fineq[-0.8ex][.65]{
 \roundgate[-1.25-3][5][1][topright][ngray][2]
  \roundgate[-1.25-4][4][1][topright][ngray][2]
  \roundgate[-1.25-5][3][1][topright][ngray][2]
  \roundgate[-1.25-6][2][1][topright][ngray][2]
 \roundgate[-1.25-7][1][1][topright][ngray][2]
\thetastatevar[-7.25-2][0][1.5][ngray]
\foreach \i in {-1,...,4}{
\cstate[\i-8.75][\i+1.5]
}
\sqrstate[-1.75-2][5.5]
\draw [decorate, decoration = {brace}, thick,shift={(0,2.)}]  (-4.6+1,3.65-1) -- (-4.6-4.25+1,3.65-4.25-1);
\node[scale=2]  at (-4.75,2.25) {\scalebox{0.7}{$x$}};
} = \frac{1}{d} \fineq[-0.8ex][.65]{
\foreach \i in {0,...,4}{
\draw [thick] (\i-8.75,\i+1.5) -- (\i-8.75+0.5,\i+1.5-0.5); 
\cstate[\i-8.75][\i+1.5]
}
} + \frac{c-1}{d^{1+x}} \fineq[-0.8ex][.65]{
\foreach \i in {0,...,4}{
\draw [thick] (\i-8.75,\i+1.5) -- (\i-8.75+0.5,\i+1.5-0.5); 
\sqrstate[\i-8.75][\i+1.5]
}
}.  
\end{aligned}
\end{equation}
This means that Eq.~\eqref{eq:amplitudeAkave} fulfils
\begin{equation}
\label{eq:recursion1}
\mathbb E\left[\mathcal A_k\right] =  \mathbb E\left[\mathcal A_{k-1}\right] +  \frac{c-1}{d^{t_0+|A'|}} \mathcal B_{{t_0+t_1}}, 
\end{equation}
where we set $t_1=\lceil |A|/2 \rceil$ and introduced 
\begin{equation}
 \mathcal B_{x} = 
\fineq[-0.8ex][.65]{
 \roundgate[1.25+2][5][1][topright][ngray][2]
 \roundgate[1.25+1][4][1][topright][ngray][2]
 \roundgate[1.25+3][4][1][topright][ngray][2]
 \roundgate[1.25][3][1][topright][ngray][2]
 \roundgate[1.25+2][3][1][topright][ngray][2]
 \roundgate[1.25+4][3][1][topright][ngray][2]
   \roundgate[1.25-1][2][1][topright][ngray][2]
  \roundgate[1.25+1][2][1][topright][ngray][2]
 \roundgate[1.25+3][2][1][topright][ngray][2]
 \roundgate[1.25+5][2][1][topright][ngray][2]
  \roundgate[1.25-2][1][1][topright][ngray][2]
 \roundgate[1.25][1][1][topright][ngray][2]
 \roundgate[1.25+2][1][1][topright][ngray][2]
 \roundgate[1.25+4][1][1][topright][ngray][2]
\roundgate[1.25+6][1][1][topright][ngray][2]
\foreach \i in {-2, ..., 3}{
\thetastatevar[2*\i+2.25][0][1.5][ngray]
}
\foreach \i in {-3,...,2}{
\sqrstate[\i+.75][\i+3.5]
}
\foreach \i in {-1,...,4}{
\cstate[7.75-\i][\i+1.5]
}
\draw [decorate, decoration = {brace}, thick,shift={(0,2.)}]  (7.5+-3.75+.05,3.75+.05) -- (7.5+-1.5+.05+3,1.5+.05-3);
\node[scale=2]  at (7.5-2.3+1.5,5-1.5) {\scalebox{0.7}{$x$}};
}.
\end{equation}
Solving Eq.~\eqref{eq:recursion1} we have 
\begin{equation}
\mathbb E\left[\mathcal A_k\right] =  \mathbb E\left[\mathcal A_{0}\right] +  \frac{c-1}{d^{t_0+|A'|}} \mathcal B_{{t_0+t_1}} k. 
\end{equation}
To make this expression explicit we use the mirrored version of Eq.~\eqref{eq:statesimplificationpc} to find 
\begin{equation}
 \mathcal B_{x}  =  \frac{1}{d}\mathcal B_{x-1} + \frac{c-1}{d^{x}} \implies  \mathcal B_{x}  =  \frac{1}{d^x} + (c-1) \frac{x}{d^x}, 
\end{equation}
where we used $\mathcal B_{1}=c/d$. Moreover, we set $\mathbb E\left[\mathcal A_{0}\right]= \mathcal C_{t_1+t_0}/{d^{|A'|}}$ where we introduced
\begin{equation}
\mathcal C_y = 
\fineq[-0.8ex][.65]{
 \roundgate[1.25+2][5][1][topright][ngray][2]
 \roundgate[1.25+1][4][1][topright][ngray][2]
 \roundgate[1.25+3][4][1][topright][ngray][2]
 \roundgate[1.25][3][1][topright][ngray][2]
 \roundgate[1.25+2][3][1][topright][ngray][2]
 \roundgate[1.25+4][3][1][topright][ngray][2]
   \roundgate[1.25-1][2][1][topright][ngray][2]
  \roundgate[1.25+1][2][1][topright][ngray][2]
 \roundgate[1.25+3][2][1][topright][ngray][2]
 \roundgate[1.25+5][2][1][topright][ngray][2]
  \roundgate[1.25-2][1][1][topright][ngray][2]
 \roundgate[1.25][1][1][topright][ngray][2]
 \roundgate[1.25+2][1][1][topright][ngray][2]
 \roundgate[1.25+4][1][1][topright][ngray][2]
\roundgate[1.25+6][1][1][topright][ngray][2]
\foreach \i in {-2, ..., 3}{
\thetastatevar[2*\i+2.25][0][1.5][ngray]
}
\foreach \i in {0,...,2}{
\sqrstate[\i+.75][\i+3.5]
}
\foreach \i in {-3,...,-1}{
\cstate[\i+.75][\i+3.5]
}
\foreach \i in {-1,...,4}{
\cstate[7.75-\i][\i+1.5]
}
\draw [decorate, decoration = {brace}, thick,shift={(0,2.)}]  (7.5+-3.75+.05,3.75+.05) -- (7.5+-1.5+.05+3,1.5+.05-3);
\node[scale=2]  at (7.5-2.3+1.5,5-1.5) {\scalebox{0.7}{$y$}};
\draw [decorate, decoration = {brace}, thick,shift={(0,2.)}]  (-2.5,-1.75+.5) -- (-3+6-3.5,4.25-3.5);
\node[scale=2]  at (-2,2) {\scalebox{0.7}{$t_0$}};
}.
\end{equation}
Using again the mirrored version of Eq.~\eqref{eq:statesimplificationpc} we find 
\begin{equation}
\mathcal C_y = \frac{1}{d}\mathcal C_{y-1} + \frac{c-1}{d^{y}} \mathcal B_{t_0} \implies \mathcal C_y = \frac{d^{t_0}}{d^y}+\frac{(c-1)(y-t_0) \mathcal B_{t_0}}{d^y},  
\end{equation}
where we used $\mathcal C_{t_0}=1$. Putting all together we then have 
\begin{equation}
\mathbb E\left[\mathcal A_k\right] = \frac{1}{d^{|A|}} + \frac{(c-1) (t_1 + k +(c-1)( t_1  t_0 + k t_0+ k t_1))}{d^{|A|+2 t_0}}.    
\end{equation}
This expression can be recast in the form of Eq.~\eqref{eq:Akconj}.

\section{Membrane calculation away from the critical point}
\label{app:membrane}

In this section we use the membrane picture to characterise $\mathcal{A}_k$. We will not review here the membrane formalism or its microscopic permutation constituents. Instead, we assume familiarity with the discussion in Ref.~\cite{foligno2023temporal}, where the same framework was applied to the influence matrix.

\begin{figure}[h!]
\centering
\includegraphics[width=\columnwidth]{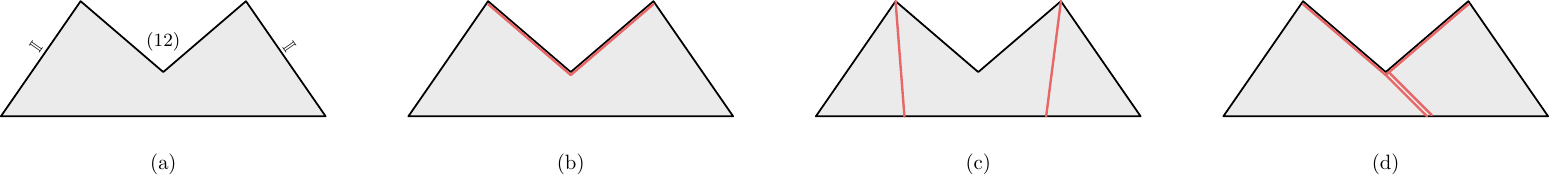}
\caption{The permutation spins and domain wall configurations for the two-peak mountain shape diagram defined in Eq.~\eqref{eq:amplitudeAk}. The size of the left ramp is $k$, the size of the right ramp is $t_1$. The total height of right peak is $t = t_1 + t_0$.  (a) Boundary permutation spins. (b) Leading contribution, in which the two domain walls follow the shortest trajectories and annihilate at the bottom of the valley. (c) Subleading contribution from two domain walls that remain separate all the way to the bottom. In this case the endpoints at the bottom can vary, leading to a degeneracy factor. (d) Subleading contribution from a magnon, namely a bound state of two domain walls. In dual unitary circuits, the magnon propagates only along the light cone direction, so there is no degeneracy factor.}
\label{fig:du_membrane}
\end{figure}

We begin by noting that, since $p_k$ is given by the difference of $\mathcal{A}_k$ and $\mathcal{A}_{k-1}$, the leading contributions cancel and the scaling of $p_k$ is controlled by the first subleading term. The membrane picture therefore gives a qualitative prediction away from the solvable critical point $p = p_c$. The boundary spin configuration is shown in Fig.~\ref{fig:du_membrane}~(a). It has the identity permutation along the two mountain slopes and the swap permutation $(12)$ along the valley ridges. Consequently, two domain walls emerge at the tips of the peaks. In our convention, these walls propagate downward toward the bottom of the valley. The amplitude $\mathcal{A}_k$ can then be viewed as a partition function that sums the weights of all such domain wall configurations.

In dual unitary circuits, the membrane tension of a domain wall takes its maximal value, $\ln d$. Once the two walls meet and annihilate, this cost disappears. The leading contribution is therefore obtained when the two domain walls follow the shortest trajectories that allow them to meet and annihilate. They move toward each other along the inner ridge and annihilate at the bottom of the valley, as shown in Fig.~\ref{fig:du_membrane}~(b). This gives a contribution of order $1/{d^{k + t_1}}$. There are perturbative corrections to this process, for instance when the domain walls meet slightly above or below the valley bottom. These trajectories are exponentially suppressed relative to $1/{d^{k + t_1}}$ as a function of the displacement from the bottom of the valley. They therefore form a geometric series and only renormalize the prefactor of $1/{d^{k + t_1}}$ without changing the exponent.

The next contribution comes from configurations in which the two domain walls survive throughout the diagram and never annihilate. The analysis is similar to that of the two stage thermalization mechanism discussed in Ref.~\cite{jonay2024physical,jonay2025twostage}, where subleading membrane contributions also appear at second order. There are two possibilities. In the first, the two domain walls wander independently all the way to the bottom as in Fig.~\ref{fig:du_membrane}~(c). The corresponding weight is of order
\begin{equation}
  \text{degeneracy}\times d^{-(k + 2t_0 + t_1)}.
\end{equation}
The prefactor is proportional to the number of degenerate paths. This degeneracy arises because, in dual unitary circuits, the membrane cost per unit time is independent of direction. 

In the second possibility, the two domain walls bind into a magnon, namely a bound state, as shown in Fig.~\ref{fig:du_membrane}~(d). In this case, the weight scales as
\begin{equation}
  d^{-(k + t_1)} d^{- r_{\rm mag} t_0}.
\end{equation}
Here $r_{\rm mag} < 1$. In a dual unitary circuit, the magnon can propagate only on the light cone, so the degeneracy factor is $1$. The crossover between the domain wall cost and the magnon cost occurs at the critical entangling power $p_c$. Therefore, for $p<p_c$ the magnon contribution dominates, whereas for $p>p_c$ the domain wall contribution dominates.
 
Numerical data for the disorder averaged $\mathcal{A}_k$ show these two subleading decay rates on the two sides of $p_c$. In the magnon phase, we find a degeneracy factor equal to $1$. In the domain wall phase, the degeneracy factor is proportional to both $t_1$ and $k$. For clean translation invariant circuits, numerical data for $\mathcal{A}_k$ still show an exponential decay in $t_0$. In part of the parameter range, the rate is consistent with the domain wall and magnon predictions, but other rates can also appear, suggesting more complicated modes in the clean limit. 

\section{The decay of the probabilities of the projections}
\label{app:pk}

In this section, we provide further numerical evidence that $p_k$ decays exponentially with $t_0$, and investigate how the decay rates are related to the emergent modes in the circuits. 

In the membrane analysis of the random averaged circuit (Sec.~\ref{app:membrane}), the rate of such exponential decay is given by either the magnon rate or domain wall rate. The transition $p_c = 1 - {1}/{d^2}$ for entangling power $p$ is obtained by comparing the averaged magnon decay rate $r_{\rm mag, avg} = -{\ln ( 1- p)}/{\ln d}$, with the rate of the two free domain walls, which is 2. For qubit systems, the subleading term of $p_k$ explores only the magnon regime because the entangling power $p$ of a single gate takes values in $[0, 2/3)$ while the transition point $p = 1 - {1}/{d^2}$ is ${3}/{4}$ for qubits. 

\begin{figure*}[h]
    \centering
    \includegraphics[width=0.3\linewidth]{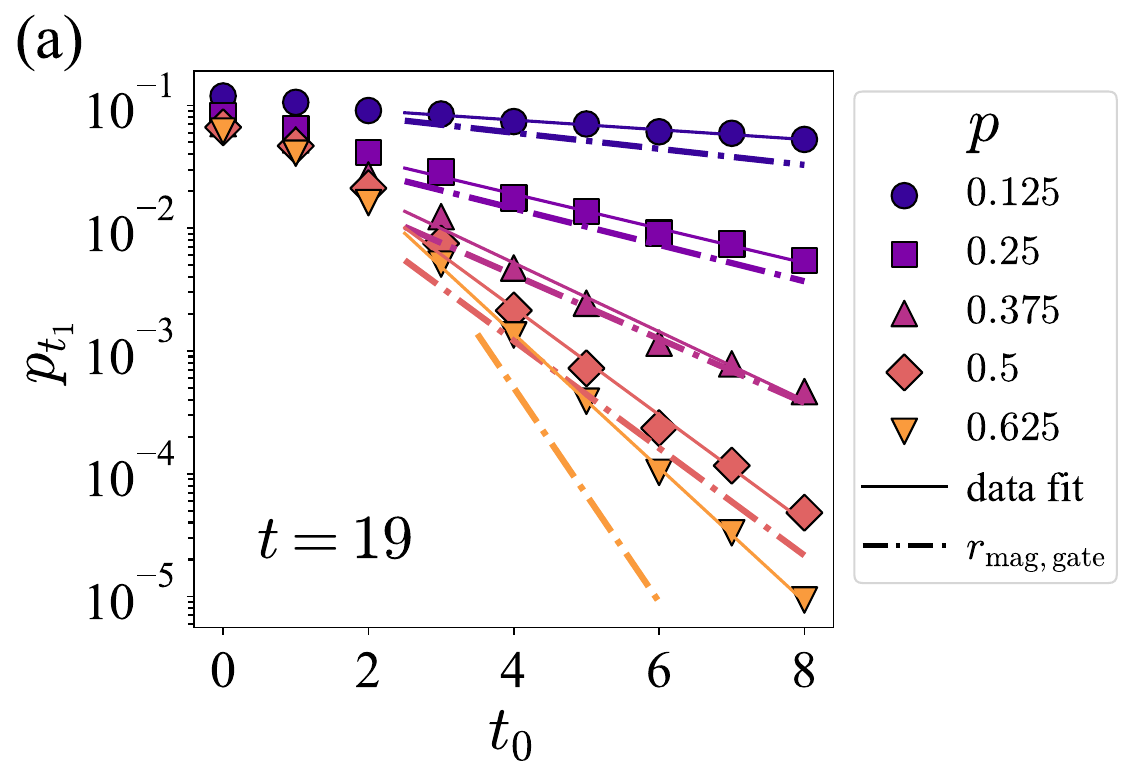} 
    \includegraphics[width=0.34\linewidth]  {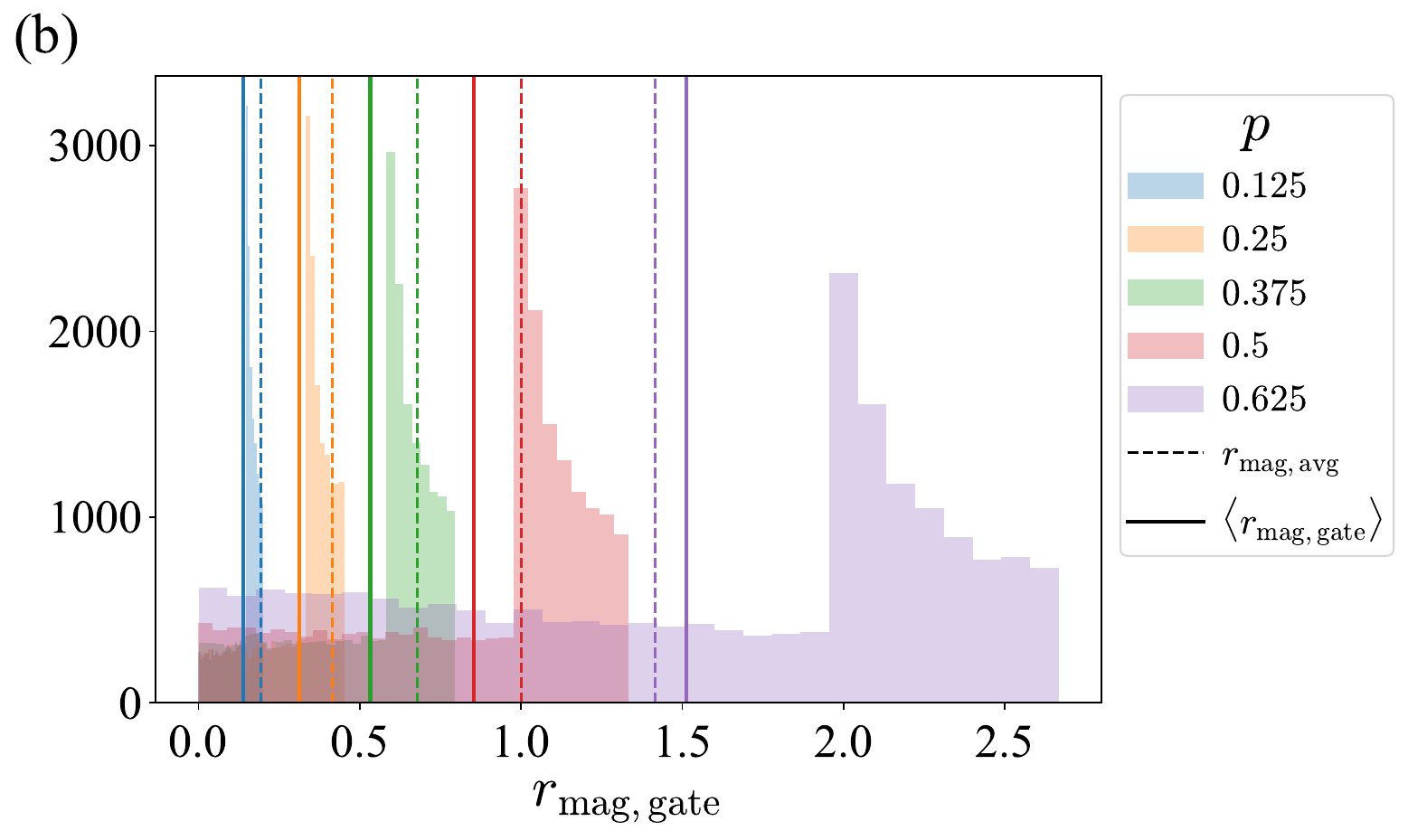}
    \includegraphics[width=0.34\linewidth]{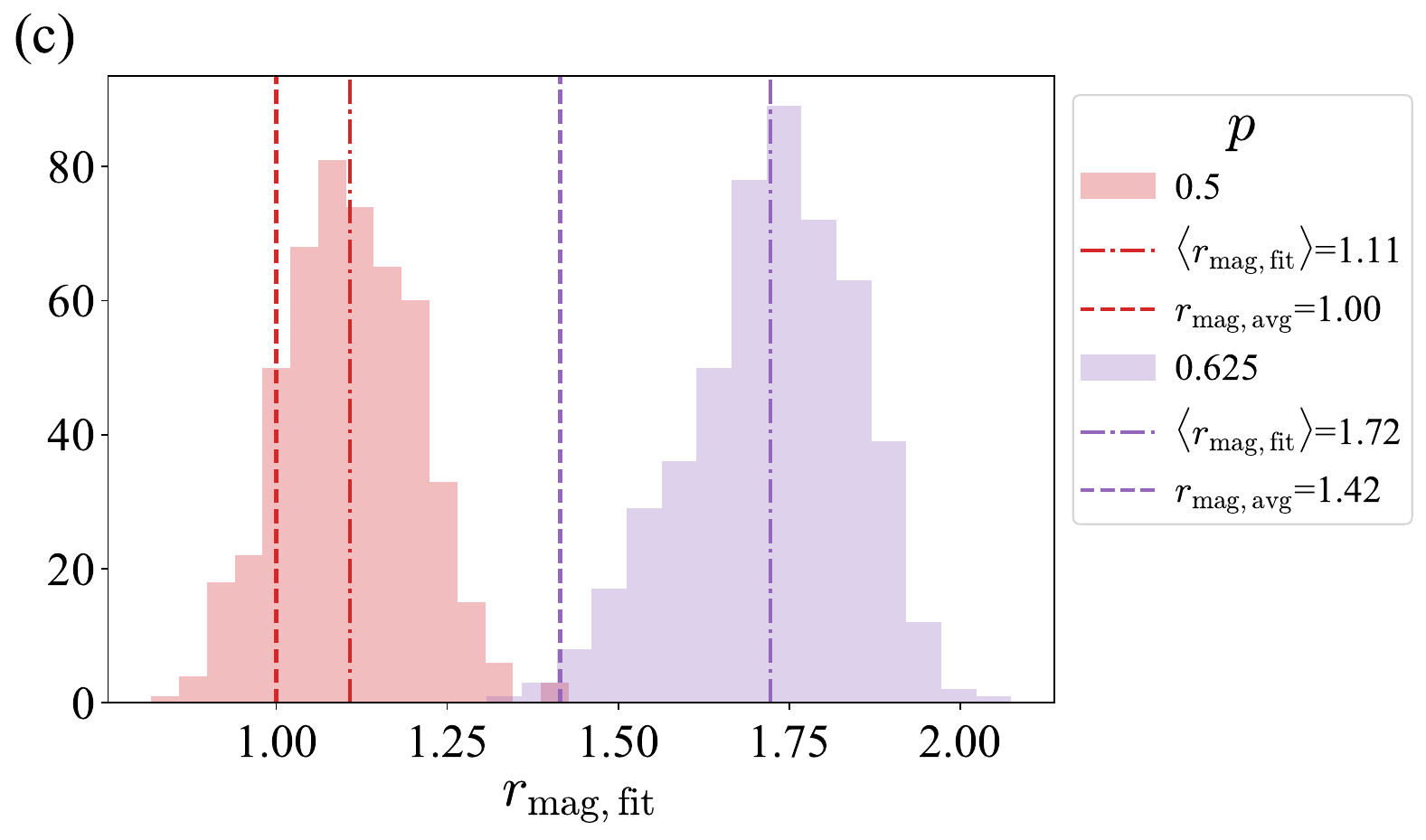}
\caption{\label{fig:pkdecay} (a) The decay of $p_k$ for the qubit dual unitary circuit at different entangling power $p \in \{0.125, 0.25, 0.375, 0.5, 0.625\}$. The dashed lines provide a guide to the eye with slopes equal to the decay rates $r_{\rm mag, gate}$ computed from the channel eigenvalue for the gate $W_{\rm sym}(p)$ used in the circuit.  (b) The distribution of magnon decay rates $r_{\rm mag, gate}$ by sampling the single qubit gates $\{u,v\}$ in $W_{\rm sym}(p)$ over the Haar ensemble. The sample size is $2\times 10^4$. The solid line marks the sample average $\langle r_{\rm mag, gate} \rangle$ and the dashed line marks the annealed average $r_{\rm mag, avg} = -\log (1-p)/ \log (2)$. (c) The distribution of the fitted decay rates $r_{\rm mag, fit}$ of $p_{t_1}$ while randomizing the four single qubit gates in $W(p)$ at entangling powers 0.5 and 0.625. The sample size is $5 \times 10^2$. The dot-dashed line marks the sample average $\langle r_{\rm mag, fit} \rangle$ and the dashed line marks the annealed average $r_{\rm mag, avg}$, same as (b). }
\end{figure*}

A magnon propagates along the light cones in a dual-unitary circuit, either to the left or to the right. The decay rate $r_{\rm mag, gate}$ of a thin magnon can be computed from the leading subleading eigenvalue of 
\begin{equation}
\fineq[-0.8ex][.65]{
 \roundgate[0][0][1][topright][bertiniorange][2]
 \cstate[-.5][.5]
 \cstate[.5][-.5]
}
\hspace{1 cm} {\rm or } \hspace{1 cm}
\fineq[-0.8ex][.65]{
 \roundgate[0][0][1][topright][bertiniorange][2]
 \cstate[-.5][-.5]
 \cstate[.5][.5]
}.
\end{equation} 
The rates for the left-moving and right-moving gates can be different. To simplify, we use a left-right symmetric gate in the form $W_{\mathrm{sym}}(p) = (u \otimes u) U(p) (v \otimes v)$ with $u,v$ chosen from the Haar ensemble and then fixed throughout the circuit simulation. We choose $p_{t_1}$ as the last element of $p_k$ to track the decay of $p_k$ for $k \ge 1$ as a function of $t_0$ (Fig.~\ref{fig:pkdecay}(a)). The rates extracted from the numerical data $r_{\rm fit}$ agree with the single gate magnon decay rates $r_{\rm mag, gate}$ rather than $r_{\rm mag, avg}$ at small $p$, while $r_{\rm fit} < r_{\rm mag, gate}$ for larger $p$.

\begin{figure*}[h!]
    \centering
    \includegraphics[width=0.34\linewidth]{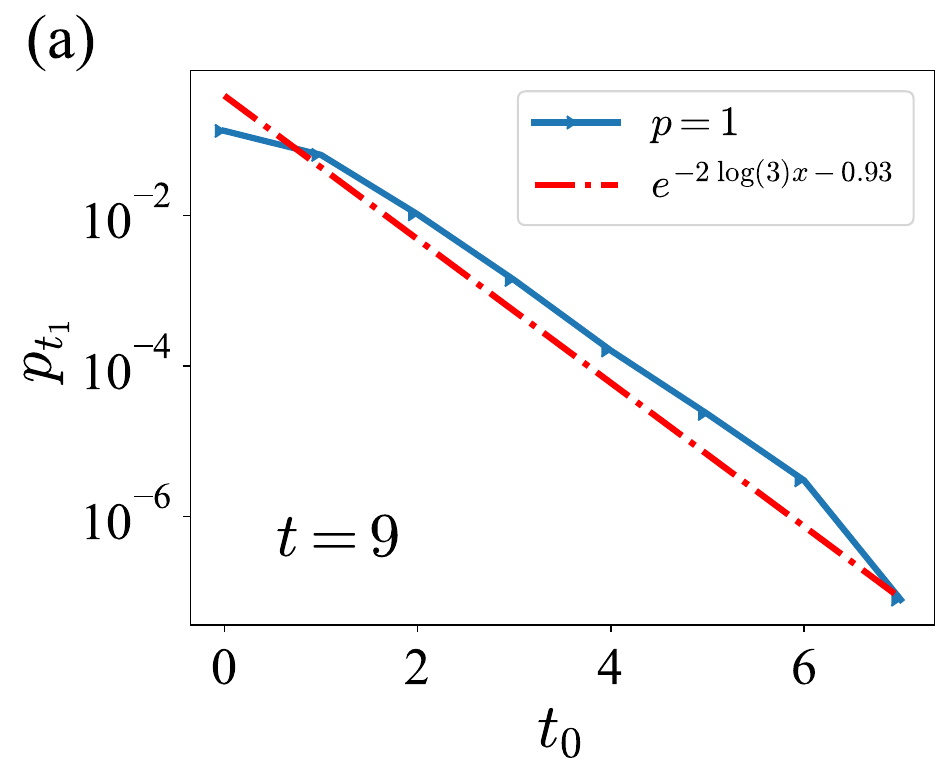}
    \hspace{0.3cm}
    \includegraphics[width=0.34\linewidth]{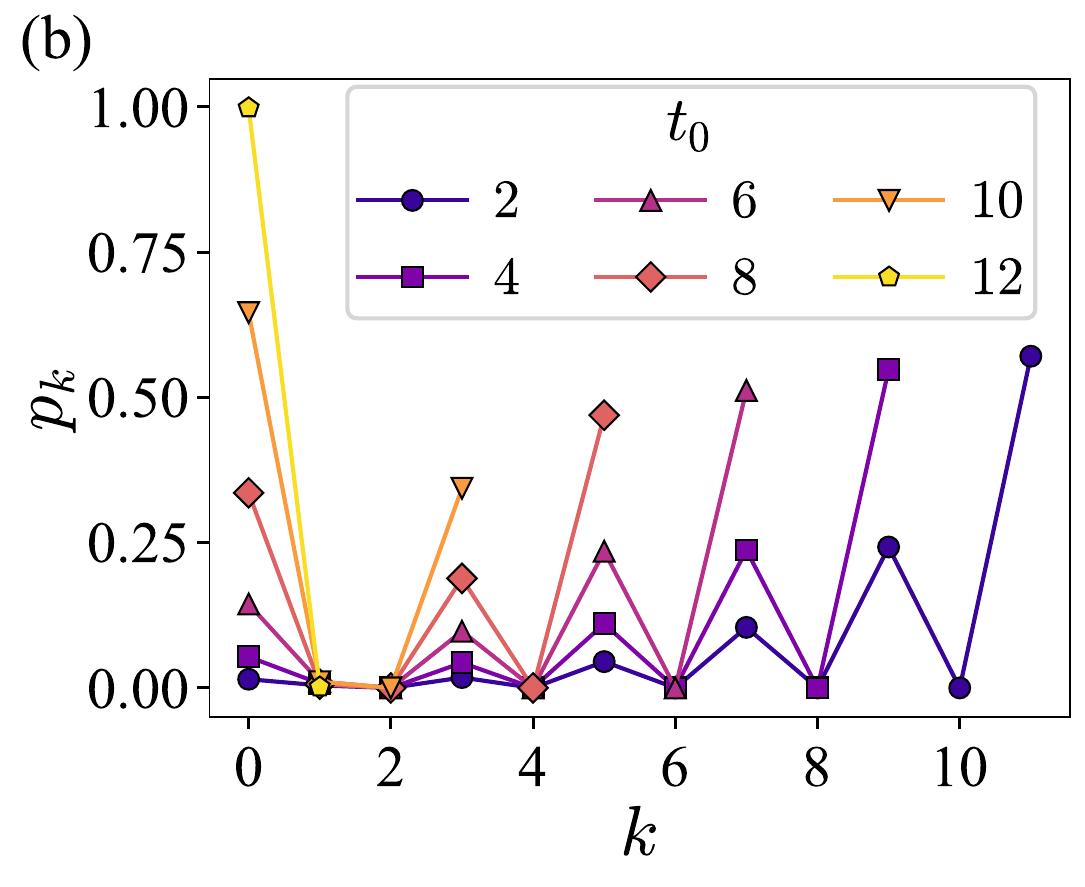}   
\caption{\label{fig:qutrit_Haar} (a) The decay of probability from the last projection operator in the dual unitary circuit for qutrits. (b) $p_k$ values for the Haar random circuit.}
\end{figure*}

To understand the mismatch, we collect the magnon rates at a fixed entangling power $p$ by sampling the $\{u,v \}$ gates over the Haar ensemble, which yields a skewed distribution across different realizations of the single-qubit random gate (Fig.~\ref{fig:pkdecay}(b)). The distribution widens as $p$ increases and the average is below the typical magnon decay rate upon picking a random dual unitary gate, in contrast to the domain wall decay rate, which remains concentrated around its random average value. This explains why $r_{\rm mag, avg}$ cannot accurately predict $r_{\rm mag, gate}$ when $p$ is large. Meanwhile, the magnon in the circuit can span more than one gate, and the rate can therefore come from channels involving multiple gates. Consequently, $r_{\rm fit} \le r_{\rm mag, gate}$. Lastly, as a consistency check, we note that $r_{\rm mag,avg}$ is the annealed average of the magnon rates over random single-qubit gates, so it sits in the left tail of the distribution rather than its mean (Fig.~\ref{fig:pkdecay}(c)), while the sample mean corresponds to the quenched average over the ensemble.

To investigate the domain wall phase of the membrane theory, we use a perfect tensor as the 2-qutrit gate, where the entangling power is 1 by definition. We observe that the decay rate of $p_{t_1}$ is indeed captured by that of two domain walls, which is 2 (Fig.~\ref{fig:qutrit_Haar}(a)).

Finally, for reference, we include the values of $p_k$ computed for a fixed realization of a Haar random circuit. In this case, the projectors act on region $A$ of the reduced transition matrix, namely on the vertical cut rather than on the reduced region $A'$ that appears after the dual unitary simplification. Here, we see that the values of $p_k$ do not decay monotonically. Therefore, the analytic $\ln(t)$ bound for dual-unitary circuits does not directly apply. However, the general technique of bounding the von Neumann entropy through the decomposition of the state still applies here. In the main text Fig.~\ref{fig:pkfixedrealisation}(c), the von Neumann entropy is still bounded above by a logarithmic dependence over the time range we can probe.

\section{Decay of the singular values of the reduced transition matrix}
\begin{figure*}[h]
    \centering
    \includegraphics[width=0.35\linewidth]{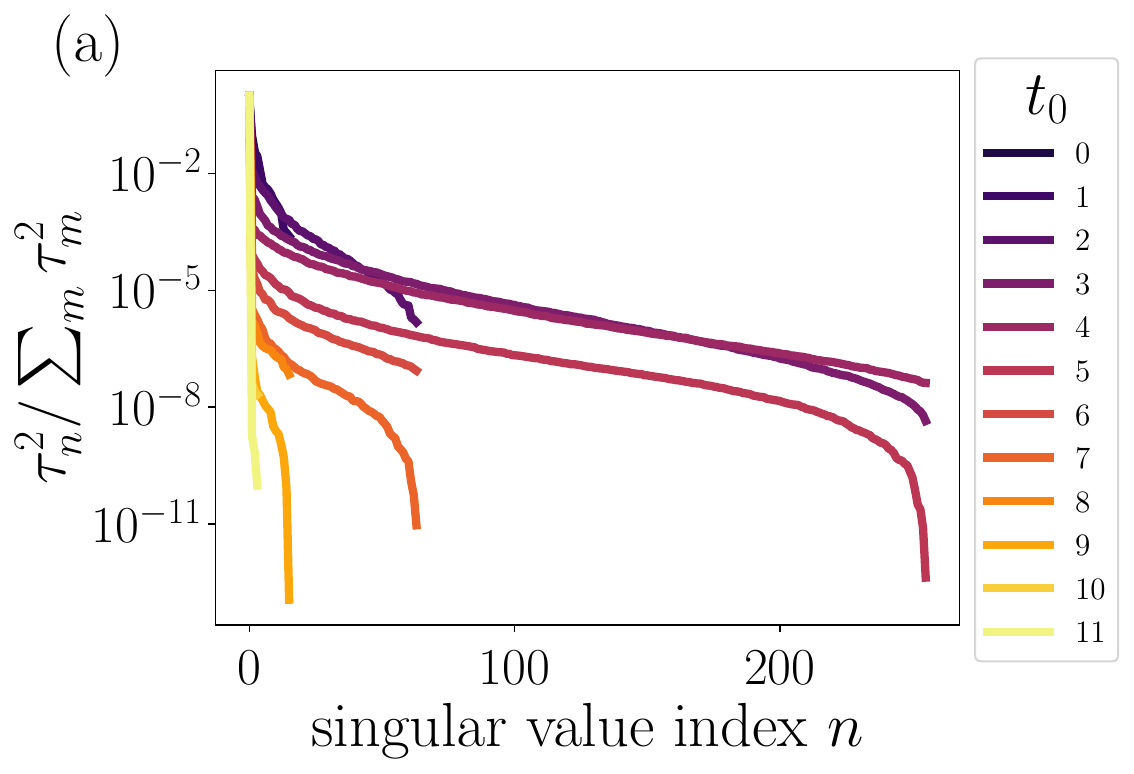} 
    \includegraphics[width=0.3\linewidth]  {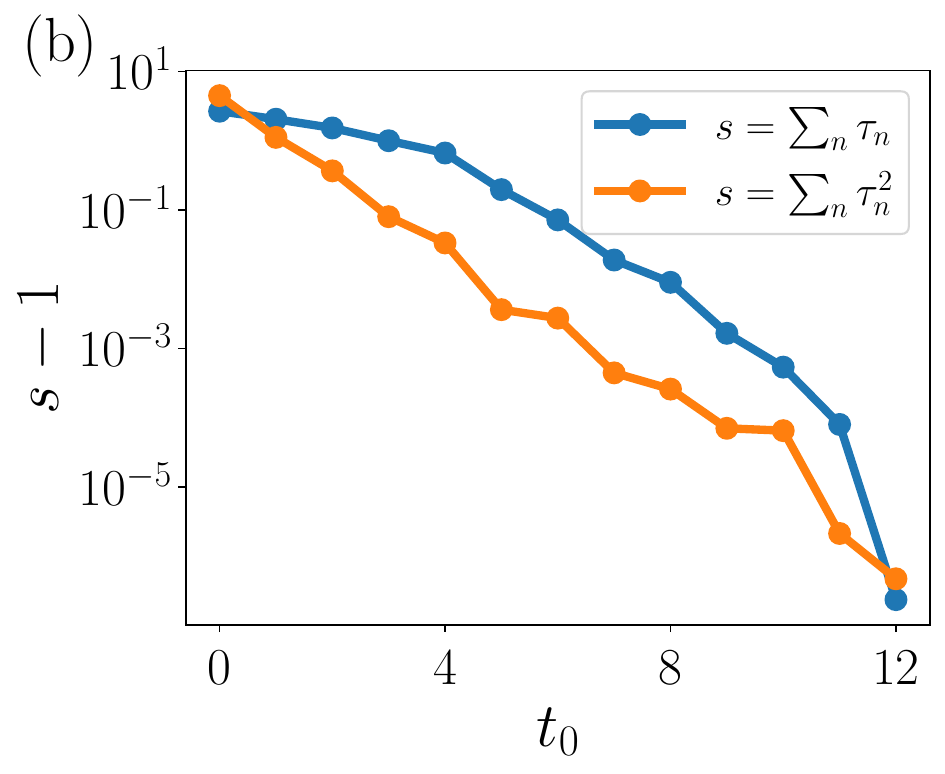}
    \includegraphics[width=0.3\linewidth]{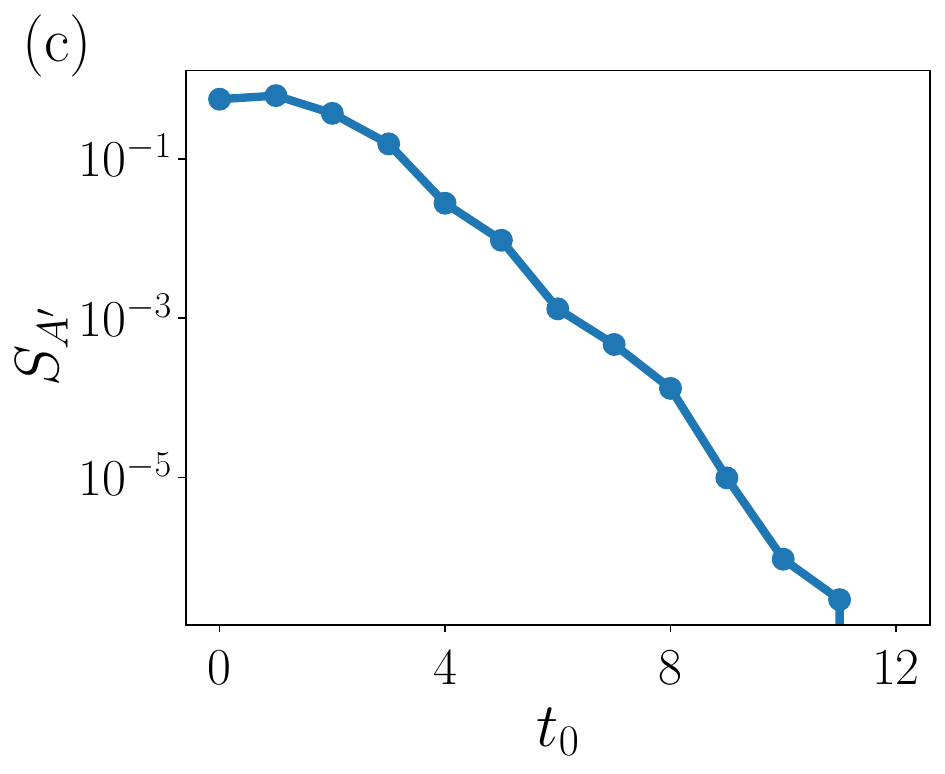}
\caption{\label{fig:Sdecay} The entangling power is set to 0.625 and $t=13$. (a) The singular values of the reduced transition matrix at $t = 13$ with different cut depths $|A|$, which translates to different $t_0$ values. The number of singular values is the minimum of $t_0 + 1$ and $t_1$. (b) The sum of the singular values and the sum of the singular values squared approach 1 exponentially. (c) The entanglement entropies computed from the singular values in (a). }
\end{figure*}

In this section we display the details of the singular values that corroborate the simulability of the reduced transition matrix. At a fixed $t$, the maximum of the entropy occurs at $t_0 = 1$, which corresponds to the tensor in Fig.~\ref{fig:Cdiag}(d) with only two bottom pairs of legs connected in our convention. For this cut of the matrix, the singular values decay exponentially as shown in Fig.~\ref{fig:Sdecay} (a). The sum of the singular values decays exponentially with $t_0$ towards 1. 

Fig.~\ref{fig:Sdecay} (c) presents the entropy $S_{A'}$ varying $t_0$. The maximum is achieved at $t_0 = 1$. This is in line with our estimate that the maximum can occur at $t_0 \lesssim O( \ln t )$.

\section{Matrix elements of the gates in the numerics}

Fig.~\ref{fig:pkfixedrealisation} is produced from a dual unitary circuit with the same two-qubit gate everywhere. The two qubit gate takes the form $W(p) =(v_+ \otimes v_-) U(p)(u_+ \otimes u_-) $ with 
$$ U(p)= \begin{pmatrix}
    e^{-i J(p)} & 0 & 0 & 0\\
    0 & 0 & -i e^{i J(p)} & 0\\
    0 & -i e^{i J(p)} & 0 & 0\\
    0 & 0 & 0 & e^{-i J(p)}
\end{pmatrix}$$ 
where $J(p) = \frac{1}{2} \arcsin{\sqrt{1- \frac 3 2 p}}$ and $p$ is the entangling power. The single qubit random matrices used are (up to 5 significant digits)
\begin{align*}
    v_+ &=
\begin{pmatrix}
-0.025047 - 0.36705 i & -0.92114 - 0.12704 i \\
0.90768 - 0.20191 i & 0.0050415 + 0.36787 i
\end{pmatrix},\quad 
v_- =
\begin{pmatrix}
0.38001 - 0.32098 i & 0.43600 + 0.74998 i \\
0.80711 + 0.31803 i & 0.26005 - 0.42404 i
\end{pmatrix},\\
u_+ &=
\begin{pmatrix}
0.20391 - 0.97064 i & -0.10791 - 0.068044 i \\
0.12500 + 0.025487 i & -0.52404 + 0.84209 i
\end{pmatrix},\quad 
u_- =
\begin{pmatrix}
-0.27904 - 0.92084 i & 0.23803 + 0.13237 i \\
-0.27185 + 0.016684 i & -0.64912 + 0.71026 i
\end{pmatrix}.
\end{align*}

\end{document}

%% file: tikz_lib.tex
\usepackage{xifthen}
\usepackage{xargs}

\newcommandx{\fineq}[5][1=-.8ex,2=1,3=1,5=0]{
	\begin{tikzpicture}[baseline={([yshift=#1]current  bounding  box.center)}, scale = #2, every node/.style={scale = #3},rotate around={#5:(0,0)},every node/.style={transform shape}]
		#4
	\end{tikzpicture}
}

\definecolor{ngray}{RGB}{255,245,186}
\definecolor{bertinired}{RGB}{232,102,102}
\definecolor{bertiniblue}{RGB}{101,147,245}
\definecolor{bertinigreyblue}{RGB}{166,218,149}
\definecolor{bertinigreyred}{RGB}{232,102,102}
\definecolor{bertinivioletc}{RGB}{45,130,60}
\definecolor{bertinigreen}{RGB}{166,218,149}
\definecolor{bertiniorange}{RGB}{255, 116, 23}
\definecolor{OliveGreen}{RGB}{85,107,47}
\definecolor{NavyBlue}{RGB}{0,0,128}
\definecolor{bertiniviolet}{RGB}{210,145,178}
\definecolor{bertinigrey1}{RGB}{98,98,98}
\definecolor{bertinigrey2}{RGB}{211,211,211}
\definecolor{bertinigrey3}{RGB}{192,192,192}
\definecolor{bertinigrey4}{RGB}{169,169,169}

\newcommandx{\cstate}[4][1=0,2=0,3= ,4=white]{
	\begin{scope}[shift={(0,0)}]
				\draw[fill=#4,thick] (#1,#2) circle (0.13);
				\node[scale=1.1] at (#1,#2) {$#3$};
\end{scope}
}
\newcommandx{\sqrstate}[4][1=0,2=0,3= ,4=white]{
	\begin{scope}[shift={(#1,#2)}]
		\draw[thick,fill=#4] (-0.13,-0.13) rectangle (0.13,0.13) ;
		\node[scale=1.1] at (0,0) {$#3$};
	\end{scope}
}

\newcommandx{\tikzdiagup}{
	\tikz {\draw[thick] (0,0)--(0.15,0.15); \draw (0,0) rectangle (0.15,0.15);}
}

\newcommandx{\gatecross}[1][1=0.5]{
	\pgfmathparse{#1/2.0}
	\let\x\pgfmathresult
	\draw[thick] (-\x,-\x) -- (\x,\x);
	\draw[thick] (\x,-\x) -- (-\x,\x);
}
\newcommandx{\gatesqu}[2][1=0.25,2=]{
	\pgfmathparse{#1/2.0}
	\let\x\pgfmathresult
	\ifthenelse{\equal{#2}{}}{
		\draw[thick, fill=white, rounded corners=2pt] (-\x,\x) rectangle (\x,-\x);
	}{
		\draw[thick, fill=#2, rounded corners=2pt] (-\x,\x) rectangle (\x,-\x);
	}
}

\newcommandx{\gatemark}[2][1=0.075,2=tr]{
	\pgfmathparse{#1}
	\let\l\pgfmathresult
	\ifthenelse{\equal{#2}{topleft}}{
		\draw[thick] (0,\l) -- ++(-\l,0) --++ (0,-\l);
	}{}
	\ifthenelse{\equal{#2}{topright}}{
		\draw[thick] (0,\l) -- ++(\l,0) --++ (0,-\l);
	}{}
	
	\ifthenelse{\equal{#2}{bottomleft}}{
		\draw[thick] (0,-\l) -- ++(-\l,0) --++ (0,\l);
	}{}
	\ifthenelse{\equal{#2}{bottomright}}{
		\draw[thick] (0,-\l) -- ++(\l,0) --++ (0,\l);
	}{}
	
}

\newcommandx{\roundgate}[6][1=0,2=0,3=1,4=topright,5=white,6=-1]{
	\pgfmathparse{#3}
	\let\l\pgfmathresult
	\begin{scope}[shift={(#1,#2)}]
		\gatecross[\l]
			\pgfmathparse{\l/2.0}
		\let\s\pgfmathresult
		\gatesqu[\s][#5]
		\pgfmathparse{\l*0.15}
		\let\m\pgfmathresult
	\ifthenelse{\equal{#6}{-1}}{		\gatemark[\m][#4]
	}{	\node at ({0},{0}) {\scalebox{1.3}{$#6$}};}
\end{scope}
}

\newcommandx{\wcirc}[2]{\begin{scope}
		\draw[fill=white] (#1,#2) circle (0.15);	\end{scope}} 
\newcommandx{\wcircc}[2]{\begin{scope}
		\draw[fill=white] (#1,#2) circle (0.13);	\end{scope}} 

\newcommandx{\wsqr}[2]{\begin{scope}
		\draw[fill=white,shift={(#1,#2)}] (-.13,.13) rectangle (.13,-.13);	\end{scope}} 
\newcommandx{\wsqrr}[2]{\begin{scope}
		\draw[fill=white,shift={(#1,#2)}] (-.11,.11) rectangle (.11,-.11);	\end{scope}} 
\newcommandx{\bcirc}[2]{\begin{scope}
		\draw[fill=black] (#1,#2) circle (0.15);	\end{scope}} 
\newcommandx{\thetastatevar}[4][1=0,2=0,3=1,4=]{
	\pgfmathparse{#3/2}
	\let\l\pgfmathresult
	\pgfmathparse{\l*0.15}
	\let\m\pgfmathresult
	\begin{scope}[shift={(#1,#2)}]
		\draw[thick] (0,0)--(0.7*\l,0.7*\l);
		\draw[thick] (0,0)--(-0.7*\l,0.7*\l);
		\ifthenelse{\equal{#4}{}}{
			\draw[fill=white] (0,0) circle (0.15);
		}{
			\draw[thick, fill=#4] (0,0) circle (0.15);
		}
	\end{scope}
}
\newcommandx{\thetastate}[4][1=0,2=0,3=1,4=]{
	\pgfmathparse{#3/2}
	\let\l\pgfmathresult
	\pgfmathparse{\l*0.15}
	\let\m\pgfmathresult
	\begin{scope}[shift={(#1,#2)}]
		\draw[thick] (0,0)--(\l,\l);
		\draw[thick] (0,0)--(-\l,\l);
		\ifthenelse{\equal{#4}{}}{
			\draw[fill=white] (0,0) circle (0.15);
		}{
			\draw[thick, fill=#4] (0,0) circle (0.15);
		}
	\end{scope}
}

\newcommandx{\thetastateflipped}[4][1=0,2=0,3=1,4=]{
	\pgfmathparse{#3/2}
	\let\l\pgfmathresult
	\pgfmathparse{\l*0.15}
	\let\m\pgfmathresult
	\begin{scope}[shift={(#1,#2)}]
		\draw[thick] (0,0)--(\l,-\l);
		\draw[thick] (0,0)--(-\l,-\l);
		\ifthenelse{\equal{#4}{}}{
			\draw[fill=white] (0,0) circle (0.15);
		}{
			\draw[thick, fill=#4] (0,0) circle (0.15);
		}
	\end{scope}
}

\newcommandx{\vertgate}[5][1=0,2=0,3=4,4=bertiniorange,5=topright]
{
	\begin{scope}[shift={(#1,#2)}]
		\ifthenelse{\equal{#3}{1}}{
			\roundgate[0][0][1][#5][#4]
		}{
			\foreach \n[evaluate=\n as \y using {2*\n-2}] in {1,...,#3}{
				\roundgate[0][\y][1][#5][#4]
			}
		}
	\end{scope}
}

\newcommandx{\tsfmatV}[8][1=0,2=0,3=l,4=4,5=tr,6=init,7=bertiniorange,8=topright]{
	\begin{scope}[shift={(#1,#2)}]
		\ifthenelse{\equal{#3}{l}}{
			\pgfmathsetmacro{\flag}{0}
		}{
			\pgfmathsetmacro{\flag}{1}
		}
		
		\foreach \y[evaluate=\y as \x using {mod(\y+\flag,2)}] in {1,...,#4}{
			\roundgate[\x][\y][1][#8][#7]
		}
		\ifthenelse{\equal{#5}{tr}}{
			\foreach \y[evaluate=\y as \x using {mod(\y+\flag,2)}] in {#4}{
				\draw [fill=white] (\x-0.5,\y+0.5) circle (0.15);
				\draw [fill=white] (\x+0.5,\y+0.5) circle (0.15);
			}
		}{}
		\ifthenelse{\equal{#6}{init}}{
			\thetastate[\flag][0][1][#7]
		}{}
	\end{scope}
}

\newcommandx{\leftriangle}[5][1=0,2=0,3=4,4=bertiniorange,5=topright]{
	\begin{scope}[shift={(#1,#2)}]
		\pgfmathsetmacro{\t}{#3}
		\pgfmathsetmacro{\steps}{ceil(\t/2)}
		\foreach \i[evaluate=\i as \x using -\t+2*\i-1, evaluate=\i as \ylim using \t-2*\i+2] in {1,...,\steps}{
			\foreach \y[evaluate=\y as \thisx using {\x+\y-1}] in {1,...,\ylim}{
				\roundgate[\thisx][\y][1][#5][#4][#5]
			}
		}
	\end{scope}
}

\newcommandx{\rightriangle}[5][1=0,2=0,3=4,4=bertiniorange,5=topright]{
	\begin{scope}[shift={(#1,#2)}]
		\pgfmathsetmacro{\t}{#3}
		\pgfmathsetmacro{\steps}{ceil(\t/2)}
		\foreach \i[evaluate=\i as \x using -\t+2*\i-1, evaluate=\i as \ylim using \t-2*\i+2] in {1,...,\steps}{
			\foreach \y[evaluate=\y as \thisx using {-\x-\y+1}] in {1,...,\ylim}{
				\roundgate[\thisx][\y][1][#5][#4][#5]
			}
		}
	\end{scope}
}

\newcommandx{\eigenVL}[8][1=0,2=0,3=l,4=5,5=tr,6=init,7=bertiniorange,8=topright]{
	\begin{scope}[shift={(#1,#2)}]
		\pgfmathsetmacro{\t}{#4}
		\leftriangle[0][0][\t][#7][#8]
		
		\ifthenelse{\equal{#6}{init}}{
			\drawinitstate[0][0][l][\t][#7]
		}{}
		
		\ifthenelse{\equal{#5}{tr}}{
			\draw[fill=white] \foreach \x in {0,...,\t} {(\x-0.5-\t,0.5+\x) circle (0.15)};
			\ifthenelse{\equal{#3}{r}}{
				\draw[fill=white] (0.5,\t+0.5) circle (0.15);
			}{}
		}{}
		\ifthenelse{\equal{#5}{parttr}}{
			\draw[fill=white] \foreach \x in {0,...,\t} {(\x-0.5-\t,0.5+\x) circle (0.15)};
		}{}
	\end{scope}
}

\newcommandx{\eigenVR}[8][1=0,2=0,3=l,4=5,5=tr,6=init,7=bertiniorange,8=topright]{
	\begin{scope}[shift={(#1,#2)}]
		\pgfmathsetmacro{\t}{#4}
		\rightriangle[0][0][\t][#7][#8]
		
		\ifthenelse{\equal{#6}{init}}{
			\drawinitstate[0][0][r][\t][#7]
		}{}
		
		\ifthenelse{\equal{#5}{tr}}{
			\draw[fill=white] \foreach \x in {0,...,\t}{(-\x+0.5+\t,0.5+\x) circle (0.15)};
			\ifthenelse{\equal{#3}{l}}{
				\draw[fill=white] (-0.5,\t+0.5) circle (0.15);
			}{}
		}{}
	\end{scope}
}

\newcommandx{\tra}[2][1]{\underset{#1}{\text{tr}}\left[#2\right]}

\newcommandx{\tsfmatDgate}[7][1=0,2=0,3=l,4=4,5=tr,6=bertiniorange,7=topright]
{
	\begin{scope}[shift={(#1,#2)}]
		\ifthenelse{\equal{#3}{l}}{
			\pgfmathsetmacro{\flag}{-1}
		}{
			\pgfmathsetmacro{\flag}{1}
		}
		\pgfmathsetmacro{\t}{#4}
		\foreach \i[evaluate=\i as \x using {\flag*\i}, evaluate=\i as \y using \i] in {1,...,\t}{
			\roundgate[\x][\y][1][#7][#6]
		}
		
		\ifthenelse{\equal{#5}{tr}}{
			\foreach \i[evaluate=\i as \x using {\flag*\i}, evaluate=\i as \y using \i] in {\t}{
				\draw [fill=white] (\x-0.5,\y+0.5) circle (0.15);
				\draw [fill=white] (\x+0.5,\y+0.5) circle (0.15);
			}  
		}{}
	\end{scope}
	
}

\newcommandx{\tsfmatD}[8][1=0,2=0,3=l,4=4,5=tr,6=init,7=bertiniorange,8=topright]{
	\begin{scope}[shift={(#1,#2)}]
		\ifthenelse{\equal{#6}{init}}{
			\thetastate[0][0][1][#7]
		}{}
		
		\ifthenelse{\equal{#3}{l}}{
			\pgfmathsetmacro{\flag}{-1}
		}{
			\pgfmathsetmacro{\flag}{1}
		}
		
		\pgfmathsetmacro{\t}{#4}
		\foreach \i[evaluate=\i as \x using {\flag*\i}, evaluate=\i as \y using \i] in {1,...,\t}{
			\roundgate[\x][\y][1][#8][#7]
		}
		
		\ifthenelse{\equal{#5}{tr}}{
			\foreach \i[evaluate=\i as \x using {\flag*\i}, evaluate=\i as \y using \i] in {\t}{
				\draw [fill=white] (\x-0.5,\y+0.5) circle (0.15);
				\draw [fill=white] (\x+0.5,\y+0.5) circle (0.15);
			}  
		}
		\ifthenelse{\equal{#5}{parttr}}{
			\foreach \i[evaluate=\i as \x using {\flag*\i}, evaluate=\i as \y using \i] in {\t}{
				\draw [fill=white] (\x+0.5*\flag,\y+0.5) circle (0.15);
			}  
		}
		{}
	\end{scope}
}

\newcommandx{\drawinitstate}[5][1=0,2=0,3=l,4=4,5=bertiniorange]{
	\pgfmathsetmacro{\t}{#4}
	\begin{scope}[shift={(#1,#2)}]
		\pgfmathsetmacro{\steps}{ceil((\t-1)/2)}
		\ifthenelse{\equal{#3}{l}}{
			\foreach \i[evaluate=\i as \x using -\t+2*\i] in {0,...,\steps}{
				\thetastate[\x][0][1][#5]
			}
		}{
			\foreach \i[evaluate=\i as \x using -\t+2*\i] in {0,...,\steps}{      
				\thetastate[-\x][0][1][#5]
			}
		}
	\end{scope}
}

\newcommandx{\drawinitstateflipped}[5][1=0,2=0,3=l,4=4,5=bertiniorange]{
	\pgfmathsetmacro{\t}{#4}
	\begin{scope}[shift={(#1,#2)}]
		\pgfmathsetmacro{\steps}{ceil((\t-1)/2)}
		\ifthenelse{\equal{#3}{l}}{
			\foreach \i[evaluate=\i as \x using -\t+2*\i] in {0,...,\steps}{
				\thetastateflipped[\x][0][1][#5]
			}
		}{
			\foreach \i[evaluate=\i as \x using -\t+2*\i] in {0,...,\steps}{      
				\thetastateflipped[-\x][0][1][#5]
			}
		}
	\end{scope}
}

\newcommandx{\eigenDL}[6][1=0,2=0,3=l,4=4,5=bertiniorange,6=topright]{
	\begin{scope}[shift={(#1,#2)}]
		\pgfmathsetmacro{\t}{#4}
		\ifthenelse{\equal{#3}{l}}{
			\eigenVL[0][0][l][\t][tr][init][#5][#6]
			\pgfmathsetmacro{\t}{#4-1}
			\rightriangle[1][0][\t][#5][#6]
			\drawinitstate[1][0][r][\t][#5]
		}{
			\begin{scope}[shift={(-0.5,0.5)}]
				\foreach \i[evaluate=\i as \x using \i, evaluate=\i as \y using \i] in {0,...,\t}{      
					\draw (\x,\y)--++(0.5,0);
					\draw[fill=white] (\x,\y) circle (0.15);
				}
			\end{scope}
		}
	\end{scope}
}

\newcommandx{\eigenDR}[6][1=0,2=0,3=l,4=4,5=bertiniorange,6=topright]{
	\begin{scope}[shift={(#1,#2)}]
		\pgfmathsetmacro{\t}{#4}
		\ifthenelse{\equal{#3}{r}}{
			\eigenVR[0][0][r][\t][tr][init][#5][#6]
			\pgfmathsetmacro{\t}{#4-1}
			\leftriangle[-1][0][\t][#5][#6]
			\drawinitstate[-1][0][l][\t][#5]
		}{
			\begin{scope}[shift={(0.5,0.5)}]
				\foreach \i[evaluate=\i as \x using \i, evaluate=\i as \y using \t-\i] in {0,...,\t}{      
					\draw (\x,\y)--++(0.5,0);
					\draw[fill=white] (\x+0.5,\y) circle (0.15);
				}
			\end{scope}
		}
	\end{scope}
}

\newcommandx{\idonpurity}[2][1=0,2=0]
{
	\begin{scope}[shift={(#1,#2)}]
		\draw[thick] (-0.5,0)--++(-0.1,0.1)--++(0,0.2)--++(0.1,-0.1);
		\draw[thick] (-0.5,0.4)--++(-0.1,0.1)--++(0,0.2)--++(0.1,-0.1);
		\draw[thick] (0.5,0)--++(0.1,0.1)--++(0,0.2)--++(-0.1,-0.1);
		\draw[thick] (0.5,0.4)--++(0.1,0.1)--++(0,0.2)--++(-0.1,-0.1);
	\end{scope}
}

\newcommandx{\swaponpurity}[2][1=0,2=0]
{
	\begin{scope}[shift={(#1,#2)}]
		\draw[thick] (-0.5,0)--++(-0.2,0.2)--++(0,0.6)--++(0.2,-0.2);
		\draw[thick] (-0.5,0.2)--++(-0.075,0.075)--++(0,0.2)--++(0.075,-0.075);
		\draw[thick] (+0.5,0)--++(+0.2,0.2)--++(0,0.6)--++(-0.2,-0.2);
		\draw[thick] (+0.5,0.2)--++(+0.075,0.075)--++(0,0.2)--++(-0.075,-0.075);
	\end{scope}
}

\newcommandx{\hook}[4][1=0,2=0,3=t,4=l]{
	\begin{scope}[shift={(#1,#2)}]
		\ifthenelse{\equal{#3}{t}}{
			\ifthenelse{\equal{#4}{l}}{\draw[thick] (0.5,-0.5) arc (45:-90:0.15);}{\draw[thick] (0.5,-0.5) arc (45:270:0.15);}
		}{\ifthenelse{\equal{#4}{l}}{\draw[ thick] (0.5,-0.5) arc (-45:90:0.15);}{\draw[ thick] (0.5,-0.5) arc (315:90:0.15);}
		}
	\end{scope}
}

\newcommandx{\hhook}[4][1=0,2=0,3=t,4=l]{
	\begin{scope}[shift={(#1,#2)}]
		\ifthenelse{\equal{#3}{t}}{
			\ifthenelse{\equal{#4}{l}}{\draw[thick] (0.5,-0.5) arc (-45:175:0.15);}{\draw[thick] (0.5,-0.5) arc (225:0:0.15);}
		}{\ifthenelse{\equal{#4}{l}}{\draw[ thick] (0.5,-0.5) arc (-45:180:-0.15);}{\draw[ thick] (0.5,-0.5) arc (45:-180:0.15);}
		}
	\end{scope}
}

\definecolor{FcolU}{rgb}{0.71,0.78,0.91}
\definecolor{colLines}{rgb}{0.31,0.31,0.31}
\definecolor{colVMPSLines}{rgb}{0.11,0.11,0.11}
\definecolor{IcolUc}{rgb}{0.71,0.41,0.42}
\definecolor{IcolU}{rgb}{0.71,0.8,0.76}
\definecolor{IcolVMPSc}{rgb}{0.73,0.69,0.7}
\definecolor{IcolVMPS}{rgb}{0.81,0.77,0.78}
\definecolor{colObs}{rgb}{1.,1.,1.}

\newcommandx{\eightlegs}[2][1=0,2=0]{
	\begin{scope}[shift={(#1,#2)}]
		\foreach \x in {1,...,8}{
			\draw (\x, 0)--++(0,0.25);
			\draw[fill] (\x,0) circle (0.05);
		}
		\foreach \x in {1,3}{
			\pgfmathsetmacro\result{2*\x-1} 
			\node () at (\result,-0.5) {$i_{\x}$};
			\pgfmathsetmacro\result{2*\x}
			\node () at (\result,-0.5) {$j_{\x}$};	
		}
		\foreach \x in {2,4}{
			\pgfmathsetmacro\result{2*\x} 
			\node () at (\result,-0.5) {$i_{\x}$};
			\pgfmathsetmacro\result{2*\x-1}
			\node () at (\result,-0.5) {$j_{\x}$};	
		}
	\end{scope}
}

\newcommandx{\idproj}[2][1=0,2=0]{
  \fineq[-0.8ex][0.8][1]{
    \begin{scope}[shift={(#1,#2)}]
      \draw (0,0) arc (-15:195:0.2);
      \draw (0,0.8) arc (360+15:165:0.2);
    \end{scope}
  }
}

\definecolor{projgold}{RGB}{212,175,55}

\newcommand{\projdiamond}[2]{%
  \draw[fill=projgold] (#1,#2+0.08)--(#1+0.08,#2)--(#1,#2-0.08)--(#1-0.08,#2)--cycle;
}

\newcommand{\rtmvisibleprojectors}{5}
\newcommand{\rtmleftx}{-0.50}
\newcommand{\rtmrightx}{0.50}
\newcommand{\rtmhalfbox}{0.10}
\newcommand{\rtmhstub}{0.24}
\newcommand{\rtmvstub}{0.06}
\newcommand{\rtmdy}{0.42}
\newcommand{\rtmblockshift}{0.30}
\newcommand{\rtmstandardmidgap}{0.62}
\newcommand{\rtmjmidsep}{0.92}
\newcommand{\rtmjdotstub}{0.12}
\newcommand{\rtmjdotraise}{2pt}
\newcommand{\rtmjfourthboxraise}{0.10}
\newcommand{\rtmjthirdboxlift}{0.60}
\newcommand{\rtmjsecondprojectoroffset}{0.50}
\newcommand{\rtmjthirdboxprojectoroffset}{0.50}
\newcommand{\rtmjthirdprojectoroffset}{0.50}
\newcommand{\rtmjsecondupstub}{0.16}
\newcommand{\rtmjthirddownstub}{0.16}

\newcommand{\rtmsetbasegeometry}{%
  \def\xl{\rtmleftx}
  \def\xr{\rtmrightx}
  \def\halfbox{\rtmhalfbox}
  \def\hstub{\rtmhstub}
  \def\vstub{\rtmvstub}
  \def\dy{\rtmdy}
  \def\blockshift{\rtmblockshift}
}

\newcommand{\rtmdrawsinglebox}[2]{%
  \draw[fill=bertiniorange] (#1-\halfbox,#2-\halfbox) rectangle (#1+\halfbox,#2+\halfbox);
}
\newcommand{\rtmdrawhline}[3]{%
  \draw (#1,#3) -- (#2,#3);
}
\newcommand{\rtmdrawvline}[3]{%
  \draw (#1,#2) -- (#1,#3);
}
\newcommand{\rtmdrawupstubat}[3]{%
  \draw (#1,#2) -- ++(0,#3);
}
\newcommand{\rtmdrawdownstubat}[3]{%
  \draw (#1,#2) -- ++(0,-#3);
}
\newcommand{\rtmdrawlefthstub}[1]{%
  \rtmdrawhline{\xl+\halfbox}{\xl+\halfbox+\hstub}{#1}
}
\newcommand{\rtmdrawrighthstub}[1]{%
  \rtmdrawhline{\xr-\halfbox}{\xr-\halfbox-\hstub}{#1}
}
\newcommand{\rtmdrawboxrow}[1]{%
  \rtmdrawsinglebox{\xl}{#1}%
  \rtmdrawsinglebox{\xr}{#1}%
  \rtmdrawlefthstub{#1}%
  \rtmdrawrighthstub{#1}%
}
\newcommand{\rtmdrawrowconnector}[1]{%
  \rtmdrawhline{\xl+\halfbox+\hstub}{\xr-\halfbox-\hstub}{#1}
}
\newcommand{\rtmdrawfullslot}[2]{%
  \rtmdrawvline{\xl}{#1+\halfbox}{#2-\halfbox}
  \rtmdrawvline{\xr}{#1+\halfbox}{#2-\halfbox}
}
\newcommand{\rtmdrawslotstubs}[2]{%
  \rtmdrawupstubat{\xl}{#1+\halfbox}{\vstub}
  \rtmdrawdownstubat{\xl}{#2-\halfbox}{\vstub}
  \rtmdrawupstubat{\xr}{#1+\halfbox}{\vstub}
  \rtmdrawdownstubat{\xr}{#2-\halfbox}{\vstub}
}
\newcommand{\rtmdrawdotgapstubs}[2]{%
  \rtmdrawupstubat{\xl}{#1+\halfbox}{\dotstub}
  \rtmdrawdownstubat{\xl}{#2-\halfbox}{\dotstub}
  \rtmdrawupstubat{\xr}{#1+\halfbox}{\dotstub}
  \rtmdrawdownstubat{\xr}{#2-\halfbox}{\dotstub}
}
\newcommand{\rtmdrawprojectorpair}[1]{%
  \ifdefined\rtmsuppressprojectordiamonds
    \ifnum\rtmsuppressprojectordiamonds=1\relax
    \else
      \projdiamond{\xl}{#1}%
      \projdiamond{\xr}{#1}%
    \fi
  \else
    \projdiamond{\xl}{#1}%
    \projdiamond{\xr}{#1}%
  \fi
}
\newcommand{\rtmdrawstandardprojectorslot}[2]{%
  \rtmdrawslotstubs{#1}{#2}%
  \pgfmathsetmacro{\py}{0.5*(#1+#2)}
  \rtmdrawprojectorpair{\py}
}
\newcommand{\rtmdrawjprojectorslot}[3]{%
  \def\rtmdrawthisprojector{1}%
  \def\rtmsuppressslotthree{0}%
  \ifdefined\rtmsuppressslotthreeprojector
    \edef\rtmsuppressslotthree{\rtmsuppressslotthreeprojector}%
  \fi
  \ifnum\nproj=3\relax
    \ifnum#1=2\relax
      \rtmdrawupstubat{\xl}{#2+\halfbox}{\rtmjsecondupstub}
      \rtmdrawupstubat{\xr}{#2+\halfbox}{\rtmjsecondupstub}
      \rtmdrawdownstubat{\xl}{#3-\halfbox}{\rtmjthirddownstub}
      \rtmdrawdownstubat{\xr}{#3-\halfbox}{\rtmjthirddownstub}
      \pgfmathsetmacro{\py}{#2+\rtmjsecondprojectoroffset*\dy}
      \pgfmathsetmacro{\pyextra}{#3-\rtmjthirdboxprojectoroffset*\dy}
      \rtmdrawprojectorpair{\pyextra}
    \else\ifnum#1=3\relax
      \ifnum\rtmsuppressslotthree=1\relax
        \rtmjdrawemptyslot{#1}{#2}{#3}%
        \def\rtmdrawthisprojector{0}%
      \else
        \rtmdrawslotstubs{#2}{#3}%
        \pgfmathsetmacro{\py}{#3-\rtmjthirdprojectoroffset*\dy}
      \fi
    \else
      \rtmdrawslotstubs{#2}{#3}%
      \pgfmathsetmacro{\py}{0.5*(#2+#3)}
    \fi\fi
  \else
    \rtmdrawslotstubs{#2}{#3}%
    \pgfmathsetmacro{\py}{0.5*(#2+#3)}
  \fi
  \ifnum\rtmdrawthisprojector=1\relax
    \rtmdrawprojectorpair{\py}
  \fi
}
\newcommand{\rtmstandarddrawrow}[2]{%
  \rtmdrawboxrow{#2}%
  \ifnum#1>\nconn\relax
  \else
    \rtmdrawrowconnector{#2}%
  \fi
}
\newcommand{\rtmstandarddrawslot}[3]{%
  \ifnum#1>\nproj\relax
    \rtmdrawfullslot{#2}{#3}%
  \else
    \rtmdrawstandardprojectorslot{#2}{#3}%
  \fi
}
\newcommand{\rtmjdrawrow}[2]{%
  \rtmdrawboxrow{#2}%
  \ifnum#1>\nconn\relax
    \ifnum#1=3\relax
      \rtmdrawrowconnector{#2}%
    \else\ifnum#1=5\relax
      \ifnum\nconnproj=5\relax
        \rtmdrawrowconnector{#2}%
      \fi
    \fi\fi
  \else
    \rtmdrawrowconnector{#2}%
  \fi
}
\newcommand{\rtmjdrawemptyslot}[3]{%
  \ifnum#1=3\relax
    \rtmdrawdotgapstubs{#2}{#3}%
  \else\ifnum#1=4\relax
    \rtmdrawdotgapstubs{#2}{#3}%
  \else
    \rtmdrawfullslot{#2}{#3}%
  \fi\fi
}
\newcommand{\rtmjdrawslot}[3]{%
  \ifnum#1>\nproj\relax
    \rtmjdrawemptyslot{#1}{#2}{#3}%
  \else
    \rtmdrawjprojectorslot{#1}{#2}{#3}%
  \fi
}
\newcommand{\rtmtensorconn}[3][1]{%
  \pgfmathsetmacro{\rtmfigscale}{0.68*(#1)}
  \pgfmathsetmacro{\rtmnodescale}{0.72*(#1)}
  \pgfmathsetlengthmacro{\rtmbaseline}{-0.95*(#1)*1ex}
  \fineq[\rtmbaseline][\rtmfigscale][\rtmnodescale]{%
    \pgfmathtruncatemacro{\nproj}{min(#2,\rtmvisibleprojectors)}
    \pgfmathtruncatemacro{\nconnproj}{min(#3,\rtmvisibleprojectors)}
    \pgfmathtruncatemacro{\nconn}{ifthenelse(\nconnproj<4,\nconnproj,\nconnproj-1)}
    \rtmsetbasegeometry
    \def\midgap{\rtmstandardmidgap}
    \pgfmathsetmacro{\yone}{\dy+\blockshift}
    \pgfmathsetmacro{\ytwo}{2*\dy+\blockshift}
    \pgfmathsetmacro{\ythree}{3*\dy+\blockshift}
    \pgfmathsetmacro{\yfour}{4*\dy+\blockshift}
    \pgfmathsetmacro{\yfive}{5*\dy+\midgap-\blockshift}
    \pgfmathsetmacro{\ysix}{6*\dy+\midgap-\blockshift}
    \pgfmathsetmacro{\yseven}{7*\dy+\midgap-\blockshift}
    \pgfmathsetmacro{\ymid}{0.5*(\yfour+\yfive)}
    \pgfmathsetmacro{\ydots}{\ymid+0.04}
    \foreach \row/\y in {1/\yone,2/\ytwo,3/\ythree,4/\ysix,5/\yseven}{
      \rtmstandarddrawrow{\row}{\y}
    }
    \foreach \slot/\ya/\yb in {
      1/\yone/\ytwo,
      2/\ytwo/\ythree,
      3/\ythree/\yfour,
      4/\yfive/\ysix,
      5/\ysix/\yseven
    }{
      \rtmstandarddrawslot{\slot}{\ya}{\yb}
    }
    \node at (\xl,\ydots) {$\vdots$};
    \node at (\xr,\ydots) {$\vdots$};
  }%
}

\newcommand{\rtmtensorjconn}[3][1]{%
  \pgfmathsetmacro{\rtmfigscale}{0.68*(#1)}
  \pgfmathsetmacro{\rtmnodescale}{0.72*(#1)}
  \pgfmathsetlengthmacro{\rtmbaseline}{-0.95*(#1)*1ex}
  \fineq[\rtmbaseline][\rtmfigscale][\rtmnodescale]{%
    \pgfmathtruncatemacro{\nproj}{min(#2,\rtmvisibleprojectors)}
    \pgfmathtruncatemacro{\nconnproj}{min(#3,\rtmvisibleprojectors)}
    \pgfmathtruncatemacro{\nconn}{ifthenelse(\nconnproj<4,\nconnproj,\nconnproj-1)}
    \rtmsetbasegeometry
    \def\dotstub{\rtmjdotstub}
    \def\dotraise{\rtmjdotraise}
    \def\midsep{\rtmjmidsep}
    \pgfmathsetmacro{\yone}{\dy+\blockshift}
    \pgfmathsetmacro{\ytwo}{2*\dy+\blockshift}
    \pgfmathsetmacro{\ythreebase}{3*\dy+\blockshift}
    \pgfmathsetmacro{\ymidrowbase}{\ythreebase+\midsep}
    \pgfmathsetmacro{\ymidrow}{\ymidrowbase}
    \ifnum\nproj=3\relax
      \pgfmathsetmacro{\ymidrow}{\ymidrowbase+\rtmjfourthboxraise}
    \fi
    \pgfmathsetmacro{\ysix}{\ymidrow+\midsep}
    \pgfmathsetmacro{\yseven}{\ysix+\dy}
    \pgfmathsetmacro{\ydotslower}{0.5*(\ythreebase+\ymidrowbase)}
    \pgfmathsetmacro{\ydotsupper}{0.5*(\ymidrow+\ysix)}
    \pgfmathsetmacro{\ythree}{\ythreebase}
    \ifnum\nproj=3\relax
      \pgfmathsetmacro{\ythree}{\ythreebase+\rtmjthirdboxlift}
      \pgfmathsetmacro{\ydotslower}{0.5*(\ytwo+\ythree)}
    \fi
    \foreach \row/\y in {1/\yone,2/\ytwo,3/\ythree,4/\ymidrow,5/\ysix,6/\yseven}{
      \rtmjdrawrow{\row}{\y}
    }
    \foreach \slot/\ya/\yb in {
      1/\yone/\ytwo,
      2/\ytwo/\ythree,
      3/\ythree/\ymidrow,
      4/\ymidrow/\ysix,
      5/\ysix/\yseven
    }{
      \rtmjdrawslot{\slot}{\ya}{\yb}
    }
    \node[inner sep=2pt,yshift=\dotraise] at (\xl,\ydotslower) {$\vdots$};
    \node[inner sep=2pt,yshift=\dotraise] at (\xr,\ydotslower) {$\vdots$};
    \node[inner sep=2pt,yshift=\dotraise] at (\xl,\ydotsupper) {$\vdots$};
    \node[inner sep=2pt,yshift=\dotraise] at (\xr,\ydotsupper) {$\vdots$};
  }%
}
\newcommand{\rtmtensorj}[2][1]{%
  \rtmtensorjconn[#1]{#2}{#2}%
}
\newcommand{\rtmtensorjcompconn}[3][1]{%
  {%
    \def\rtmsuppressslotthreeprojector{1}%
    \rtmtensorjconn[#1]{#2}{#3}%
  }%
}

\newcommand{\rtmtensorjcompnodiamondconn}[3][1]{%
  {%
    \def\rtmsuppressprojectordiamonds{1}%
    \rtmtensorjcompconn[#1]{#2}{#3}%
  }%
}

%% file: main.bbl
\begin{thebibliography}{56}%
\makeatletter
\providecommand \@ifxundefined [1]{%
 \@ifx{#1\undefined}
}%
\providecommand \@ifnum [1]{%
 \ifnum #1\expandafter \@firstoftwo
 \else \expandafter \@secondoftwo
 \fi
}%
\providecommand \@ifx [1]{%
 \ifx #1\expandafter \@firstoftwo
 \else \expandafter \@secondoftwo
 \fi
}%
\providecommand \natexlab [1]{#1}%
\providecommand \enquote  [1]{``#1''}%
\providecommand \bibnamefont  [1]{#1}%
\providecommand \bibfnamefont [1]{#1}%
\providecommand \citenamefont [1]{#1}%
\providecommand \href@noop [0]{\@secondoftwo}%
\providecommand \href [0]{\begingroup \@sanitize@url \@href}%
\providecommand \@href[1]{\@@startlink{#1}\@@href}%
\providecommand \@@href[1]{\endgroup#1\@@endlink}%
\providecommand \@sanitize@url [0]{\catcode `\\12\catcode `\$12\catcode
  `\&12\catcode `\#12\catcode `\^12\catcode `\_12\catcode `\%12\relax}%
\providecommand \@@startlink[1]{}%
\providecommand \@@endlink[0]{}%
\providecommand \url  [0]{\begingroup\@sanitize@url \@url }%
\providecommand \@url [1]{\endgroup\@href {#1}{\urlprefix }}%
\providecommand \urlprefix  [0]{URL }%
\providecommand \Eprint [0]{\href }%
\providecommand \doibase [0]{https://doi.org/}%
\providecommand \selectlanguage [0]{\@gobble}%
\providecommand \bibinfo  [0]{\@secondoftwo}%
\providecommand \bibfield  [0]{\@secondoftwo}%
\providecommand \translation [1]{[#1]}%
\providecommand \BibitemOpen [0]{}%
\providecommand \bibitemStop [0]{}%
\providecommand \bibitemNoStop [0]{.\EOS\space}%
\providecommand \EOS [0]{\spacefactor3000\relax}%
\providecommand \BibitemShut  [1]{\csname bibitem#1\endcsname}%
\let\auto@bib@innerbib\@empty
\bibitem [{\citenamefont {Vidal}(2003)}]{vidal2003efficient}%
  \BibitemOpen
  \bibfield  {author} {\bibinfo {author} {\bibfnamefont {G.}~\bibnamefont
  {Vidal}},\ }\bibfield  {title} {\bibinfo {title} {Efficient classical
  simulation of slightly entangled quantum computations},\ }\href
  {https://doi.org/10.1103/PhysRevLett.91.147902} {\bibfield  {journal}
  {\bibinfo  {journal} {Phys. Rev. Lett.}\ }\textbf {\bibinfo {volume} {91}},\
  \bibinfo {pages} {147902} (\bibinfo {year} {2003})}\BibitemShut {NoStop}%
\bibitem [{\citenamefont {Schollw{\"o}ck}(2011)}]{schollwock2011density}%
  \BibitemOpen
  \bibfield  {author} {\bibinfo {author} {\bibfnamefont {U.}~\bibnamefont
  {Schollw{\"o}ck}},\ }\bibfield  {title} {\bibinfo {title} {The density-matrix
  renormalization group in the age of matrix product states},\ }\href
  {https://doi.org/10.1016/j.aop.2010.09.012} {\bibfield  {journal} {\bibinfo
  {journal} {Ann. Phys.}\ }\textbf {\bibinfo {volume} {326}},\ \bibinfo {pages}
  {96} (\bibinfo {year} {2011})}\BibitemShut {NoStop}%
\bibitem [{\citenamefont {Cirac}\ \emph {et~al.}(2021)\citenamefont {Cirac},
  \citenamefont {P\'erez-Garc\'{\i}a}, \citenamefont {Schuch},\ and\
  \citenamefont {Verstraete}}]{cirac2021matrix}%
  \BibitemOpen
  \bibfield  {author} {\bibinfo {author} {\bibfnamefont {J.~I.}\ \bibnamefont
  {Cirac}}, \bibinfo {author} {\bibfnamefont {D.}~\bibnamefont
  {P\'erez-Garc\'{\i}a}}, \bibinfo {author} {\bibfnamefont {N.}~\bibnamefont
  {Schuch}},\ and\ \bibinfo {author} {\bibfnamefont {F.}~\bibnamefont
  {Verstraete}},\ }\bibfield  {title} {\bibinfo {title} {Matrix product states
  and projected entangled pair states: Concepts, symmetries, theorems},\ }\href
  {https://doi.org/10.1103/RevModPhys.93.045003} {\bibfield  {journal}
  {\bibinfo  {journal} {Rev. Mod. Phys.}\ }\textbf {\bibinfo {volume} {93}},\
  \bibinfo {pages} {045003} (\bibinfo {year} {2021})}\BibitemShut {NoStop}%
\bibitem [{\citenamefont {Vidal}(2004)}]{vidal2004efficient}%
  \BibitemOpen
  \bibfield  {author} {\bibinfo {author} {\bibfnamefont {G.}~\bibnamefont
  {Vidal}},\ }\bibfield  {title} {\bibinfo {title} {Efficient simulation of
  one-dimensional quantum many-body systems},\ }\href
  {https://doi.org/10.1103/PhysRevLett.93.040502} {\bibfield  {journal}
  {\bibinfo  {journal} {Phys. Rev. Lett.}\ }\textbf {\bibinfo {volume} {93}},\
  \bibinfo {pages} {040502} (\bibinfo {year} {2004})}\BibitemShut {NoStop}%
\bibitem [{\citenamefont {Daley}\ \emph {et~al.}(2004)\citenamefont {Daley},
  \citenamefont {Kollath}, \citenamefont {Schollw\"ock},\ and\ \citenamefont
  {Vidal}}]{daley2004time}%
  \BibitemOpen
  \bibfield  {author} {\bibinfo {author} {\bibfnamefont {A.~J.}\ \bibnamefont
  {Daley}}, \bibinfo {author} {\bibfnamefont {C.}~\bibnamefont {Kollath}},
  \bibinfo {author} {\bibfnamefont {U.}~\bibnamefont {Schollw\"ock}},\ and\
  \bibinfo {author} {\bibfnamefont {G.}~\bibnamefont {Vidal}},\ }\bibfield
  {title} {\bibinfo {title} {Time-dependent density-matrix
  renormalization-group using adaptive effective hilbert spaces},\ }\href
  {https://doi.org/10.1088/1742-5468/2004/04/P04005} {\bibfield  {journal}
  {\bibinfo  {journal} {Journal of Statistical Mechanics: Theory and
  Experiment}\ }\textbf {\bibinfo {volume} {2004}},\ \bibinfo {pages} {P04005}
  (\bibinfo {year} {2004})}\BibitemShut {NoStop}%
\bibitem [{\citenamefont {White}\ and\ \citenamefont
  {Feiguin}(2004)}]{white2004real}%
  \BibitemOpen
  \bibfield  {author} {\bibinfo {author} {\bibfnamefont {S.~R.}\ \bibnamefont
  {White}}\ and\ \bibinfo {author} {\bibfnamefont {A.~E.}\ \bibnamefont
  {Feiguin}},\ }\bibfield  {title} {\bibinfo {title} {Real-time evolution using
  the density matrix renormalization group},\ }\href
  {https://doi.org/10.1103/PhysRevLett.93.076401} {\bibfield  {journal}
  {\bibinfo  {journal} {Phys. Rev. Lett.}\ }\textbf {\bibinfo {volume} {93}},\
  \bibinfo {pages} {076401} (\bibinfo {year} {2004})}\BibitemShut {NoStop}%
\bibitem [{\citenamefont {Potter}\ and\ \citenamefont
  {Vasseur}(2022)}]{potter2022entanglement}%
  \BibitemOpen
  \bibfield  {author} {\bibinfo {author} {\bibfnamefont {A.~C.}\ \bibnamefont
  {Potter}}\ and\ \bibinfo {author} {\bibfnamefont {R.}~\bibnamefont
  {Vasseur}},\ }\bibinfo {title} {Entanglement dynamics in hybrid quantum
  circuits},\ in\ \href {https://doi.org/10.1007/978-3-031-03998-0_9} {\emph
  {\bibinfo {booktitle} {Entanglement in Spin Chains: From Theory to Quantum
  Technology Applications}}},\ \bibinfo {editor} {edited by\ \bibinfo {editor}
  {\bibfnamefont {A.}~\bibnamefont {Bayat}}, \bibinfo {editor} {\bibfnamefont
  {S.}~\bibnamefont {Bose}},\ and\ \bibinfo {editor} {\bibfnamefont
  {H.}~\bibnamefont {Johannesson}}}\ (\bibinfo  {publisher} {Springer
  International Publishing},\ \bibinfo {address} {Cham},\ \bibinfo {year}
  {2022})\ pp.\ \bibinfo {pages} {211--249}\BibitemShut {NoStop}%
\bibitem [{\citenamefont {Fisher}\ \emph {et~al.}(2023)\citenamefont {Fisher},
  \citenamefont {Khemani}, \citenamefont {Nahum},\ and\ \citenamefont
  {Vijay}}]{fisher2022random}%
  \BibitemOpen
  \bibfield  {author} {\bibinfo {author} {\bibfnamefont {M.~P.~A.}\
  \bibnamefont {Fisher}}, \bibinfo {author} {\bibfnamefont {V.}~\bibnamefont
  {Khemani}}, \bibinfo {author} {\bibfnamefont {A.}~\bibnamefont {Nahum}},\
  and\ \bibinfo {author} {\bibfnamefont {S.}~\bibnamefont {Vijay}},\ }\bibfield
   {title} {\bibinfo {title} {Random quantum circuits},\ }\href
  {https://doi.org/10.1146/annurev-conmatphys-031720-030658} {\bibfield
  {journal} {\bibinfo  {journal} {Annu. Rev. Condens. Matter Phys.}\ }\textbf
  {\bibinfo {volume} {14}},\ \bibinfo {pages} {335} (\bibinfo {year}
  {2023})}\BibitemShut {NoStop}%
\bibitem [{\citenamefont {Bertini}\ \emph {et~al.}(2026)\citenamefont
  {Bertini}, \citenamefont {Claeys},\ and\ \citenamefont
  {Prosen}}]{bertini2025exactly}%
  \BibitemOpen
  \bibfield  {author} {\bibinfo {author} {\bibfnamefont {B.}~\bibnamefont
  {Bertini}}, \bibinfo {author} {\bibfnamefont {P.~W.}\ \bibnamefont
  {Claeys}},\ and\ \bibinfo {author} {\bibfnamefont {T.}~\bibnamefont
  {Prosen}},\ }\bibfield  {title} {\bibinfo {title} {Exactly solvable quantum
  many-body dynamics from space-time duality},\ }\href
  {https://doi.org/10.1103/yx73-dk86} {\bibfield  {journal} {\bibinfo
  {journal} {Rev. Mod. Phys.}\ }\textbf {\bibinfo {volume} {98}},\ \bibinfo
  {pages} {025001} (\bibinfo {year} {2026})}\BibitemShut {NoStop}%
\bibitem [{\citenamefont {Ba{\~n}uls}\ \emph {et~al.}(2009)\citenamefont
  {Ba{\~n}uls}, \citenamefont {Hastings}, \citenamefont {Verstraete},\ and\
  \citenamefont {Cirac}}]{banuls2009matrix}%
  \BibitemOpen
  \bibfield  {author} {\bibinfo {author} {\bibfnamefont {M.~C.}\ \bibnamefont
  {Ba{\~n}uls}}, \bibinfo {author} {\bibfnamefont {M.~B.}\ \bibnamefont
  {Hastings}}, \bibinfo {author} {\bibfnamefont {F.}~\bibnamefont
  {Verstraete}},\ and\ \bibinfo {author} {\bibfnamefont {J.~I.}\ \bibnamefont
  {Cirac}},\ }\bibfield  {title} {\bibinfo {title} {Matrix {{Product States}}
  for dynamical simulation of infinite chains},\ }\href
  {https://doi.org/10.1103/PhysRevLett.102.240603} {\bibfield  {journal}
  {\bibinfo  {journal} {Physical Review Letters}\ }\textbf {\bibinfo {volume}
  {102}},\ \bibinfo {pages} {240603} (\bibinfo {year} {2009})},\ \Eprint
  {https://arxiv.org/abs/0904.1926} {arXiv:0904.1926 [cond-mat,
  physics:quant-ph]} \BibitemShut {NoStop}%
\bibitem [{\citenamefont {M{\"u}ller-Hermes}\ \emph {et~al.}(2012)\citenamefont
  {M{\"u}ller-Hermes}, \citenamefont {Cirac},\ and\ \citenamefont
  {Ba{\~n}uls}}]{muellerhermes2012tensor}%
  \BibitemOpen
  \bibfield  {author} {\bibinfo {author} {\bibfnamefont {A.}~\bibnamefont
  {M{\"u}ller-Hermes}}, \bibinfo {author} {\bibfnamefont {J.~I.}\ \bibnamefont
  {Cirac}},\ and\ \bibinfo {author} {\bibfnamefont {M.~C.}\ \bibnamefont
  {Ba{\~n}uls}},\ }\bibfield  {title} {\bibinfo {title} {Tensor network
  techniques for the computation of dynamical observables in one-dimensional
  quantum spin systems},\ }\href
  {https://doi.org/10.1088/1367-2630/14/7/075003} {\bibfield  {journal}
  {\bibinfo  {journal} {New J. Phys.}\ }\textbf {\bibinfo {volume} {14}},\
  \bibinfo {pages} {075003} (\bibinfo {year} {2012})}\BibitemShut {NoStop}%
\bibitem [{\citenamefont {Hastings}\ and\ \citenamefont
  {Mahajan}(2015)}]{hastings2015connecting}%
  \BibitemOpen
  \bibfield  {author} {\bibinfo {author} {\bibfnamefont {M.~B.}\ \bibnamefont
  {Hastings}}\ and\ \bibinfo {author} {\bibfnamefont {R.}~\bibnamefont
  {Mahajan}},\ }\bibfield  {title} {\bibinfo {title} {Connecting entanglement
  in time and space: Improving the folding algorithm},\ }\href
  {https://doi.org/10.1103/PhysRevA.91.032306} {\bibfield  {journal} {\bibinfo
  {journal} {Phys. Rev. A}\ }\textbf {\bibinfo {volume} {91}},\ \bibinfo
  {pages} {032306} (\bibinfo {year} {2015})}\BibitemShut {NoStop}%
\bibitem [{\citenamefont {Lerose}\ \emph
  {et~al.}(2021{\natexlab{a}})\citenamefont {Lerose}, \citenamefont {Sonner},\
  and\ \citenamefont {Abanin}}]{lerose2021influence}%
  \BibitemOpen
  \bibfield  {author} {\bibinfo {author} {\bibfnamefont {A.}~\bibnamefont
  {Lerose}}, \bibinfo {author} {\bibfnamefont {M.}~\bibnamefont {Sonner}},\
  and\ \bibinfo {author} {\bibfnamefont {D.~A.}\ \bibnamefont {Abanin}},\
  }\bibfield  {title} {\bibinfo {title} {Influence matrix approach to many-body
  {F}loquet dynamics},\ }\href {https://doi.org/10.1103/PhysRevX.11.021040}
  {\bibfield  {journal} {\bibinfo  {journal} {Phys. Rev. X}\ }\textbf {\bibinfo
  {volume} {11}},\ \bibinfo {pages} {021040} (\bibinfo {year}
  {2021}{\natexlab{a}})}\BibitemShut {NoStop}%
\bibitem [{\citenamefont {Lerose}\ \emph {et~al.}(2023)\citenamefont {Lerose},
  \citenamefont {Sonner},\ and\ \citenamefont {Abanin}}]{lerose2023overcoming}%
  \BibitemOpen
  \bibfield  {author} {\bibinfo {author} {\bibfnamefont {A.}~\bibnamefont
  {Lerose}}, \bibinfo {author} {\bibfnamefont {M.}~\bibnamefont {Sonner}},\
  and\ \bibinfo {author} {\bibfnamefont {D.~A.}\ \bibnamefont {Abanin}},\
  }\bibfield  {title} {\bibinfo {title} {Overcoming the entanglement barrier in
  quantum many-body dynamics via space-time duality},\ }\href
  {https://doi.org/10.1103/PhysRevB.107.L060305} {\bibfield  {journal}
  {\bibinfo  {journal} {Phys. Rev. B}\ }\textbf {\bibinfo {volume} {107}},\
  \bibinfo {pages} {L060305} (\bibinfo {year} {2023})}\BibitemShut {NoStop}%
\bibitem [{\citenamefont {Feynman}\ and\ \citenamefont
  {Vernon}(1963)}]{feynman1963the}%
  \BibitemOpen
  \bibfield  {author} {\bibinfo {author} {\bibfnamefont {R.}~\bibnamefont
  {Feynman}}\ and\ \bibinfo {author} {\bibfnamefont {F.}~\bibnamefont
  {Vernon}},\ }\bibfield  {title} {\bibinfo {title} {The theory of a general
  quantum system interacting with a linear dissipative system},\ }\href
  {https://doi.org/https://doi.org/10.1016/0003-4916(63)90068-X} {\bibfield
  {journal} {\bibinfo  {journal} {Annals of Physics}\ }\textbf {\bibinfo
  {volume} {24}},\ \bibinfo {pages} {118} (\bibinfo {year} {1963})}\BibitemShut
  {NoStop}%
\bibitem [{\citenamefont {Leggett}\ \emph {et~al.}(1987)\citenamefont
  {Leggett}, \citenamefont {Chakravarty}, \citenamefont {Dorsey}, \citenamefont
  {Fisher}, \citenamefont {Garg},\ and\ \citenamefont
  {Zwerger}}]{leggett1987dynamics}%
  \BibitemOpen
  \bibfield  {author} {\bibinfo {author} {\bibfnamefont {A.~J.}\ \bibnamefont
  {Leggett}}, \bibinfo {author} {\bibfnamefont {S.}~\bibnamefont
  {Chakravarty}}, \bibinfo {author} {\bibfnamefont {A.~T.}\ \bibnamefont
  {Dorsey}}, \bibinfo {author} {\bibfnamefont {M.~P.~A.}\ \bibnamefont
  {Fisher}}, \bibinfo {author} {\bibfnamefont {A.}~\bibnamefont {Garg}},\ and\
  \bibinfo {author} {\bibfnamefont {W.}~\bibnamefont {Zwerger}},\ }\bibfield
  {title} {\bibinfo {title} {Dynamics of the dissipative two-state system},\
  }\href {https://doi.org/10.1103/RevModPhys.59.1} {\bibfield  {journal}
  {\bibinfo  {journal} {Rev. Mod. Phys.}\ }\textbf {\bibinfo {volume} {59}},\
  \bibinfo {pages} {1} (\bibinfo {year} {1987})}\BibitemShut {NoStop}%
\bibitem [{\citenamefont {Foligno}\ \emph {et~al.}(2023)\citenamefont
  {Foligno}, \citenamefont {Zhou},\ and\ \citenamefont
  {Bertini}}]{foligno2023temporal}%
  \BibitemOpen
  \bibfield  {author} {\bibinfo {author} {\bibfnamefont {A.}~\bibnamefont
  {Foligno}}, \bibinfo {author} {\bibfnamefont {T.}~\bibnamefont {Zhou}},\ and\
  \bibinfo {author} {\bibfnamefont {B.}~\bibnamefont {Bertini}},\ }\bibfield
  {title} {\bibinfo {title} {Temporal entanglement in chaotic quantum
  circuits},\ }\href {https://doi.org/10.1103/PhysRevX.13.041008} {\bibfield
  {journal} {\bibinfo  {journal} {Phys. Rev. X}\ }\textbf {\bibinfo {volume}
  {13}},\ \bibinfo {pages} {041008} (\bibinfo {year} {2023})}\BibitemShut
  {NoStop}%
\bibitem [{\citenamefont {Lerose}\ \emph
  {et~al.}(2021{\natexlab{b}})\citenamefont {Lerose}, \citenamefont {Sonner},\
  and\ \citenamefont {Abanin}}]{lerose2021scaling}%
  \BibitemOpen
  \bibfield  {author} {\bibinfo {author} {\bibfnamefont {A.}~\bibnamefont
  {Lerose}}, \bibinfo {author} {\bibfnamefont {M.}~\bibnamefont {Sonner}},\
  and\ \bibinfo {author} {\bibfnamefont {D.~A.}\ \bibnamefont {Abanin}},\
  }\bibfield  {title} {\bibinfo {title} {Scaling of temporal entanglement in
  proximity to integrability},\ }\href
  {https://doi.org/10.1103/PhysRevB.104.035137} {\bibfield  {journal} {\bibinfo
   {journal} {Phys. Rev. B}\ }\textbf {\bibinfo {volume} {104}},\ \bibinfo
  {pages} {035137} (\bibinfo {year} {2021}{\natexlab{b}})}\BibitemShut
  {NoStop}%
\bibitem [{\citenamefont {Klobas}\ \emph {et~al.}(2021)\citenamefont {Klobas},
  \citenamefont {Bertini},\ and\ \citenamefont {Piroli}}]{klobas2021exact}%
  \BibitemOpen
  \bibfield  {author} {\bibinfo {author} {\bibfnamefont {K.}~\bibnamefont
  {Klobas}}, \bibinfo {author} {\bibfnamefont {B.}~\bibnamefont {Bertini}},\
  and\ \bibinfo {author} {\bibfnamefont {L.}~\bibnamefont {Piroli}},\
  }\bibfield  {title} {\bibinfo {title} {Exact thermalization dynamics in the
  ``{R}ule 54'' quantum cellular automaton},\ }\href
  {https://doi.org/10.1103/PhysRevLett.126.160602} {\bibfield  {journal}
  {\bibinfo  {journal} {Phys. Rev. Lett.}\ }\textbf {\bibinfo {volume} {126}},\
  \bibinfo {pages} {160602} (\bibinfo {year} {2021})}\BibitemShut {NoStop}%
\bibitem [{\citenamefont {Klobas}\ and\ \citenamefont
  {Bertini}(2021)}]{klobas2021exactrelaxation}%
  \BibitemOpen
  \bibfield  {author} {\bibinfo {author} {\bibfnamefont {K.}~\bibnamefont
  {Klobas}}\ and\ \bibinfo {author} {\bibfnamefont {B.}~\bibnamefont
  {Bertini}},\ }\bibfield  {title} {\bibinfo {title} {Exact relaxation to
  {Gibbs} and non-equilibrium steady states in the quantum cellular automaton
  {Rule} 54},\ }\href {https://doi.org/10.21468/SciPostPhys.11.6.106}
  {\bibfield  {journal} {\bibinfo  {journal} {SciPost Phys.}\ }\textbf
  {\bibinfo {volume} {11}},\ \bibinfo {pages} {106} (\bibinfo {year}
  {2021})}\BibitemShut {NoStop}%
\bibitem [{\citenamefont {Giudice}\ \emph {et~al.}(2022)\citenamefont
  {Giudice}, \citenamefont {Giudici}, \citenamefont {Sonner}, \citenamefont
  {Thoenniss}, \citenamefont {Lerose}, \citenamefont {Abanin},\ and\
  \citenamefont {Piroli}}]{giudice2022temporal}%
  \BibitemOpen
  \bibfield  {author} {\bibinfo {author} {\bibfnamefont {G.}~\bibnamefont
  {Giudice}}, \bibinfo {author} {\bibfnamefont {G.}~\bibnamefont {Giudici}},
  \bibinfo {author} {\bibfnamefont {M.}~\bibnamefont {Sonner}}, \bibinfo
  {author} {\bibfnamefont {J.}~\bibnamefont {Thoenniss}}, \bibinfo {author}
  {\bibfnamefont {A.}~\bibnamefont {Lerose}}, \bibinfo {author} {\bibfnamefont
  {D.~A.}\ \bibnamefont {Abanin}},\ and\ \bibinfo {author} {\bibfnamefont
  {L.}~\bibnamefont {Piroli}},\ }\bibfield  {title} {\bibinfo {title} {Temporal
  entanglement, quasiparticles, and the role of interactions},\ }\href
  {https://doi.org/10.1103/PhysRevLett.128.220401} {\bibfield  {journal}
  {\bibinfo  {journal} {Phys. Rev. Lett.}\ }\textbf {\bibinfo {volume} {128}},\
  \bibinfo {pages} {220401} (\bibinfo {year} {2022})}\BibitemShut {NoStop}%
\bibitem [{\citenamefont {Bertini}\ \emph
  {et~al.}(2019{\natexlab{a}})\citenamefont {Bertini}, \citenamefont {Kos},\
  and\ \citenamefont {Prosen}}]{bertini2019entanglement}%
  \BibitemOpen
  \bibfield  {author} {\bibinfo {author} {\bibfnamefont {B.}~\bibnamefont
  {Bertini}}, \bibinfo {author} {\bibfnamefont {P.}~\bibnamefont {Kos}},\ and\
  \bibinfo {author} {\bibfnamefont {T.}~\bibnamefont {Prosen}},\ }\bibfield
  {title} {\bibinfo {title} {Entanglement spreading in a minimal model of
  maximal many-body quantum chaos},\ }\href
  {https://doi.org/10.1103/PhysRevX.9.021033} {\bibfield  {journal} {\bibinfo
  {journal} {Phys. Rev. X}\ }\textbf {\bibinfo {volume} {9}},\ \bibinfo {pages}
  {021033} (\bibinfo {year} {2019}{\natexlab{a}})}\BibitemShut {NoStop}%
\bibitem [{\citenamefont {Piroli}\ \emph {et~al.}(2020)\citenamefont {Piroli},
  \citenamefont {Bertini}, \citenamefont {Cirac},\ and\ \citenamefont
  {Prosen}}]{piroli2020exact}%
  \BibitemOpen
  \bibfield  {author} {\bibinfo {author} {\bibfnamefont {L.}~\bibnamefont
  {Piroli}}, \bibinfo {author} {\bibfnamefont {B.}~\bibnamefont {Bertini}},
  \bibinfo {author} {\bibfnamefont {J.~I.}\ \bibnamefont {Cirac}},\ and\
  \bibinfo {author} {\bibfnamefont {T.}~\bibnamefont {Prosen}},\ }\bibfield
  {title} {\bibinfo {title} {Exact dynamics in dual-unitary quantum circuits},\
  }\href {https://doi.org/10.1103/PhysRevB.101.094304} {\bibfield  {journal}
  {\bibinfo  {journal} {Phys. Rev. B}\ }\textbf {\bibinfo {volume} {101}},\
  \bibinfo {pages} {094304} (\bibinfo {year} {2020})}\BibitemShut {NoStop}%
\bibitem [{\citenamefont {Foligno}\ \emph {et~al.}(2025)\citenamefont
  {Foligno}, \citenamefont {Calabrese},\ and\ \citenamefont
  {Bertini}}]{foligno2025nonequilibrium}%
  \BibitemOpen
  \bibfield  {author} {\bibinfo {author} {\bibfnamefont {A.}~\bibnamefont
  {Foligno}}, \bibinfo {author} {\bibfnamefont {P.}~\bibnamefont {Calabrese}},\
  and\ \bibinfo {author} {\bibfnamefont {B.}~\bibnamefont {Bertini}},\
  }\bibfield  {title} {\bibinfo {title} {Nonequilibrium dynamics of charged
  dual-unitary circuits},\ }\href {https://doi.org/10.1103/PRXQuantum.6.010324}
  {\bibfield  {journal} {\bibinfo  {journal} {PRX Quantum}\ }\textbf {\bibinfo
  {volume} {6}},\ \bibinfo {pages} {010324} (\bibinfo {year}
  {2025})}\BibitemShut {NoStop}%
\bibitem [{\citenamefont {Segal}\ \emph {et~al.}(2010)\citenamefont {Segal},
  \citenamefont {Millis},\ and\ \citenamefont
  {Reichman}}]{segal2010numerically}%
  \BibitemOpen
  \bibfield  {author} {\bibinfo {author} {\bibfnamefont {D.}~\bibnamefont
  {Segal}}, \bibinfo {author} {\bibfnamefont {A.~J.}\ \bibnamefont {Millis}},\
  and\ \bibinfo {author} {\bibfnamefont {D.~R.}\ \bibnamefont {Reichman}},\
  }\bibfield  {title} {\bibinfo {title} {Numerically exact path-integral
  simulation of nonequilibrium quantum transport and dissipation},\ }\href
  {https://doi.org/10.1103/PhysRevB.82.205323} {\bibfield  {journal} {\bibinfo
  {journal} {Phys. Rev. B}\ }\textbf {\bibinfo {volume} {82}},\ \bibinfo
  {pages} {205323} (\bibinfo {year} {2010})}\BibitemShut {NoStop}%
\bibitem [{\citenamefont {Segal}\ \emph {et~al.}(2011)\citenamefont {Segal},
  \citenamefont {Millis},\ and\ \citenamefont
  {Reichman}}]{segal2011nonequilibrium}%
  \BibitemOpen
  \bibfield  {author} {\bibinfo {author} {\bibfnamefont {D.}~\bibnamefont
  {Segal}}, \bibinfo {author} {\bibfnamefont {A.~J.}\ \bibnamefont {Millis}},\
  and\ \bibinfo {author} {\bibfnamefont {D.~R.}\ \bibnamefont {Reichman}},\
  }\bibfield  {title} {\bibinfo {title} {Nonequilibrium transport in quantum
  impurity models: exact path integral simulations},\ }\href
  {https://doi.org/10.1039/C1CP20702D} {\bibfield  {journal} {\bibinfo
  {journal} {Phys. Chem. Chem. Phys.}\ }\textbf {\bibinfo {volume} {13}},\
  \bibinfo {pages} {14378} (\bibinfo {year} {2011})}\BibitemShut {NoStop}%
\bibitem [{\citenamefont {Magazz{\`u}}\ and\ \citenamefont
  {Grifoni}(2022)}]{magazzu2022feynman}%
  \BibitemOpen
  \bibfield  {author} {\bibinfo {author} {\bibfnamefont {L.}~\bibnamefont
  {Magazz{\`u}}}\ and\ \bibinfo {author} {\bibfnamefont {M.}~\bibnamefont
  {Grifoni}},\ }\bibfield  {title} {\bibinfo {title} {Feynman-vernon influence
  functional approach to quantum transport in interacting nanojunctions: An
  analytical hierarchical study},\ }\href
  {https://doi.org/10.1103/PhysRevB.105.125417} {\bibfield  {journal} {\bibinfo
   {journal} {Phys. Rev. B}\ }\textbf {\bibinfo {volume} {105}},\ \bibinfo
  {pages} {125417} (\bibinfo {year} {2022})}\BibitemShut {NoStop}%
\bibitem [{\citenamefont {Thoenniss}\ \emph
  {et~al.}(2023{\natexlab{a}})\citenamefont {Thoenniss}, \citenamefont
  {Lerose},\ and\ \citenamefont {Abanin}}]{thoenniss2023nonequilibrium}%
  \BibitemOpen
  \bibfield  {author} {\bibinfo {author} {\bibfnamefont {J.}~\bibnamefont
  {Thoenniss}}, \bibinfo {author} {\bibfnamefont {A.}~\bibnamefont {Lerose}},\
  and\ \bibinfo {author} {\bibfnamefont {D.~A.}\ \bibnamefont {Abanin}},\
  }\bibfield  {title} {\bibinfo {title} {Nonequilibrium quantum impurity
  problems via matrix-product states in the temporal domain},\ }\href
  {https://doi.org/10.1103/PhysRevB.107.195101} {\bibfield  {journal} {\bibinfo
   {journal} {Phys. Rev. B}\ }\textbf {\bibinfo {volume} {107}},\ \bibinfo
  {pages} {195101} (\bibinfo {year} {2023}{\natexlab{a}})}\BibitemShut
  {NoStop}%
\bibitem [{\citenamefont {Thoenniss}\ \emph
  {et~al.}(2023{\natexlab{b}})\citenamefont {Thoenniss}, \citenamefont
  {Sonner}, \citenamefont {Lerose},\ and\ \citenamefont
  {Abanin}}]{thoenniss2023efficient}%
  \BibitemOpen
  \bibfield  {author} {\bibinfo {author} {\bibfnamefont {J.}~\bibnamefont
  {Thoenniss}}, \bibinfo {author} {\bibfnamefont {M.}~\bibnamefont {Sonner}},
  \bibinfo {author} {\bibfnamefont {A.}~\bibnamefont {Lerose}},\ and\ \bibinfo
  {author} {\bibfnamefont {D.~A.}\ \bibnamefont {Abanin}},\ }\bibfield  {title}
  {\bibinfo {title} {Efficient method for quantum impurity problems out of
  equilibrium},\ }\href {https://doi.org/10.1103/PhysRevB.107.L201115}
  {\bibfield  {journal} {\bibinfo  {journal} {Phys. Rev. B}\ }\textbf {\bibinfo
  {volume} {107}},\ \bibinfo {pages} {L201115} (\bibinfo {year}
  {2023}{\natexlab{b}})}\BibitemShut {NoStop}%
\bibitem [{\citenamefont {Ng}\ \emph {et~al.}(2023)\citenamefont {Ng},
  \citenamefont {Park}, \citenamefont {Millis}, \citenamefont {Chan},\ and\
  \citenamefont {Reichman}}]{ng2023realtime}%
  \BibitemOpen
  \bibfield  {author} {\bibinfo {author} {\bibfnamefont {N.}~\bibnamefont
  {Ng}}, \bibinfo {author} {\bibfnamefont {G.}~\bibnamefont {Park}}, \bibinfo
  {author} {\bibfnamefont {A.~J.}\ \bibnamefont {Millis}}, \bibinfo {author}
  {\bibfnamefont {G.~K.-L.}\ \bibnamefont {Chan}},\ and\ \bibinfo {author}
  {\bibfnamefont {D.~R.}\ \bibnamefont {Reichman}},\ }\bibfield  {title}
  {\bibinfo {title} {Real-time evolution of anderson impurity models via tensor
  network influence functionals},\ }\href
  {https://doi.org/10.1103/PhysRevB.107.125103} {\bibfield  {journal} {\bibinfo
   {journal} {Phys. Rev. B}\ }\textbf {\bibinfo {volume} {107}},\ \bibinfo
  {pages} {125103} (\bibinfo {year} {2023})}\BibitemShut {NoStop}%
\bibitem [{\citenamefont {Chen}\ \emph
  {et~al.}(2024{\natexlab{a}})\citenamefont {Chen}, \citenamefont {Xu},\ and\
  \citenamefont {Guo}}]{chen2024grassmann}%
  \BibitemOpen
  \bibfield  {author} {\bibinfo {author} {\bibfnamefont {R.}~\bibnamefont
  {Chen}}, \bibinfo {author} {\bibfnamefont {X.}~\bibnamefont {Xu}},\ and\
  \bibinfo {author} {\bibfnamefont {C.}~\bibnamefont {Guo}},\ }\bibfield
  {title} {\bibinfo {title} {Grassmann time-evolving matrix product operators
  for quantum impurity models},\ }\href
  {https://doi.org/10.1103/PhysRevB.109.045140} {\bibfield  {journal} {\bibinfo
   {journal} {Phys. Rev. B}\ }\textbf {\bibinfo {volume} {109}},\ \bibinfo
  {pages} {045140} (\bibinfo {year} {2024}{\natexlab{a}})}\BibitemShut
  {NoStop}%
\bibitem [{\citenamefont {Chen}\ \emph
  {et~al.}(2024{\natexlab{b}})\citenamefont {Chen}, \citenamefont {Xu},\ and\
  \citenamefont {Guo}}]{chen2024realtime}%
  \BibitemOpen
  \bibfield  {author} {\bibinfo {author} {\bibfnamefont {R.}~\bibnamefont
  {Chen}}, \bibinfo {author} {\bibfnamefont {X.}~\bibnamefont {Xu}},\ and\
  \bibinfo {author} {\bibfnamefont {C.}~\bibnamefont {Guo}},\ }\bibfield
  {title} {\bibinfo {title} {Real-time impurity solver using grassmann
  time-evolving matrix product operators},\ }\href
  {https://doi.org/10.1103/PhysRevB.109.165113} {\bibfield  {journal} {\bibinfo
   {journal} {Phys. Rev. B}\ }\textbf {\bibinfo {volume} {109}},\ \bibinfo
  {pages} {165113} (\bibinfo {year} {2024}{\natexlab{b}})}\BibitemShut
  {NoStop}%
\bibitem [{\citenamefont {Park}\ \emph {et~al.}(2024)\citenamefont {Park},
  \citenamefont {Ng}, \citenamefont {Reichman},\ and\ \citenamefont
  {Chan}}]{park2024continuous}%
  \BibitemOpen
  \bibfield  {author} {\bibinfo {author} {\bibfnamefont {G.}~\bibnamefont
  {Park}}, \bibinfo {author} {\bibfnamefont {N.}~\bibnamefont {Ng}}, \bibinfo
  {author} {\bibfnamefont {D.~R.}\ \bibnamefont {Reichman}},\ and\ \bibinfo
  {author} {\bibfnamefont {G.~K.-L.}\ \bibnamefont {Chan}},\ }\bibfield
  {title} {\bibinfo {title} {Tensor network influence functionals in the
  continuous-time limit: Connections to quantum embedding, bath discretization,
  and higher-order time propagation},\ }\href
  {https://doi.org/10.1103/PhysRevB.110.045104} {\bibfield  {journal} {\bibinfo
   {journal} {Phys. Rev. B}\ }\textbf {\bibinfo {volume} {110}},\ \bibinfo
  {pages} {045104} (\bibinfo {year} {2024})}\BibitemShut {NoStop}%
\bibitem [{\citenamefont {Nayak}\ \emph {et~al.}(2025)\citenamefont {Nayak},
  \citenamefont {Thoenniss}, \citenamefont {Sonner}, \citenamefont {Abanin},\
  and\ \citenamefont {Werner}}]{nayak2025steadystate}%
  \BibitemOpen
  \bibfield  {author} {\bibinfo {author} {\bibfnamefont {M.}~\bibnamefont
  {Nayak}}, \bibinfo {author} {\bibfnamefont {J.}~\bibnamefont {Thoenniss}},
  \bibinfo {author} {\bibfnamefont {M.}~\bibnamefont {Sonner}}, \bibinfo
  {author} {\bibfnamefont {D.~A.}\ \bibnamefont {Abanin}},\ and\ \bibinfo
  {author} {\bibfnamefont {P.}~\bibnamefont {Werner}},\ }\bibfield  {title}
  {\bibinfo {title} {Steady-state dynamical mean field theory based on
  influence functional matrix product states},\ }\href
  {https://doi.org/10.1103/xsbn-jk16} {\bibfield  {journal} {\bibinfo
  {journal} {Phys. Rev. B}\ }\textbf {\bibinfo {volume} {112}},\ \bibinfo
  {pages} {035103} (\bibinfo {year} {2025})}\BibitemShut {NoStop}%
\bibitem [{\citenamefont {Sonner}\ \emph {et~al.}(2025)\citenamefont {Sonner},
  \citenamefont {Link},\ and\ \citenamefont {Abanin}}]{sonner2025semigroup}%
  \BibitemOpen
  \bibfield  {author} {\bibinfo {author} {\bibfnamefont {M.}~\bibnamefont
  {Sonner}}, \bibinfo {author} {\bibfnamefont {V.}~\bibnamefont {Link}},\ and\
  \bibinfo {author} {\bibfnamefont {D.~A.}\ \bibnamefont {Abanin}},\ }\bibfield
   {title} {\bibinfo {title} {Semigroup influence matrices for nonequilibrium
  quantum impurity models},\ }\href {https://doi.org/10.1103/5gfn-l7w7}
  {\bibfield  {journal} {\bibinfo  {journal} {Phys. Rev. Lett.}\ }\textbf
  {\bibinfo {volume} {135}},\ \bibinfo {pages} {170402} (\bibinfo {year}
  {2025})}\BibitemShut {NoStop}%
\bibitem [{\citenamefont {Vilkoviskiy}\ \emph {et~al.}(2025)\citenamefont
  {Vilkoviskiy}, \citenamefont {Sonner}, \citenamefont {Huang}, \citenamefont
  {Ho}, \citenamefont {Lerose},\ and\ \citenamefont
  {Abanin}}]{vilkoviskiy2025temporal}%
  \BibitemOpen
  \bibfield  {author} {\bibinfo {author} {\bibfnamefont {I.}~\bibnamefont
  {Vilkoviskiy}}, \bibinfo {author} {\bibfnamefont {M.}~\bibnamefont {Sonner}},
  \bibinfo {author} {\bibfnamefont {Q.~C.}\ \bibnamefont {Huang}}, \bibinfo
  {author} {\bibfnamefont {W.~W.}\ \bibnamefont {Ho}}, \bibinfo {author}
  {\bibfnamefont {A.}~\bibnamefont {Lerose}},\ and\ \bibinfo {author}
  {\bibfnamefont {D.~A.}\ \bibnamefont {Abanin}},\ }\href
  {https://arxiv.org/abs/2511.03846} {\bibinfo {title} {Temporal entanglement
  transition in chaotic quantum many-body dynamics}} (\bibinfo {year} {2025}),\
  \Eprint {https://arxiv.org/abs/2511.03846} {arXiv:2511.03846 [quant-ph]}
  \BibitemShut {NoStop}%
\bibitem [{\citenamefont {O'Donovan}\ \emph {et~al.}(2026)\citenamefont
  {O'Donovan}, \citenamefont {Dowling}, \citenamefont {Modi},\ and\
  \citenamefont {Mitchison}}]{odonovan2026diagnosing}%
  \BibitemOpen
  \bibfield  {author} {\bibinfo {author} {\bibfnamefont {P.}~\bibnamefont
  {O'Donovan}}, \bibinfo {author} {\bibfnamefont {N.}~\bibnamefont {Dowling}},
  \bibinfo {author} {\bibfnamefont {K.}~\bibnamefont {Modi}},\ and\ \bibinfo
  {author} {\bibfnamefont {M.~T.}\ \bibnamefont {Mitchison}},\ }\bibfield
  {title} {\bibinfo {title} {Diagnosing chaos with projected ensembles of
  process tensors},\ }\href {https://doi.org/10.1103/fgc4-hgk1} {\bibfield
  {journal} {\bibinfo  {journal} {PRX Quantum}\ }\textbf {\bibinfo {volume}
  {7}},\ \bibinfo {pages} {020322} (\bibinfo {year} {2026})}\BibitemShut
  {NoStop}%
\bibitem [{\citenamefont {Schuch}\ \emph {et~al.}(2008)\citenamefont {Schuch},
  \citenamefont {Wolf}, \citenamefont {Verstraete},\ and\ \citenamefont
  {Cirac}}]{schuch2008entropy}%
  \BibitemOpen
  \bibfield  {author} {\bibinfo {author} {\bibfnamefont {N.}~\bibnamefont
  {Schuch}}, \bibinfo {author} {\bibfnamefont {M.~M.}\ \bibnamefont {Wolf}},
  \bibinfo {author} {\bibfnamefont {F.}~\bibnamefont {Verstraete}},\ and\
  \bibinfo {author} {\bibfnamefont {J.~I.}\ \bibnamefont {Cirac}},\ }\bibfield
  {title} {\bibinfo {title} {Entropy scaling and simulability by matrix product
  states},\ }\href {https://doi.org/10.1103/PhysRevLett.100.030504} {\bibfield
  {journal} {\bibinfo  {journal} {Phys. Rev. Lett.}\ }\textbf {\bibinfo
  {volume} {100}},\ \bibinfo {pages} {030504} (\bibinfo {year}
  {2008})}\BibitemShut {NoStop}%
\bibitem [{\citenamefont {Carignano}\ \emph {et~al.}(2024)\citenamefont
  {Carignano}, \citenamefont {Ramos~Marim{\'o}n},\ and\ \citenamefont
  {Tagliacozzo}}]{carignano2023temporal}%
  \BibitemOpen
  \bibfield  {author} {\bibinfo {author} {\bibfnamefont {S.}~\bibnamefont
  {Carignano}}, \bibinfo {author} {\bibfnamefont {C.}~\bibnamefont
  {Ramos~Marim{\'o}n}},\ and\ \bibinfo {author} {\bibfnamefont
  {L.}~\bibnamefont {Tagliacozzo}},\ }\bibfield  {title} {\bibinfo {title}
  {Temporal entropy and the complexity of computing the expectation value of
  local operators after a quench},\ }\href
  {https://doi.org/10.1103/PhysRevResearch.6.033021} {\bibfield  {journal}
  {\bibinfo  {journal} {Phys. Rev. Res.}\ }\textbf {\bibinfo {volume} {6}},\
  \bibinfo {pages} {033021} (\bibinfo {year} {2024})}\BibitemShut {NoStop}%
\bibitem [{\citenamefont {Carignano}(2026)}]{carignano2026itransverse}%
  \BibitemOpen
  \bibfield  {author} {\bibinfo {author} {\bibfnamefont {S.}~\bibnamefont
  {Carignano}},\ }\bibfield  {title} {\bibinfo {title} {The {ITransverse}.jl
  library for transverse tensor network contractions},\ }\href
  {https://doi.org/10.21468/SciPostPhysCodeb.63} {\bibfield  {journal}
  {\bibinfo  {journal} {SciPost Phys. Codebases}\ }\textbf {\bibinfo {volume}
  {63}},\ \bibinfo {pages} {63} (\bibinfo {year} {2026})},\ \Eprint
  {https://arxiv.org/abs/2509.03699} {arXiv:2509.03699 [quant-ph]} \BibitemShut
  {NoStop}%
\bibitem [{\citenamefont {Carignano}\ \emph {et~al.}(2025)\citenamefont
  {Carignano}, \citenamefont {Lami}, \citenamefont {Nardis},\ and\
  \citenamefont {Tagliacozzo}}]{carignano2025overcoming}%
  \BibitemOpen
  \bibfield  {author} {\bibinfo {author} {\bibfnamefont {S.}~\bibnamefont
  {Carignano}}, \bibinfo {author} {\bibfnamefont {G.}~\bibnamefont {Lami}},
  \bibinfo {author} {\bibfnamefont {J.~D.}\ \bibnamefont {Nardis}},\ and\
  \bibinfo {author} {\bibfnamefont {L.}~\bibnamefont {Tagliacozzo}},\ }\href
  {https://arxiv.org/abs/2505.09714} {\bibinfo {title} {Overcoming the
  entanglement barrier with sampled tensor networks}} (\bibinfo {year}
  {2025}),\ \Eprint {https://arxiv.org/abs/2505.09714} {arXiv:2505.09714
  [quant-ph]} \BibitemShut {NoStop}%
\bibitem [{\citenamefont {Yadalam}\ and\ \citenamefont
  {Mitchison}(2026)}]{yadalam2026process}%
  \BibitemOpen
  \bibfield  {author} {\bibinfo {author} {\bibfnamefont {H.~K.}\ \bibnamefont
  {Yadalam}}\ and\ \bibinfo {author} {\bibfnamefont {M.~T.}\ \bibnamefont
  {Mitchison}},\ }\href {https://arxiv.org/abs/2603.28894} {\bibinfo {title}
  {Process-tensor approach to full counting statistics of charge transport in
  quantum many-body circuits}} (\bibinfo {year} {2026}),\ \Eprint
  {https://arxiv.org/abs/2603.28894} {arXiv:2603.28894 [quant-ph]} \BibitemShut
  {NoStop}%
\bibitem [{\citenamefont {Mirsky}(1960)}]{mirsky1960symmetric}%
  \BibitemOpen
  \bibfield  {author} {\bibinfo {author} {\bibfnamefont {L.}~\bibnamefont
  {Mirsky}},\ }\bibfield  {title} {\bibinfo {title} {Symmetric gauge functions
  and unitarily invariant norms},\ }\href@noop {} {\bibfield  {journal}
  {\bibinfo  {journal} {The quarterly journal of mathematics}\ }\textbf
  {\bibinfo {volume} {11}},\ \bibinfo {pages} {50} (\bibinfo {year}
  {1960})}\BibitemShut {NoStop}%
\bibitem [{\citenamefont {Nielsen}\ and\ \citenamefont
  {Chuang}(2010)}]{nielsen2010quantum}%
  \BibitemOpen
  \bibfield  {author} {\bibinfo {author} {\bibfnamefont {M.}~\bibnamefont
  {Nielsen}}\ and\ \bibinfo {author} {\bibfnamefont {I.}~\bibnamefont
  {Chuang}},\ }\href {https://books.google.co.uk/books?id=-s4DEy7o-a0C} {\emph
  {\bibinfo {title} {Quantum Computation and Quantum Information: 10th
  Anniversary Edition}}}\ (\bibinfo  {publisher} {Cambridge University Press},\
  \bibinfo {year} {2010})\BibitemShut {NoStop}%
\bibitem [{\citenamefont {Bertini}\ \emph
  {et~al.}(2019{\natexlab{b}})\citenamefont {Bertini}, \citenamefont {Kos},\
  and\ \citenamefont {Prosen}}]{bertini2019exact}%
  \BibitemOpen
  \bibfield  {author} {\bibinfo {author} {\bibfnamefont {B.}~\bibnamefont
  {Bertini}}, \bibinfo {author} {\bibfnamefont {P.}~\bibnamefont {Kos}},\ and\
  \bibinfo {author} {\bibfnamefont {T.}~\bibnamefont {Prosen}},\ }\bibfield
  {title} {\bibinfo {title} {Exact correlation functions for dual-unitary
  lattice models in $1+1$ dimensions},\ }\href
  {https://doi.org/10.1103/PhysRevLett.123.210601} {\bibfield  {journal}
  {\bibinfo  {journal} {Phys. Rev. Lett.}\ }\textbf {\bibinfo {volume} {123}},\
  \bibinfo {pages} {210601} (\bibinfo {year} {2019}{\natexlab{b}})}\BibitemShut
  {NoStop}%
\bibitem [{Note50()}]{Note50}%
  \BibitemOpen
  \bibinfo {note} {See Supplemental Material at [URL will be inserted by
  publisher] for the derivation of the entropy bounds for dual-unitary
  circuits, the proof of the conjectured form of $p_k$ for random dual-unitary
  circuits, the membrane calculation away from the critical point, additional
  numerical data for the decay of $p_k$ and the singular values of the reduced
  transition matrix, and the explicit gate matrices used in the
  numerics.}\BibitemShut {Stop}%
\bibitem [{\citenamefont {Jonay}\ and\ \citenamefont
  {Zhou}(2024)}]{jonay2024physical}%
  \BibitemOpen
  \bibfield  {author} {\bibinfo {author} {\bibfnamefont {C.}~\bibnamefont
  {Jonay}}\ and\ \bibinfo {author} {\bibfnamefont {T.}~\bibnamefont {Zhou}},\
  }\bibfield  {title} {\bibinfo {title} {Physical theory of two-stage
  thermalization},\ }\href {https://doi.org/10.1103/PhysRevB.110.L020306}
  {\bibfield  {journal} {\bibinfo  {journal} {Phys. Rev. B}\ }\textbf {\bibinfo
  {volume} {110}},\ \bibinfo {pages} {L020306} (\bibinfo {year}
  {2024})}\BibitemShut {NoStop}%
\bibitem [{\citenamefont {Jonay}\ \emph {et~al.}(2025)\citenamefont {Jonay},
  \citenamefont {Li},\ and\ \citenamefont {Zhou}}]{jonay2025twostage}%
  \BibitemOpen
  \bibfield  {author} {\bibinfo {author} {\bibfnamefont {C.}~\bibnamefont
  {Jonay}}, \bibinfo {author} {\bibfnamefont {C.}~\bibnamefont {Li}},\ and\
  \bibinfo {author} {\bibfnamefont {T.}~\bibnamefont {Zhou}},\ }\bibfield
  {title} {\bibinfo {title} {Two-stage relaxation of operators through domain
  wall and magnon dynamics},\ }\href
  {https://doi.org/10.1103/PhysRevB.111.224304} {\bibfield  {journal} {\bibinfo
   {journal} {Phys. Rev. B}\ }\textbf {\bibinfo {volume} {111}},\ \bibinfo
  {pages} {224304} (\bibinfo {year} {2025})}\BibitemShut {NoStop}%
\bibitem [{\citenamefont {Bertini}\ and\ \citenamefont
  {Piroli}(2020)}]{bertini2020scrambling}%
  \BibitemOpen
  \bibfield  {author} {\bibinfo {author} {\bibfnamefont {B.}~\bibnamefont
  {Bertini}}\ and\ \bibinfo {author} {\bibfnamefont {L.}~\bibnamefont
  {Piroli}},\ }\bibfield  {title} {\bibinfo {title} {Scrambling in random
  unitary circuits: Exact results},\ }\href
  {https://doi.org/10.1103/PhysRevB.102.064305} {\bibfield  {journal} {\bibinfo
   {journal} {Phys. Rev. B}\ }\textbf {\bibinfo {volume} {102}},\ \bibinfo
  {pages} {064305} (\bibinfo {year} {2020})}\BibitemShut {NoStop}%
\bibitem [{\citenamefont {Foligno}\ and\ \citenamefont
  {Bertini}(2023)}]{foligno2022growth}%
  \BibitemOpen
  \bibfield  {author} {\bibinfo {author} {\bibfnamefont {A.}~\bibnamefont
  {Foligno}}\ and\ \bibinfo {author} {\bibfnamefont {B.}~\bibnamefont
  {Bertini}},\ }\bibfield  {title} {\bibinfo {title} {Growth of entanglement of
  generic states under dual-unitary dynamics},\ }\href
  {https://doi.org/10.1103/PhysRevB.107.174311} {\bibfield  {journal} {\bibinfo
   {journal} {Phys. Rev. B}\ }\textbf {\bibinfo {volume} {107}},\ \bibinfo
  {pages} {174311} (\bibinfo {year} {2023})}\BibitemShut {NoStop}%
\bibitem [{Note1()}]{Note1}%
  \BibitemOpen
  \bibinfo {note} {The entangling power $p$ measures the ability of the local
  gate to entangle two qubits in a random product state~\cite
  {zanardi2000entangling}.}\BibitemShut {Stop}%
\bibitem [{\citenamefont {Jonay}\ \emph {et~al.}(2018)\citenamefont {Jonay},
  \citenamefont {Huse},\ and\ \citenamefont {Nahum}}]{jonay2018coarsegrained}%
  \BibitemOpen
  \bibfield  {author} {\bibinfo {author} {\bibfnamefont {C.}~\bibnamefont
  {Jonay}}, \bibinfo {author} {\bibfnamefont {D.~A.}\ \bibnamefont {Huse}},\
  and\ \bibinfo {author} {\bibfnamefont {A.}~\bibnamefont {Nahum}},\ }\bibfield
   {title} {\bibinfo {title} {Coarse-grained dynamics of operator and state
  entanglement},\ }\Eprint {https://arxiv.org/abs/arXiv:1803.00089}
  {arXiv:1803.00089}  (\bibinfo {year} {2018})\BibitemShut {NoStop}%
\bibitem [{\citenamefont {Zhou}\ and\ \citenamefont
  {Nahum}(2020)}]{zhou2020entanglement}%
  \BibitemOpen
  \bibfield  {author} {\bibinfo {author} {\bibfnamefont {T.}~\bibnamefont
  {Zhou}}\ and\ \bibinfo {author} {\bibfnamefont {A.}~\bibnamefont {Nahum}},\
  }\bibfield  {title} {\bibinfo {title} {Entanglement membrane in chaotic
  many-body systems},\ }\href {https://doi.org/10.1103/PhysRevX.10.031066}
  {\bibfield  {journal} {\bibinfo  {journal} {Phys. Rev. X}\ }\textbf {\bibinfo
  {volume} {10}},\ \bibinfo {pages} {031066} (\bibinfo {year}
  {2020})}\BibitemShut {NoStop}%
\bibitem [{\citenamefont {Carignano}\ and\ \citenamefont
  {Tagliacozzo}(2025)}]{carignano2025loschmidt}%
  \BibitemOpen
  \bibfield  {author} {\bibinfo {author} {\bibfnamefont {S.}~\bibnamefont
  {Carignano}}\ and\ \bibinfo {author} {\bibfnamefont {L.}~\bibnamefont
  {Tagliacozzo}},\ }\href {https://doi.org/10.48550/arXiv.2405.14706} {\bibinfo
  {title} {Loschmidt echo, emerging dual unitarity and scaling of generalized
  temporal entropies after quenches to the critical point}} (\bibinfo {year}
  {2025}),\ \Eprint {https://arxiv.org/abs/2405.14706} {arXiv:2405.14706
  [cond-mat]} \BibitemShut {NoStop}%
\bibitem [{\citenamefont {Zanardi}\ \emph {et~al.}(2000)\citenamefont
  {Zanardi}, \citenamefont {Zalka},\ and\ \citenamefont
  {Faoro}}]{zanardi2000entangling}%
  \BibitemOpen
  \bibfield  {author} {\bibinfo {author} {\bibfnamefont {P.}~\bibnamefont
  {Zanardi}}, \bibinfo {author} {\bibfnamefont {C.}~\bibnamefont {Zalka}},\
  and\ \bibinfo {author} {\bibfnamefont {L.}~\bibnamefont {Faoro}},\ }\bibfield
   {title} {\bibinfo {title} {Entangling power of quantum evolutions},\ }\href
  {https://doi.org/10.1103/PhysRevA.62.030301} {\bibfield  {journal} {\bibinfo
  {journal} {Phys. Rev. A}\ }\textbf {\bibinfo {volume} {62}},\ \bibinfo
  {pages} {030301} (\bibinfo {year} {2000})}\BibitemShut {NoStop}%
\bibitem [{\citenamefont {Suzuki}\ \emph {et~al.}(2025)\citenamefont {Suzuki},
  \citenamefont {Katsura}, \citenamefont {Mitsuhashi}, \citenamefont {Soejima},
  \citenamefont {Eisert},\ and\ \citenamefont {Yoshioka}}]{suzuki2025global}%
  \BibitemOpen
  \bibfield  {author} {\bibinfo {author} {\bibfnamefont {R.}~\bibnamefont
  {Suzuki}}, \bibinfo {author} {\bibfnamefont {H.}~\bibnamefont {Katsura}},
  \bibinfo {author} {\bibfnamefont {Y.}~\bibnamefont {Mitsuhashi}}, \bibinfo
  {author} {\bibfnamefont {T.}~\bibnamefont {Soejima}}, \bibinfo {author}
  {\bibfnamefont {J.}~\bibnamefont {Eisert}},\ and\ \bibinfo {author}
  {\bibfnamefont {N.}~\bibnamefont {Yoshioka}},\ }\href
  {https://arxiv.org/abs/2410.24127} {\bibinfo {title} {More global randomness
  from less random local gates}} (\bibinfo {year} {2025}),\ \Eprint
  {https://arxiv.org/abs/2410.24127} {arXiv:2410.24127 [quant-ph]} \BibitemShut
  {NoStop}%
\end{thebibliography}%
